\documentclass[nofootinbib,aps,11pt]{revtex4}
\pdfoutput=1
\usepackage{graphicx}
\usepackage{cancel}
\usepackage{amssymb}
\usepackage{textcomp}
\usepackage{amsmath}
\usepackage{bm}
\usepackage{times}
\usepackage{epsfig}
\usepackage{color}
\def\beq{\begin{equation}}
\def\eeq{\end{equation}}
\def\bea{\begin{eqnarray}}
\def\eea{\end{eqnarray}}

	\newcommand{\abs}[1]{ \mathopen{}\left| {#1}\right| }

\newcommand{\ZBL}{Z_{BL}}
\newcommand{\mzbl}{m_{Z_{BL}}}

\def\mm{\mathcal{M}}

\begin{document}

\title{\Large  {{\bf{THE MINIMAL THEORY FOR R-PARITY VIOLATION AT THE LHC}}}}
\bigskip
\author{\large Pavel Fileviez P{\'e}rez}
\address{Center for Cosmology and Particle Physics (CCPP) \\
New York University, 4 Washington Place, NY 10003, USA }
\author{\large Sogee Spinner}
\address{International School for Advanced Studies (SISSA) \\ Via Bonomea 265, 34136 Trieste, Italy}
\date{\today}

\begin{abstract}
We investigate the simplest gauge theory for spontaneous R-parity breaking and its testability at the LHC. 
This theory, based on a local B-L gauge symmetry, can be considered as the simplest framework for understanding the origin of the R-parity 
violating interactions, giving rise to potential lepton number violating signals and suppressed baryon number violating operators.
The full spectrum of the theory and the constraints coming from neutrino masses are analyzed in detail.
We discuss the proton decay issue and the possible dark matter candidates. In order to assess the testability of the theory 
we study the properties of the new gauge boson, the neutralino decays and the main production channels for the charged sleptons at the LHC. 
We find that final states with four charged leptons, three of them with the same-sign,  and four jets are the most striking signals for the testability 
of the lepton number violation associated with spontaneous R-parity violation at the LHC.
\end{abstract}

\maketitle


\section{Introduction}
The Large Hadron Collider (LHC) will hopefully soon discover the underlying theory for the TeV scale and might allow us to understand a more fundamental law of nature.
For more than three decades the idea of Supersymmetry has attracted the attention of many experts in the particle physics community 
and the minimal supersymmetric extension of the Standard Model~\cite{MSSM1,MSSM2,MSSM3} (MSSM) is still considered one of the most appealing candidates for the theory of 
particle physics at the TeV scale. It is well known that the MSSM provides an understanding of why the SM-like Higgs boson is light, contains a 
cold dark matter candidate, allows for the unification of the gauge couplings and allows for the mechanism 
of electroweak baryogenesis to explain the baryon asymmetry in the universe.

There are several open issues in the MSSM, one of them being the origin of the discrete symmetry R-parity~\cite{R1,R2}.
This symmetry plays a major role in the MSSM and it is defined as $R=(-1)^{3(B-L)+2S}$, where $B$, $L$ and $S$ stand for baryon number, 
lepton number and spin, respectively. In many MSSM studies it is assumed that R-parity is conserved or explicitly broken 
without understanding the origin of this symmetry. However, the fate of R-parity 
is crucial for the discovery of supersymmetry since, as is well known, R-parity conservation give rise to channels with multi-jets, multi-leptons and missing 
energy at the LHC, while signatures of broken R-parity are multi-leptons, multi-jets, and missing energy due to the SM neutrinos only.

The simplest and most elegant framework for the origin of R-parity is based on local B-L symmetry. This connection was explored for the first time in Ref.~\cite{Hayashi}, and in Ref.~\cite{Mohapatra} a simpler scenario was studied.
See also Ref.~\cite{Martin} for a complete discussion of how to gauge R-parity.\footnote{It is important to mention that the breaking of B-L in the context of the MSSM was studied for the first time in Ref.~\cite{AM}. 
See also Refs.~\cite{Martin2,Masiero,Goran1,Goran2} for the study of R-parity in other models.} 
Recently, we have investigated the simplest B-L models in 
Refs.~\cite{Rp1,Rp2,Rp5,Rp6} and found the following main result

\begin{center}
{\it{The simplest theories based on local B-L make the following prediction:  \\
R-parity must be spontaneously broken at the TeV scale and  \\ one expects to observe lepton number violation at the LHC ! }}
\end{center}
In this letter we study in detail the theory proposed in Ref.~\cite{Rp2} which can be considered as the simplest gauge theory for R-parity violation.
In this context the only way to break local B-L and obtain the MSSM after symmetry breaking is to give a vacuum expectation value to one of 
the right-handed sneutrino required by anomaly cancellation. One of the most 
important features of this theory is that the B-L and R-parity breaking scales are determined by the soft supersymmetric breaking scale. 
This idea was studied for the first time in Ref.~\cite{Rp1} which defined the simplest left-right symmetric model.

We review the theory and symmetry breaking mechanism in Sections~II and III. The full spectrum of the theory~\cite{Rp2} is discussed in Section~IV and the constraints coming from neutrino masses in Section~VI. We discuss the proton decay issue and the possible dark matter candidates in Section~V. In order to understand the testability of the theory 
we study the properties of the new gauge boson, the neutralino decays in Section~VII and in Section~VIII the main production channels for the charged sleptons at the Large Hadron Collider. We find that the channels with four charged leptons (three with the same electric charge) and four jets give us the most striking signals 
for the testability of lepton number violation at the LHC.

\section{The Minimal Gauge Theory for Spontaneous R-parity Violation}
The simplest gauge theory for spontaneous R-parity breaking was proposed in Ref.~\cite{Rp2}. In this context one can understand dynamically the origin of the 
R-parity violating terms in the MSSM. Here we discuss the structure of the theory and the full spectrum.
\begin{itemize}
\item {\it \underline{Gauge Symmetry and Matter Fields}}: This theory is based on the gauge group $$SU(3)_C \otimes SU(2)_L \otimes U(1)_Y \otimes U(1)_{B-L},$$ 
and the different matter chiral superfields are given by  
\begin{equation}
\hat{Q} = \left(
\begin{array} {c}
\hat{u} \\ \hat{d}
\end{array}
\right) \ \sim \ (2,1/3,1/3),
\ \ 
\hat{L} = \left(
\begin{array} {c}
 \hat{\nu} \\ \hat{e}
\end{array}
\right) \ \sim \ (2,-1,-1),
\end{equation}
\begin{equation}
\hat{u}^c \ \sim \ (1,-4/3,-1/3),
\ \ 
\hat{d}^c \ \sim \ (1,2/3,-1/3),
\ \ 
\hat{e}^c \ \sim \ (1,2,1).
\end{equation}
In order to cancel the $B-L$ anomalies one introduces three chiral superfields for the right-handed neutrinos:
\begin{equation}
\hat{\nu}^c \ \sim \ (1,0,1).
\end{equation}
\item {\it \underline{Higgs Sector}}: The Higgs sector is composed of two Higgs chiral superfields as in the MSSM 
\begin{equation}
\hat{H}_u = \left(
\begin{array} {c}
	\hat{H}_u^+
\\
	\hat{H}_u^0
\end{array}
\right) \ \sim \ (2, 1, 0),
\ \ \
\hat{H}_d = \left(
\begin{array} {c}
	\hat{H}_d^0
\\
	\hat{H}_d^-
\end{array}
\right) \ \sim \ (2, -1, 0).
\end{equation}

\item {\it \underline{Interactions}:} With this field content the superpotential reads as
\begin{equation}
{\cal W}_{BL}={\cal W}_{MSSM} \ + \ Y_\nu \ \hat{L}^T \ i \sigma_2 \ \hat{H}_u \ \hat{\nu}^c,
\end{equation}
where
\begin{eqnarray}
	{\cal W}_{MSSM} &=& Y_u \ \hat{Q}^T \ i \sigma_2 \ \hat{H}_u \ \hat{u}^c
\ + \ Y_d \ \hat{Q}^T \ i \sigma_2 \ \hat{H}_d \ \hat{d}^c
\ + \ Y_e \ \hat{L}^T \ i \sigma_2 \ \hat{H}_d \  \hat{e}^c
\ + \ \mu \ \hat{H}_u^T \ i \sigma_2 \  \hat{H}_d. \nonumber \\
\end{eqnarray}
In addition to the superpotential, the model is also specified by the soft terms:
\begin{eqnarray}
\nonumber
V_{soft} & = & m_{\tilde \nu^c}^2 \abs{\tilde{\nu}^c}^2 \ + \ m_{\tilde L}^2 \ \abs{\tilde L}^2 \ + \ m_{\tilde e^c}^2 \ \abs{\tilde e^c}^2
\ + \ m_{H_u}^2 \abs{H_u}^2 + m_{H_d}^2 \abs{H_d}^2 \ + \ \left( \frac{1}{2} M_{BL} \tilde{B^{'}} \tilde{B^{'}} \right.
  	\nonumber
	\\
	& + &  \left. A_\nu \ \tilde{L}^T \ i \sigma_2 \ H_u \ \tilde{\nu}^c  \  + \  B\mu \ H_u^T \ i \sigma_2 \ H_d
		\ + \  \mathrm{h.c.} \right) \ + \ V_{soft}^{MSSM},
\label{soft}
\end{eqnarray}
where the terms not shown here correspond to terms in the soft MSSM potential.

Since we have a new gauge symmetry in the theory we need to modify the kinetic terms for all MSSM matter superfields, 
and include the kinetic term for right-handed neutrino superfields
\begin{equation}
{\cal{L}}_{Kin} (\nu^c) = \int d^2 \theta d^2 \bar{\theta}  \   (\hat{\nu}^c)^\dagger  e^{g_{BL} \hat{V}_{BL}}  \hat{\nu}^c.
\end{equation}   
Here $\hat{V}_{BL}$ is the B-L vector superfield. Using these interactions we can study the full spectrum of the theory.

\end{itemize}

\section{Electroweak and B-L Symmetry Breaking}
As in the MSSM, electroweak symmetry is broken by the vevs of $H_u^0$ and $H_d^0$, while $U(1)_{B-L}$ 
is broken due to the vev of right-handed sneutrinos. Notice that this is the only field which can break local $B-L$ and 
give mass to the new neutral gauge boson in the theory. Therefore, the theory predicts spontaneous R-parity violation.
It is important to mention that the $B-L$ and R-parity breaking scales are determined by the soft supersymmetric breaking scale, and one must 
expect lepton number violation at the LHC. 

The neutral fields are defined as
\begin{eqnarray}
H_u^0 &=& \frac{1}{\sqrt{2}} \left(  v_u \ + \ h_u \right) \ + \ \frac{i}{\sqrt{2}} A_u, \\
H_d^0 &=& \frac{1}{\sqrt{2}} \left(  v_d \ + \ h_d \right) \ + \ \frac{i}{\sqrt{2}} A_d, \\
\tilde{\nu}^i &=& \frac{1}{\sqrt{2}} \left(  v_{L}^{i} \ + \  h_L^{i}   \right) \ + \ \frac{i}{\sqrt{2}} A_L^{i} , \\
\tilde{\nu}^c_i &=& \frac{1}{\sqrt{2}} \left(  v_{R}^{i} \ + \  h_R^{i}   \right) \ + \ \frac{i} {\sqrt{2}} A_R^{i}, 
\end{eqnarray}
and the relevant scalar potential reads as
\begin{eqnarray}
V &=& V_F \ + \ V_D \ + \ V_{soft}, \\
V_F &=&  |\mu|^2 |H_u^0|^2 \ + \  | - \mu H_d^0 + \tilde{\nu}_i Y_\nu^{ij} \tilde{\nu}^c_j |^2 \ + \ \sum_{i}  | Y_\nu^{ij} \tilde{\nu}^c_j|^2 |H_u^0|^2  \ + \ \sum_{j}  |\tilde{\nu}_i   Y_\nu^{ij}|^2 |H_u^0|^2, \\
V_D &=& \frac{(g_1^2 + g_2^2)}{8} \left(  |H_u^0|^2 - |H_d^0|^2 - \sum_{i} |\tilde{\nu}_i|^2 \right)^2 \ + \ \frac{g_{BL}^2}{8} \left( \sum_{i} ( |\tilde{\nu}^c_i|^2 - |\tilde{\nu}_i|^2 ) \right)^2, \\
V_{soft} &=& (\tilde{\nu}^c_i)^\dagger m_{\tilde{\nu}^c_{ij}}^2 \tilde{\nu}^c_j \ + \  \tilde{\nu}_i^\dagger m_{\tilde{L}_{ij}}^2 \tilde{\nu}_j \ + \  m_{H_u}^2 |H_u^0|^2 \ + \ m_{H_d}^2 |H_d^0|^2 
\ + \ \left( \tilde{\nu}_i  a_\nu^{ij} \tilde{\nu}_j^c H_u^0 -  B \mu H_u^0 H_d^0 \ + \  \rm{h.c.}\right). \nonumber \\
\end{eqnarray}
Using the above scalar potential and assuming that all parameters are real we can find the minimization conditions
\begin{align}
	& v_u \left[  \mu^2 \ + \  \frac{1}{2} Y_\nu^{ij}  v_R^j Y_\nu^{ik} v_R^k \ + \  \frac{1}{2} v_L^i Y_\nu^{ij} v_L^k Y_\nu^{kj}
	\ + \ \frac{g_1^2 + g_2^2}{8} \left( v_u^2 - v_d^2 - v_L^i v_L^i \right) \ + \ m_{H_u}^2  \right]
\nonumber
\\
	\label{Eq1}
	& +  \frac{1}{\sqrt{2}} v_L^i a_\nu^{ij} v_R^j \ - \ B \mu v_d =0,
\\
\label{Eq2}
	& v_d \left[  \mu^2 \ - \ \frac{(g_1^2 + g_2^2)}{8} \left( v_u^2 - v_d^2 - v_L^i v_L^i \right) + m_{H_d}^2 \right] 
	\ - \ \frac{1}{\sqrt{2}} \mu v_L^i Y_\nu^{ij} v_R^j \ - \ B \mu v_u =0,
\end{align}
\begin{align}
	& \frac{1}{2} v_L^i Y_\nu^{ij} v_R^j v_L^m Y_\nu^{mk} \ - \ \frac{1}{\sqrt{2}} \mu v_d v_L^i Y_\nu^{ik} \ + \ \frac{1}{2}
	v_u^2  Y_\nu^{ij} v_R^j Y_\nu^{ik} \ + \ \frac{g_{BL}^2}{8} \left( v_R^i v_R^i - v_L^i v_L^i \right) v_R^k
\nonumber
\\
	\label{Eq3}
	& \frac{1}{2} v_R^i \left[ (m_{\tilde{\nu}^c}^2)_{ki} + (m_{\tilde{\nu}^c}^2)_{ik} \right] \ + \
	\frac{1}{\sqrt{2}} v_L^i a_\nu^{ik} v_u =0,
\\
	& \frac{1}{2} v_L^i Y_\nu^{ij} v_R^j Y_\nu^{km} v_R^m \ - \ \frac{1}{\sqrt{2}} \mu v_d Y_\nu^{kj} v_R^j  \ + \ \frac{1}{2}
	v_u^2 v_L^i Y_\nu^{ij} Y_\nu^{kj} - \frac{(g_1^2 + g_2^2)}{8} \left( v_u^2-v_d^2 - v_L^i v_L^i \right) v_L^k
\nonumber
\\
	\label{Eq4}
	& -  \frac{g_{BL}^2}{8} \left( v_R^i v_R^i - v_L^i v_L^i \right) v_L^k \ + \ \frac{1}{2} v_L^i \left[ (m_{\tilde{L}}^2)_{ki}
	+ (m_{\tilde{L}}^2)_{ik} \right] \ + \ \frac{1}{\sqrt{2}} a_\nu^{kj} v_R^j v_u=0.
\end{align}
In order to have phenomenological allowed solutions the $v_L^i$ have to be small, and the $v_R^i$ have 
to be much larger than $v_u, v_d$ and $v_L^i$. Up to negligibly small terms\footnote{The size would go as $\frac{Y_\nu \mu v_d v_L}{m_{\tilde \nu^c}^2} < 10^{-10}$. The maximum values for $Y_\nu$ and $v_L$ are about $10^{-6}$ and $10^{-2}$ GeV, respectively, see Fig.~\ref{vy}.} the right-handed sneutrino acquire a vev in only one family. A possible solution and the one used throughout this paper is $v_R^i=(0,0,v_R)$. In this case:
\begin{eqnarray}
\label{vR.sln}
v_R^2 &\approx& -\frac{8 (m_{\tilde{\nu}^c}^2)_{33}}{g_{BL}^2}, \\
v_L^k &\approx & \frac{v_R}{\sqrt{2}} \frac{\left( \mu v_d Y_\nu^{k3} - a_\nu^{k3} v_u \right)}{\left[ (m_{\tilde{L}}^2)_{kk} - \frac{(g_1^2 + g_2^2)}{8} (v_u^2 - v_d^2) - \frac{g_{BL}^2}{8} v_R^2\right]}.
\end{eqnarray}
Notice that the minimization conditions for $v_u$, Eq.~(\ref{Eq1}), and $v_d$, Eq.~(\ref{Eq2}) are not greatly altered from their MSSM equivalents since the extra terms are very small.

\subsection{Radiative Symmetry Breaking}
In the MSSM, the large top Yukawa coupling drives the up-type soft Higgs mass squared parameter to negative values for generic boundary conditions leading to radiative electroweak symmetry breaking~\cite{Ibanez:1982fr}; a celebrated success of the MSSM. A valid question is then if the same success is possible in achieving a tachyonic right-handed sneutrino mass in this $B-L$ model as required by Eq.~(\ref{vR.sln}).  Unfortunately, this is not possible through a large Yukawa coupling since the Yukawa couplings of the right-handed neutrino are all dictated to be small by neutrino masses.  However, there is an alternate possibility whereby a positive mass squared parameter for the right-handed sneutrino at the high scale will run to a tachyonic value at the low scale. This is due to the presence of the so-called $S$-term (due to $D$-term contributions to the RGE) in the soft mass RGE, as discussed for this $B-L$ model in~\cite{Ambroso:2009jd, Ambroso:2009sc, Ambroso:2010pe}. A short outline of the mechanism follows.

The RGE for the right-handed sneutrino soft mass squared parameter is
\begin{equation}
	16 \pi^2 \frac{d m_{\tilde \nu^c}^2}{d t} = - 3 g_{BL}^2 \left|M_{BL}\right|^2 + \frac{3}{4} g_{BL}^2 S_{BL},
\end{equation}
with
\begin{equation}
	S_{BL} = \text{Tr}\left(2 m_{\tilde Q}^2 - m_{\tilde u^c}^2 - m_{\tilde d^c}^2 - 2 m_{\tilde L}^2 + m_{\tilde e^c}^2 + m_{\tilde \nu^c}^2\right),
\end{equation}
where the trace is over the three generations of the fermions and the soft mass parameters in the trace are for the squark doublet, right-handed up squark, right-handed down squark, slepton doublet, right-handed charged slepton and right-handed sneutrino, respectively. The gaugino mass term always drives the sneutrino mass parameter positive at the low scale but if the overall sign of the $S$-term is positive, it could lead to the opposite effect. Such an effect would require a non-zero $S$-term at the high scale, which is not possible if the soft masses are universal across the generations of each flavor. An example of a suitable boundary condition with minimal variation from the popular MSUGRA Ansatz is universal boundary conditions for all sfermions except for the right-handed sneutrinos, which might have the boundary conditions
\begin{equation}
	m_{\tilde \nu^c_1}^2 = m_{\tilde \nu^c_2}^2 = P \, m_0^2,
	\quad \quad m_{\tilde \nu^c_3}^2 = Q \, m_0^2,
\end{equation}
where $m_0$ is the universal mass and $P > 1$ and $Q<1$. The boundary condition for $S_{BL}$ is then
\begin{equation}
	S_{BL} = (2 P + Q - 1)m_0^2,
\end{equation}
having the necessary sign to contribute negatively to the sneutrino soft mass parameter as it is evolved from the high scale down. The necessary sizes of $P$ and $Q$ depend on the size of the gaugino mass parameter which has the opposite effect, see~\cite{Ambroso:2009jd, Ambroso:2009sc, Ambroso:2010pe} for more details. So while, the traditional radiative symmetry breaking from universal boundary conditions is not possible in these models, it is possible to radiatively break $B-L$ through this $S$-term starting from a positive value. For the implementation of the radiative mechanism in the non-minimal model see Ref.~\cite{Fate}.

\section{Mass Spectrum and Lepton Number Violation}

\subsection{ R-Parity Violating Interactions}
After symmetry breaking lepton number is spontaneously broken in the form of bilinear R-parity violating interactions. There are no trilinear R-parity violating interactions at the renormalizable level. These bilinear interactions mix the leptons with the Higgsinos and gauginos:
	$$\frac{1}{2} g_{BL} v_{R} ( \nu_3^c \tilde B^{'} ), \quad 
	\frac{1}{2} g_{2} v_L^i ( \nu_i \tilde W^0 ), \quad
	\frac{1}{\sqrt{2}} g_{2} v_L^i (e_i \tilde W^+ ),$$
	$$\frac{1}{2} g_{1} v_L ^i( \nu_i \tilde B ), \quad
	\frac{1}{\sqrt{2}} Y_\nu^{i3} v_R ( L^T_i i\sigma_2 \ \tilde{H}_u ), \quad
	\frac{1}{\sqrt{2}} Y_\nu^{i3} v_L^i ( \tilde{H}_u^0 \ \nu^c_3  ), \quad
	\frac{1}{\sqrt{2}} Y_e^i v_L^i ( \tilde{H}_d^- \ e^c_i ).$$
The first term is new and is the only term not suppressed by neutrino
masses. The fifth term corresponds to the so-called $\epsilon$ term,
and second, third and fourth terms are small but important for the decay
of neutralinos and charginos. See Section~V for the discussion of the baryon number violating operators.

There are also lepton number violating interactions coming from the soft terms and the B-L D-term.
From $V_{soft}$ one gets
$$ A_\nu^{i3} \frac{v_R}{\sqrt{2}} \tilde{L}_i^T \ i \sigma_2 \ H_u,$$
while from the D-term one finds
$$g_{BL}^2 v_R \ \tilde{\nu}^c \left( \tilde{q}^\dagger \frac{1}{6} \tilde{q} \ - \  \tilde{l}^\dagger \frac{1}{2} \tilde{l} \right).$$
As one can expect these terms are important to understand the scalar sector of the theory.

\subsection{Mass Spectrum}

\underline{Gauge Boson:}
%
The neutral gauge boson associated to the $B-L$ gauge group is $Z_{BL}$.
Using the covariant derivate for the right-handed sneutrinos, $D_\mu \tilde{\nu}^c = \partial_\mu \tilde{\nu}^c + i \frac{g_{BL}}{2} B_\mu^{'} \tilde{\nu}^c$, 
the mass term for $Z_{BL}$ is:
\begin{equation}
	M_{Z_{BL}}=\frac{g_{BL}}{2} v_R.
\end{equation}
Now, using the experimental collider constraint~\cite{Carena:2004xs}:
\begin{equation}
\label{ZBL.const}
	\frac{M_{Z_{BL}}}{g_{BL}} \geq 3 \text{ TeV},
\end{equation}
and Eq. (21) one finds the condition
\begin{equation}
	| (m_{\tilde{\nu}^c)_{33}} | > 2.12 \ g_{BL} \ \rm{TeV}.
\end{equation}
Then, if $g_{BL} =0.1$ the soft mass above has to be larger than 200 GeV.  
This condition can be easily satisfied without assuming a very heavy spectrum for the supersymmetric particles. 

\begin{figure}[h] 
	\includegraphics[scale=0.8]{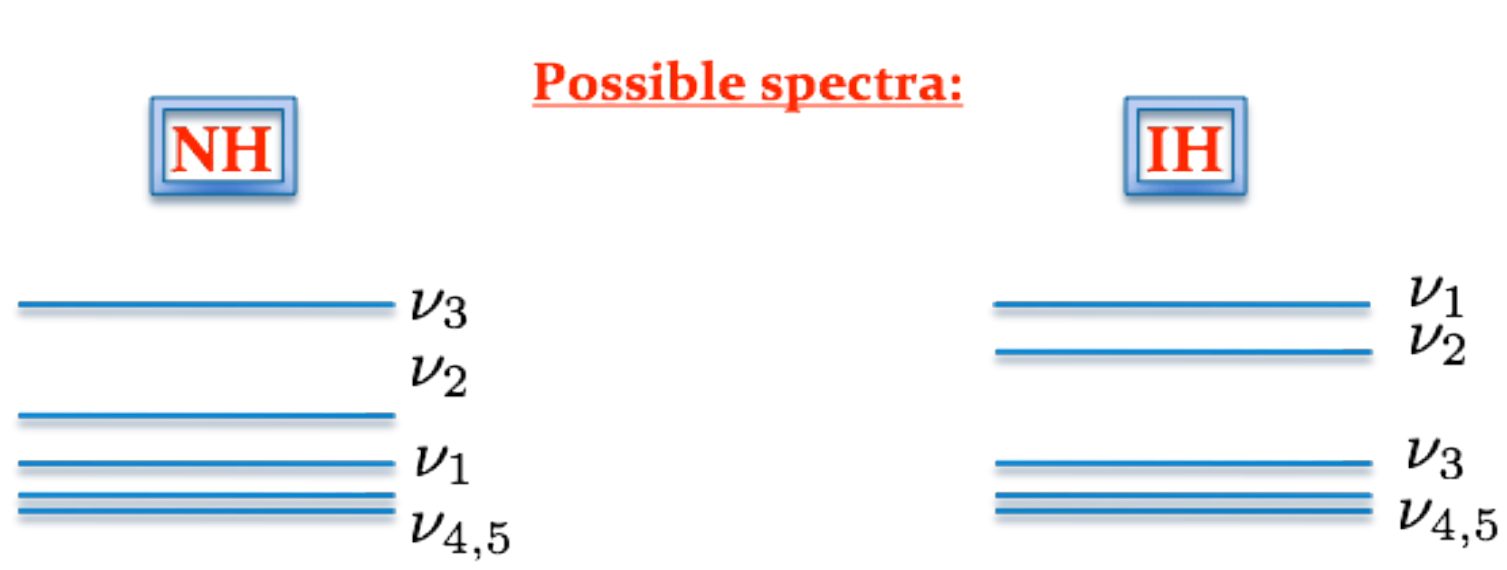}
	\caption{Sample spectra for neutrino masses in the normal and inverted hierarchies.}
\label{nu.spec}
\end{figure}

\underline{Neutralinos and Neutrinos:}
%
As in any supersymmetric theory where $R$-parity is broken all the fermions with the same quantum numbers mix and 
form physical states which are linear combinations of the original fields. The neutralinos in this theory are a linear combination 
of the  fields, $\left(\nu_i, \ \nu^c_j, \ \tilde B', \ \tilde B, \ \tilde W^0, \ \tilde H_d^0, \ \tilde H_u^0\right)$. Then, the neutralino mass matrix is given by
\begin{equation}
	{\cal M}_{N} =
	\begin{pmatrix}
			0
		&
			\frac{1}{\sqrt{2}} \ Y_\nu^{ij} v_u
		&
			-\frac{1}{2} g_{BL} \ v_L^i
		&
			-\frac{1}{2} g_1 \ v_L^i
		&
			\frac{1}{2} g_2 \ v_L^i
		&
			0
		&
			\frac{1}{\sqrt{2}} \ Y_\nu^{ij} \ v_R^j
	\\
			\frac{1}{\sqrt{2}} \ Y_\nu^{ij} \ v_u
		&
			0
		&
			\frac{1}{2} g_{BL} \ v_R^j
		&
			0
		&
			0
		&
			0
		&
			\frac{1}{\sqrt{2}} \ Y_\nu^{ij} \ v_L^i
	\\
			-\frac{1}{2} g_{BL} \ v_L^i
		&
			\frac{1}{2} g_{BL} \ v_R^j
		&
			M_{BL}
		&
			0
		&
			0
		&
			0
		&
			0
	\\
			-\frac{1}{2} g_1 \ v_L^i
		&
			0
		&
			0
		&
			M_1
		&
			0
		&
			-\frac{1}{2} g_1 v_d
		&
			\frac{1}{2} g_1 v_u
	\\
			\frac{1}{2} g_2 \ v_L^i
		&
			0
		&
			0
		&
			0
		&
			M_2
		&
			\frac{1}{2} g_2 v_d
		&
			-\frac{1}{2} g_2 v_u
	\\
			0
		&
			0
		&
			0
		&
			-\frac{1}{2} g_1 v_d
		&
			\frac{1}{2} g_2 v_d
		&
			0
		&
			-\mu
	\\
			\frac{1}{\sqrt{2}} \ Y_\nu^{ij} \ v_R^j
		&
			\frac{1}{\sqrt{2}} \ Y_\nu^{ij} \ v_L^i
		&
			0
		&
			\frac{1}{2} g_1 v_u
		&
			-\frac{1}{2} g_2 v_u
		&
			-\mu
		&
			0
	\end{pmatrix}.
\label{neutralino}
\end{equation}
We have discussed above that only one right-handed sneutrinos get a vev, $v_R^i=(0,0,v_R)$.
Now, integrating out the neutralinos one can find the mass matrix for the light neutrinos. In this case 
one has three active neutrinos and two sterile neutrinos, and the mass matrix in the basis 
$(\nu_e, \nu_\mu, \nu_\tau, \nu^c_e, \nu^c_\mu)$ is given by
\begin{widetext}
\begin{equation}
\label{Mnu}
M_\nu =
\begin{pmatrix}
	A \ v_{L}^i v_{L}^j
	+ B \ \left[Y_\nu^{i3} v_{L}^j + Y_\nu^{j3} v_{L}^i \right]
	+ C \ Y_\nu^{i3} Y_\nu^{j3} 
	&
	\frac{1}{\sqrt{2}} v_u Y_\nu^{i \beta} 
\\
	\frac{1}{\sqrt{2}} v_u Y_\nu^{\alpha j}
	&
	O_{2\times2}
\end{pmatrix},
\end{equation}
where
\begin{align}
	A & = \frac{2 \mu^2}{\tilde m^3}, \ \ \
	B = \left(\frac{v_u}{\sqrt{2} v_R}  + \frac{\sqrt{2} \mu v_d v_R}{\tilde m^3}\right),
	\ \ \
	C = \left(\frac{2 M_{BL} v_u^2}{g_{BL}^2 v_R^2} + \frac{v_d^2 v_R^2}{\tilde m^3}\right), 
\end{align}
\begin{align}	
	\tilde m^{3} & = \frac
		{
			4
			\left[
				\mu v_u v_d \left(g_1^2 M_2 + g_2^2 M_1 \right)
				- 2 M_1 M_2 \mu^2
			\right]
		}
		{g_1^2 M_2 + g_2^2 M_1}
		.
\end{align}
\end{widetext}
Here $\alpha$ and $\beta$ take only the values 1 and 2. From the experimental limits on active neutrino masses we obtain $(Y_\nu)_{i \alpha} \lesssim 10^{-12}$.
This can be compared to $(Y_\nu)_{i 3} \lesssim 10^{-5}$, which is less constrained because of the TeV scale
seesaw suppression. It has been pointed out 
in Ref.~\cite{Barger:2010iv, Ghosh:2010hy} (and earlier in a different context~\cite{Mohapatra}) that this theory predicts the existence of two light sterile neutrinos which 
are degenerate or lighter than the active neutrinos, a so-called $3+2$ neutrino model. A sample possible spectrum is displayed in Fig.~\ref{nu.spec}.

Recently, it has been shown in~\cite{Hamann:2010bk} that precision cosmology and big-bang nucleosynthesis mildly favor extra radiation in the universe beyond photons and ordinary neutrinos, lending support to the existence of sub-eV sterile neutrinos.

\underline{Charginos and Charged Leptons:}
%
In this theory the chargino mass matrix, in the basis $\left(e^c_j, \ \tilde W^+, \ \tilde H_u^+\right)$ and $\left(e_i, \ \tilde W^-, \ \tilde H_d^-\right)$, is given by
\begin{equation}
\label{chargino}
{\cal M}_{{\tilde \chi}^{\pm}}	=	
	\left(
	\begin{array}{cc}
		0
	&
		M_C
	\\
		M_C^T
	&	0
	\end{array}
	\right),
\end{equation}
with
\begin{equation}
	M_C =
	\begin{pmatrix}
			-\frac{1}{\sqrt{2}} Y_e^{ij} v_d
		&
			0
		&
			\frac{1}{\sqrt{2}} Y_e^{ij}  v_L^j
	\\
			\frac{1}{\sqrt{2}} \ g_2  v_L^i
		&
			M_2
		&
			\frac{1}{\sqrt{2}} g_2 v_d
	\\
			-\frac{1}{\sqrt{2}} Y_\nu^{ij} v_R^j
		&
			\frac{1}{\sqrt{2}} g_2 v_u
		&
			\mu
	\end{pmatrix}.
\end{equation}

\underline{Squarks and Sleptons:}
%
In the sfermion sector, the mass matrices $\mathcal{M}_{\tilde u}^2$, and $\mathcal{M}_{\tilde d}^2$ for squarks, 
and $\mathcal{M}_{\tilde e}^2$ for charged sleptons, in the basis $\left(\tilde f, \  {\tilde f}^{c*} \right)$, are given by
\begin{eqnarray}
\mathcal{M}_{\tilde u}^2&=&\left(\begin{array}{cc}
		m_{\tilde Q}^2
		\ + \ m_{u}^2
		\ + \  \left(\frac{1}{2} \ - \  \frac{2}{3} s_W^2 \right) \ M_Z^2 \ c_{2\beta} 
		\ + \ \frac{1}{3} D_{BL}
		&
		\frac{1}{\sqrt 2} \left(a_u \ v_u - Y_u \ \mu \ v_d\right)
	\\
		\frac{1}{\sqrt 2} \left(a_u \ v_u - Y_u \ \mu \ v_d\right)
		&
		m_{\tilde u^c}^2
		\ + \ m_{u}^2
		\ + \ \frac{2}{3} M_Z^2 \ c_{2\beta}\ s_W^2 
		\ - \ \frac{1}{3} D_{BL}
	\end{array}\right),
	\nonumber \\
\end{eqnarray}	
\begin{eqnarray}	
\mathcal{M}_{\tilde d}^2 &=& \left(\begin{array}{cc}
		m_{\tilde Q}^2
		\ + \ m_{d}^2
		\ - \  \left(\frac{1}{2}  \ - \ \frac{1}{3} \  s^2_W \right) M_Z^2 \ c_{2 \beta}
		\ + \ \frac{1}{3} D_{BL}
		&
		\frac{1}{\sqrt 2} \left(Y_d \ \mu \ v_u - a_d \ v_d\right)
	\\
		\frac{1}{\sqrt 2} \left(Y_d \ \mu \ v_u - a_d \ v_d\right)
		&
		m_{\tilde d^c}^2
		\ + \ m_{d}^2
		\ - \  \frac{1}{3} \ M_Z^2  \ c_{2\beta} \ s^2_W
		\ - \ \frac{1}{3} D_{BL}
	\end{array}\right),
	\nonumber \\
	\\
\mathcal{M}_{\tilde e}^2 &=& \left(\begin{array}{cc}
\label{Selectron.Mass}
		m_{\tilde L}^2
		\ + \ m_{e}^2
		\ - \ \left( \frac{1}{2} \ - s_W^2 \right) M_Z^2 \ c_{2\beta} 
		\ - \ D_{BL}
		&
		\frac{1}{\sqrt 2} \left(Y_e \ \mu \ v_u - a_e \ v_d\right)
	\\
		\frac{1}{\sqrt 2} \left(Y_e \ \mu \ v_u - a_e \ v_d\right)
		&
		m_{\tilde e^c}^2
		\ + \ m_{e}^2
		\ - \  M_Z^2 \ c_{2\beta}\ s_W^2
		\ + \ D_{BL}
	\end{array}\right), \nonumber \\ 
\end{eqnarray}
where $c_{2\beta} = \cos 2\beta$, $s_W = \sin\theta_W$ and
\begin{equation}
D_{BL} \equiv \frac{1}{8} \ g_{BL}^2  v^2_R= \frac{1}{2} M_{Z_{BL}}^2.
\end{equation}
$m_u, \ m_d$ and $m_e$ are the respective fermion masses and $a_u, \ a_d$ and $a_e$
are the trilinear $a$-terms corresponding to the Yukawa couplings $Y_u, \ Y_d$ and $Y_e$.
Typically, it is assumed that substantial left-right mixing occurs only in the third generation.  
Regardless, the physical states are related to the gauge states by
\begin{align}
\label{eq_squarkmixing}
	\begin{pmatrix}
		\tilde f_1
		\\
		\tilde f_2
	\end{pmatrix}
	& =
	\begin{pmatrix}
		\cos \theta_{\tilde f}
		&
		\sin \theta_{\tilde f}
		\\
		- \sin \theta_{\tilde f}
		&
		\cos \theta_{\tilde f}
	\end{pmatrix}
	\begin{pmatrix}
		\tilde f
		\\
		\tilde f^{c*}
	\end{pmatrix}.
\end{align}
The masses in the sneutrino sector are given by
\begin{eqnarray}
M_{\tilde{\nu}_i}^2 &=& m_{\tilde{L}_i}^2 \ + \ \frac{1}{2} M_{Z}^2 \cos 2 \beta - \frac{1}{2} M_{Z_{BL}}^2, \\
M_{\tilde{\nu}_3^c}^2 &=& M_{Z_{BL}}^2, \\
M_{\tilde{\nu}_\alpha^c}^2 &=& m_{\tilde{\nu}^c_\alpha}^2 \ + \ D_{BL},
\end{eqnarray}
and $\alpha=1..2$. For simplicity we listed the mass matrices in 
the limit $v_L^i, a_\nu, Y_\nu \to 0$. For the most general expressions see Appendix~A.
It is important to mention that all sfermion masses are modified due to the existence of the B-L D-term. 

%
In order to understand the properties of the spectrum we assume a simplified spectrum for the superpartners.
In the case of the sfermions we will assume for simplicity the same value for all soft masses. In this case if we neglect the left-right 
mixing the full spectrum of sfermions will be defined by $M_{SUSY}$ (the universal soft supersymmetry breaking mass), $\tan \beta$ and the mass of the B-L
 gauge boson. Using this simplified spectrum we show in Fig.1 the values for the sfermion masses for different values of $M_{Z_{BL}}$.
 Notice that the condition that left-handed slepton masses have to be positive 
 impose a bound on the $M_{Z_{BL}}$ for a given value of $M_{SUSY}$, i.e. $M_{Z_{BL}} < \sqrt{2} M_{\tilde{L}}$. In this way, we can appreciate that the spectrum 
 can be very constrained. As it is well-known, in the general case one cannot predict the spectrum since the soft masses are unknown.

\begin{figure}[h] 
\includegraphics[scale=0.6]{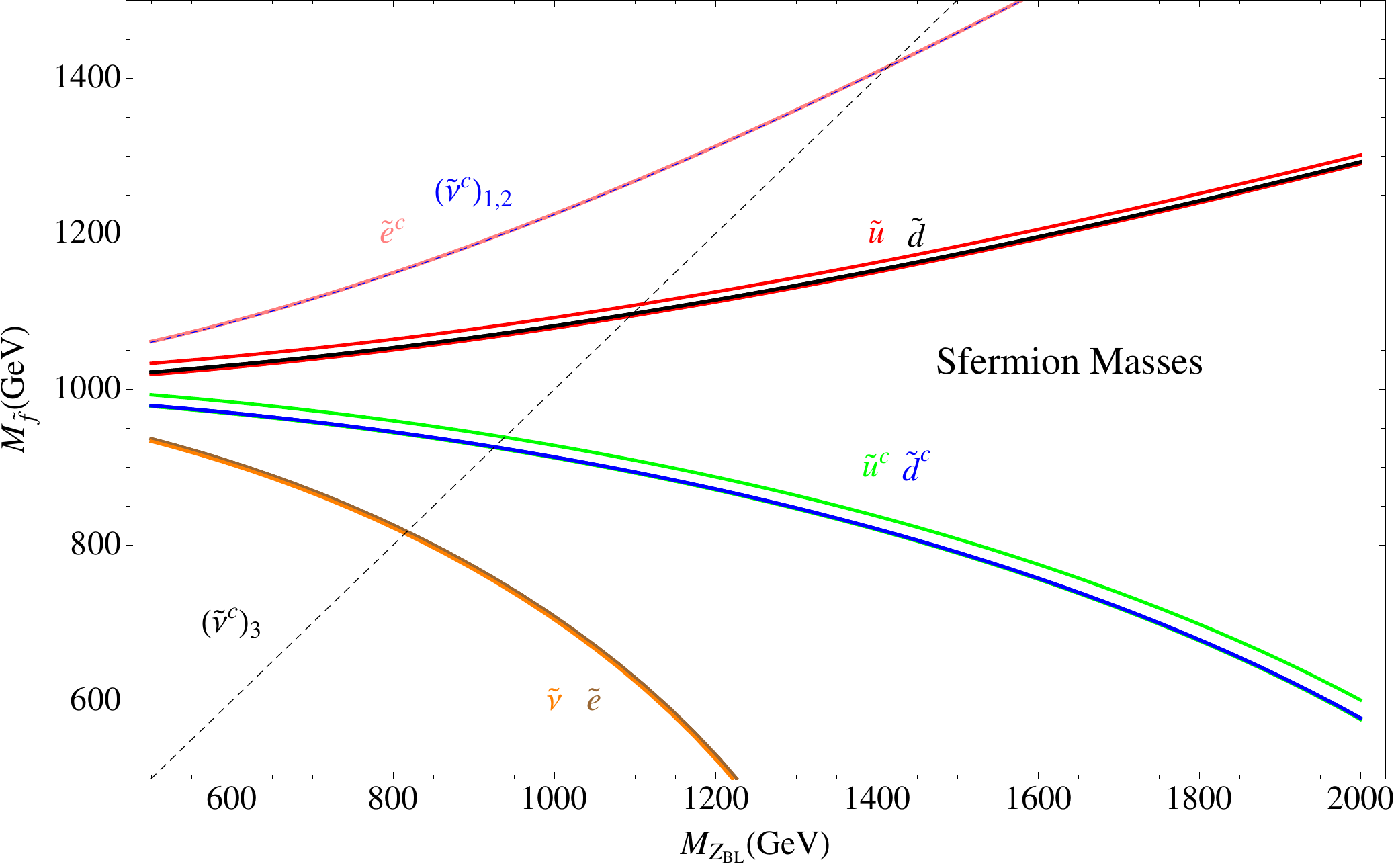}
\caption{Spectrum for sfermion masses assuming the same value for all soft masses, $M_{SUSY}=1$ TeV and $\tan \beta=5$.}
\label{spectrum}
\end{figure}

\section{Nucleon Stability and Dark Matter}
%
It is well-known that when R-parity is broken in a supersymmetric theory one has to understand the possible constraints coming from proton decay~\cite{proton}.
In the MSSM one has several interactions which could mediate proton decay at tree level and one-loop level. At the renormalizable level one has the lepton and baryon 
number violating interactions 
\begin{equation}
{\cal{W}}_{RpV}=\epsilon \hat{L} \hat{H}_u \ + \  \lambda \hat{L} \hat{L} \hat{e}^c \ + \   \lambda^{'} \hat{Q} \hat{L} \hat{d}^c \ + \  \lambda^{''} \hat{u}^c \hat{d}^c \hat{d}^c,
\end{equation}
which are not allowed in our theory before B-L breaking, and in general one has the dimension five operators
\begin{equation}
{\cal{W}}_{RpC}^5= \frac{\lambda_\nu}{\Lambda} \hat{L} \hat{L} \hat{H}_u \hat{H}_u \ + \  \frac{\lambda_L}{\Lambda} \hat{Q} \hat{Q} \hat{Q} \hat{L} \ + \  \frac{\lambda_R}{\Lambda} \hat{u}^c \hat{d}^c \hat{u}^c \hat{e}^c 
\ + \  \frac{\lambda_{\nu^c}}{\Lambda} \hat{u}^c \hat{d}^c \hat{d}^c \hat{\nu}^c.
\end{equation}
Notice that the first term in the above equation is not allowed in our theory, but the last terms can mediate proton decay. The operators $QQQL$ and $u^c d^c u^c e^c$ mediate proton decay 
at one-loop level and typically the scale $\Lambda$ should be larger than $10^{17}$ GeV in order to satisfy the experimental bounds on proton decay. For a detailed discussion 
see Ref.~\cite{proton}. Once B-L is broken by the vev of the right-handed sneutrinos one finds new contributions to proton decay at tree level. 
From the Yukawa coupling $Y_\nu \hat{L} \hat{H}_u \hat{\nu}^c$ one gets the lepton number violating interaction $L \tilde{H}_u$ and from the last term, 
$\hat{u}^c \hat{d}^c \hat{d}^c \hat{\nu}^c$, in the above equation one gets the interaction $\tilde{u}^c d^c s^c$. Using these interactions and integrating 
out the neutralinos and squarks we find the following constraint
\begin{equation}
\frac{\lambda^{1123}_{\nu^c}}{\Lambda} \frac{Y_u Y_\nu^{i3}}{M_{\tilde{u}}^2} \frac{v_R^2}{2 M_{\tilde{\chi}^0}} \ < \ 10^{-30} \ \rm{GeV}^{-2},
\end{equation}
from the channel $p \to K^+ \nu$. Then, assuming that $\lambda^{1123}_{\nu^c} \sim 1$, and $Y_{\nu}^{i3} \sim 10^{-6}$ (see Fig.~\ref{vy})
one gets $\Lambda > 10^{17}$ GeV. This constrain is similar to the one we get from the dimension five operators.
Therefore, one can say that if the above couplings are order one the cutoff of the theory has to be large. Also one can think about 
possible scenarios where the couplings are small, see for example~\cite{Yuval}.

At first glance, finding a dark matter in $R$-parity violating theories seems hopeless.  But while the traditional neutralino LSP case is no longer valid, the situation is not lost. As first discussed in~\cite{Takayama:2000uz}, such models can have an unstable LSP gravitino, with a lifetime longer than the age of the universe.  The strong suppression on its lifetime is due to both the planck mass ($M_P$) suppression associated with its interaction strength and bilinear $R$-parity violation which is small due to neutrino masses and must facilitate the decay of the LSP. In the mass insertion approximation, this can be understood as the gravitino going into a photon and neutralino which then has some small mixing with the neutrino due to $R$-parity violation ($m_{\chi \, \nu}$), thereby allowing $\tilde G \to \gamma \nu$ as in Fig.~3. Adopting approximations made in~\cite{Takayama:2000uz}, the lifetime for the gravitino decaying into a photon and neutrino (in years) is about
\begin{figure}[h] 
	\includegraphics[scale=0.4]{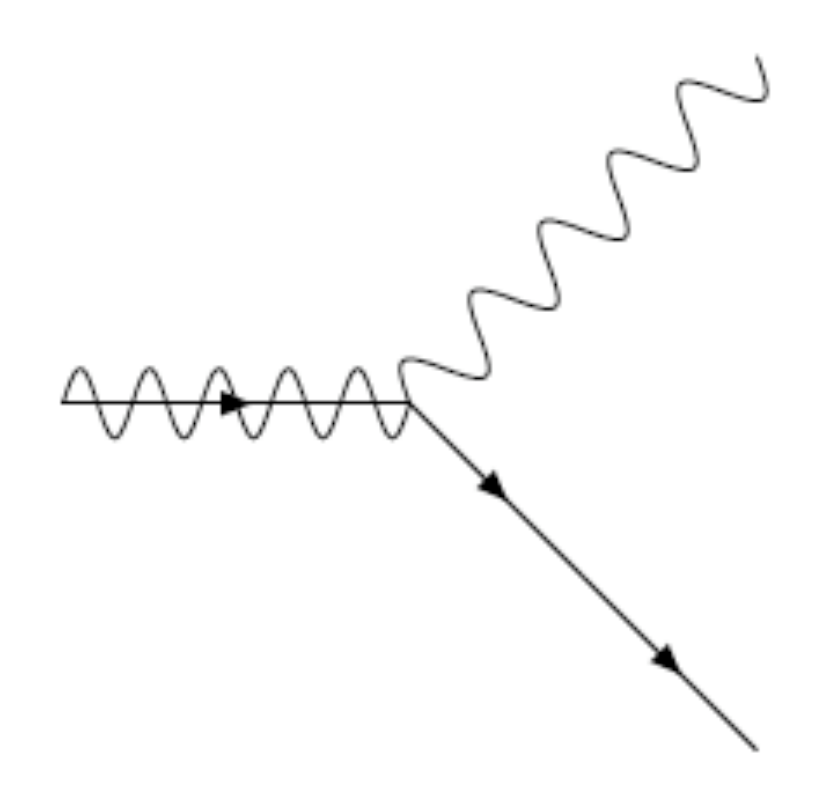}
	\put(-30,20){$\tilde G$}
	\put(-15,25){$\gamma$}
	\put(-12.3, 8){\Large+}
	\put(-18, 8.5){$\chi$}
	\put(-8, 0){$\nu$}
	\put(-9.8, 10.4){$m_{\chi \, \nu}$}
\caption{Gravitino decay into a photon and a neutrino. }
\label{gravitino}
\end{figure}
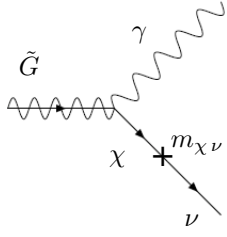
\begin{equation}
	\tau(\tilde G \to \gamma \nu) \sim 2 \times 10^{10}
	\left(\frac{m_{3/2}}{100 \text{ GeV}} \right)^{-3}
	\left(\frac{m_{\chi \nu}/m_\chi}{10^{-6}} \right)^{-2}
	\text{ years},
\end{equation}
which for appropriate values of the gravitino mass leads to a long enough life time. Unlike in $R$-parity conserving models with a gravitino LSP, there are no issues with big bang nucleosynthesis from slow NLSP decay since the NLSP decays more promptly through $R$-parity violating interactions. Several interesting studies have been conducted on the signatures and constraints of unstable gravitino dark matter, see for example~\cite{Takayama:2000uz,Buchmuller:2007ui}.

\section{Experimental Constraints}
%
In order to understand the different lepton number violating decays in the theory we need to understand which are the main constraints from neutrino experiments.
Today, we know well the numerical values for two of the neutrino mixings and the mass squared differences. The neutrino mixing matrix $V_{PMNS}$ is defined as
\begin{equation}
	V_{PMNS} = 
	\begin{pmatrix}
		c_{12} \, c_{13}
		&
		c_{13} \, s_{12}
		&
		s_{13}
		\\
		-c_{23} \, s_{12}  -   c_{12} \, s_{13} \, s_{23}
		&
		c_{12} \, c_{23} - s_{12} \, s_{13} \, s_{23}
		&
		c_{13} \, s_{23}
		\\
		s_{12} \, s_{23}   -    c_{12} \, c_{23} \, s_{13}
		&
		-c_{12} \, s_{23} - c_{23} \, s_{12} \, s_{13}
		&
		c_{13} \, c_{23}
	\end{pmatrix},
\end{equation}
where $c_{ij} = \cos \theta_{ij}$ and $s_{ij} = \sin \theta_{ij}$ with $0 \leq \theta_{ij} \leq \pi/2$. 
The physical neutrino masses are contained in $m_\nu=diag(m_{\nu_1},m_{\nu_2},m_{\nu_3})$. 
As it is well-known, there are two possible neutrino spectra:
\begin{align}
\begin{split}
	\text{Normal Hierarchy (NH):} & \quad
		m_{\nu_1}, \ \
		m_{\nu_2} = \sqrt{ m_{\nu_1}^2 + \Delta m_{21}^2}, \ \
		m_{\nu_3} = \sqrt{ m_{\nu_1}^2 + |\Delta m_{31}^2| },
	\\
	\text{Inverted Hierarchy (IH):} & \quad
		m_{\nu_1} = \sqrt{m_{\nu_3}^2 + |\Delta m_{31}^2|}, \ \
		m_{\nu_2} = \sqrt{ m_{\nu_1}^2 + \Delta m_{21}^2}, \ \
		m_{\nu_3},
\end{split}
\end{align}
where~\cite{arXiv:1103.0734}
\begin{align}
	7.27 \times 10^{-5} \text{eV}^2 \leq & \, \Delta m_{21}^2 \, \leq 8.03 \times 10^{-5} \text{ eV}^2, 
\\ 
	2.17 \times 10^{-3} \text{ eV}^2 < & \, |\Delta m_{31}^2| \, < 2.54 \times 10^{-3} \text{ eV}^2,
\end{align}
are the solar and atmospheric mass squared differences, respectively. 

\begin{figure}[h!] 
	\includegraphics[scale=0.75]{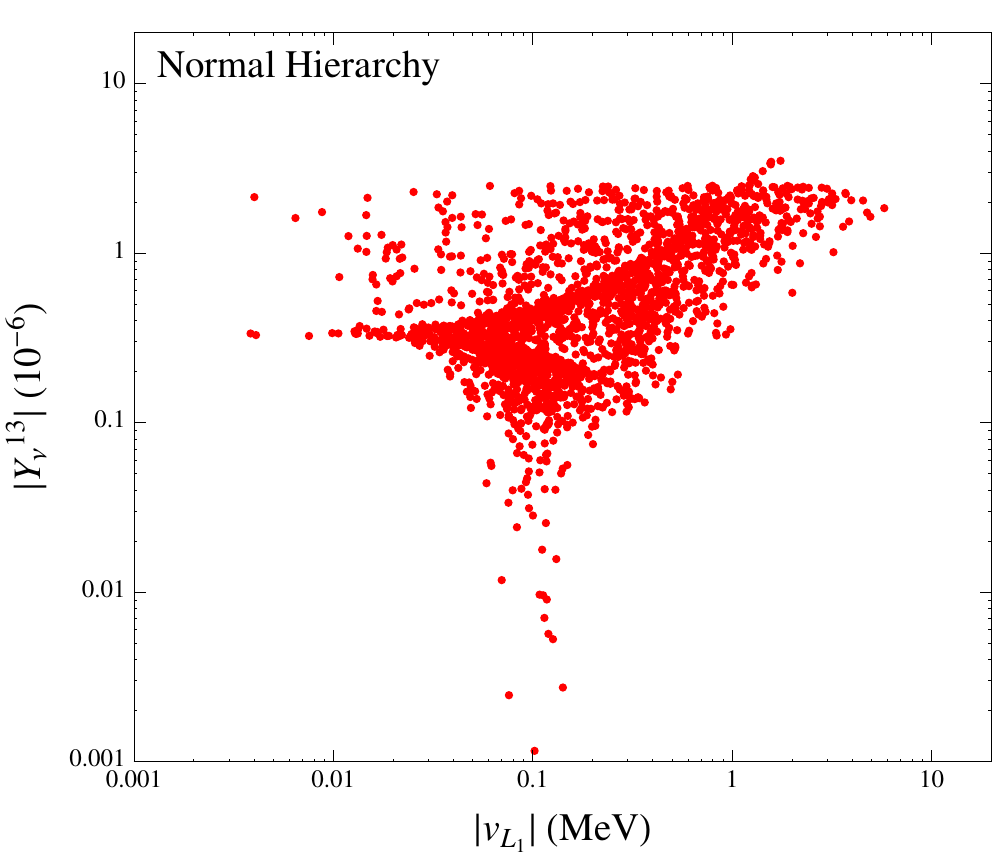}
	\put(-40,-4){(a)}
	\includegraphics[scale=0.75]{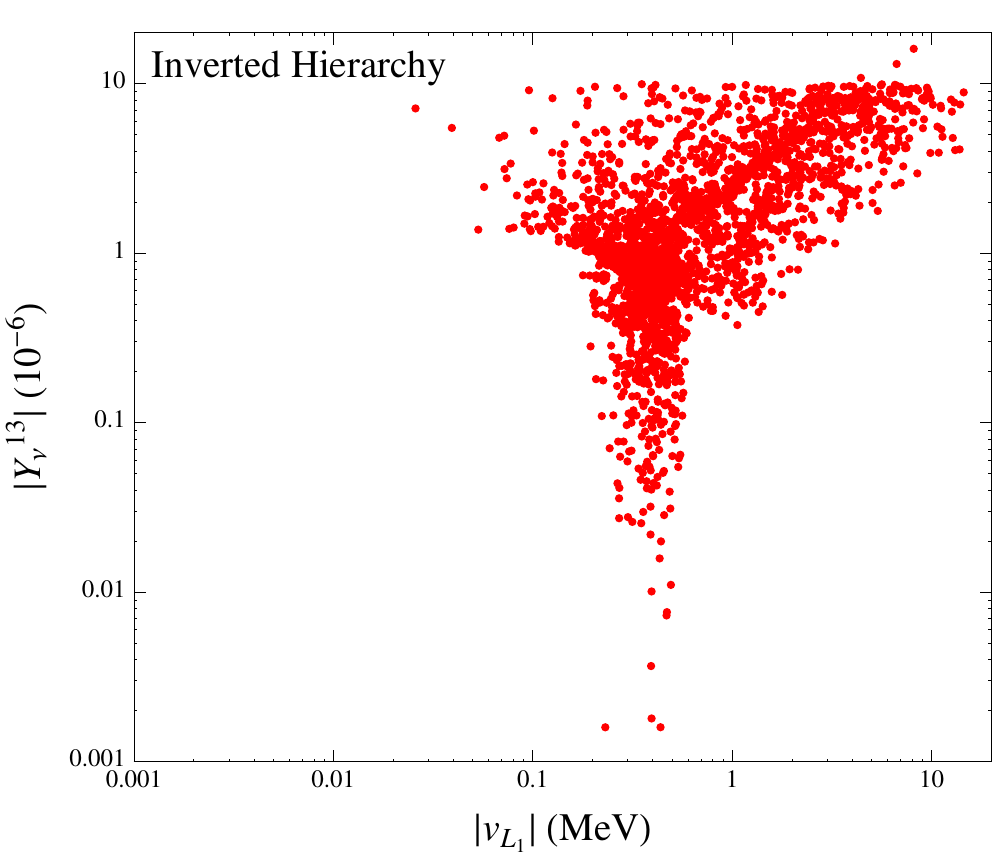}
	\put(-40,-4){(b)}
\\
	\includegraphics[scale=0.75]{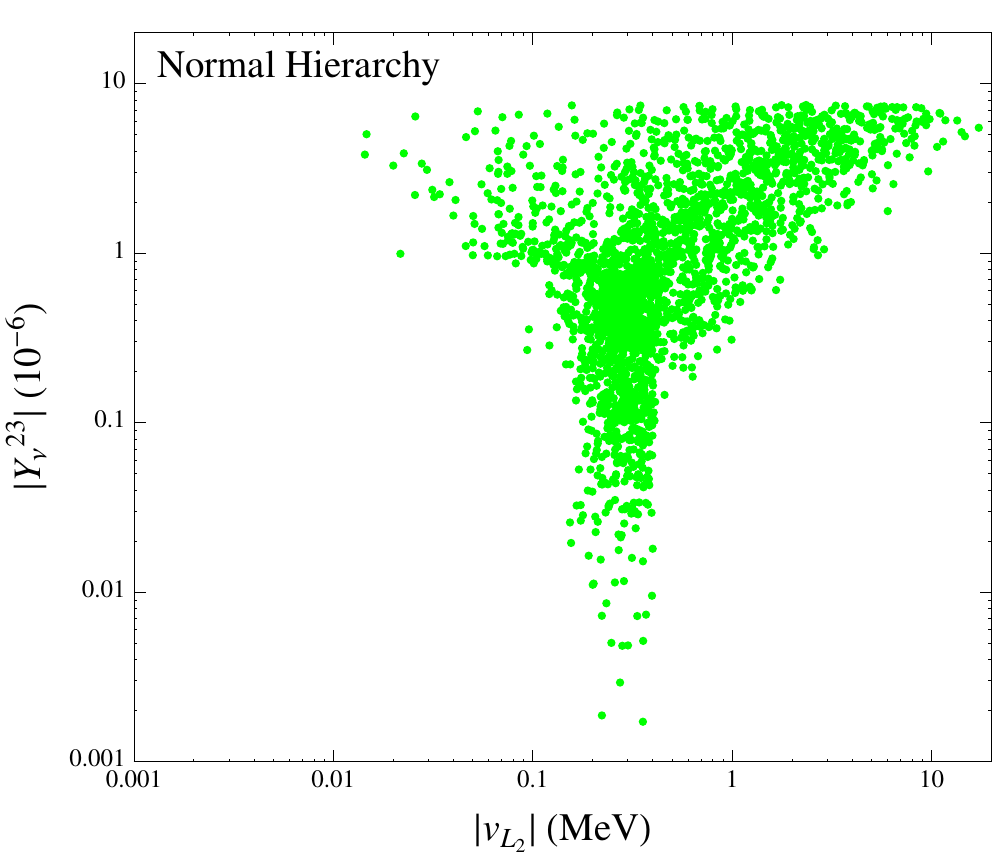}
	\put(-40,-4){(c)}
	\includegraphics[scale=0.75]{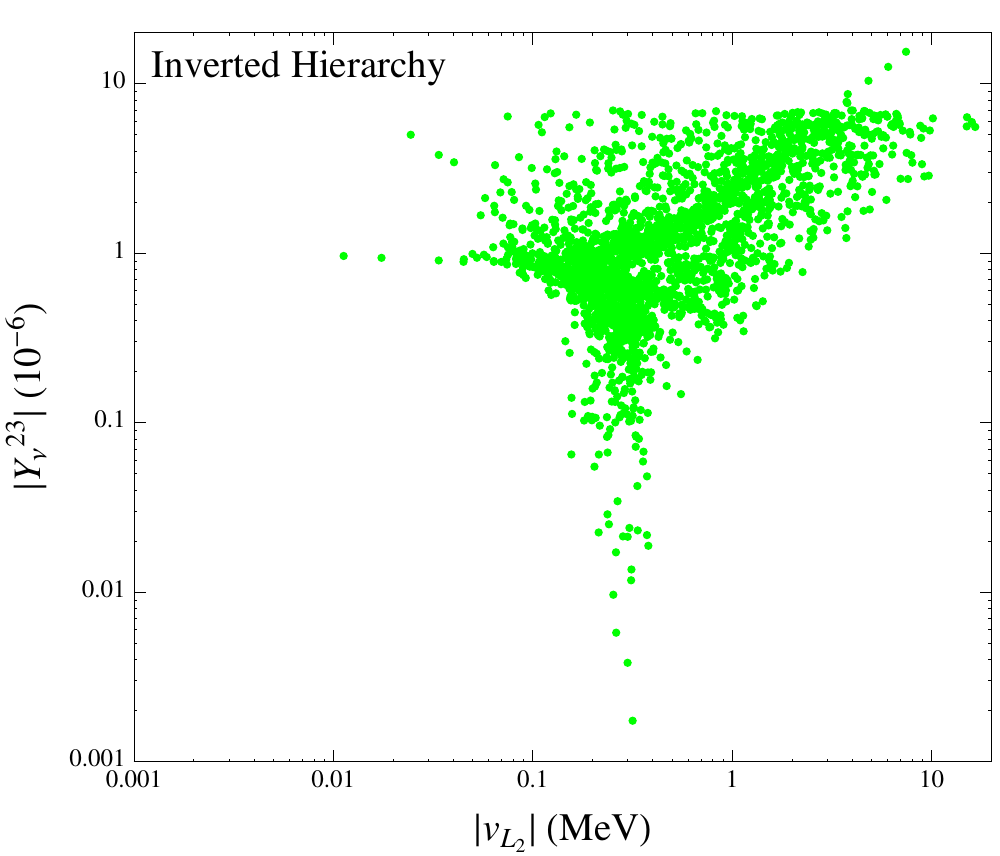}
	\put(-40,-4){(d)}
\\
	\includegraphics[scale=0.75]{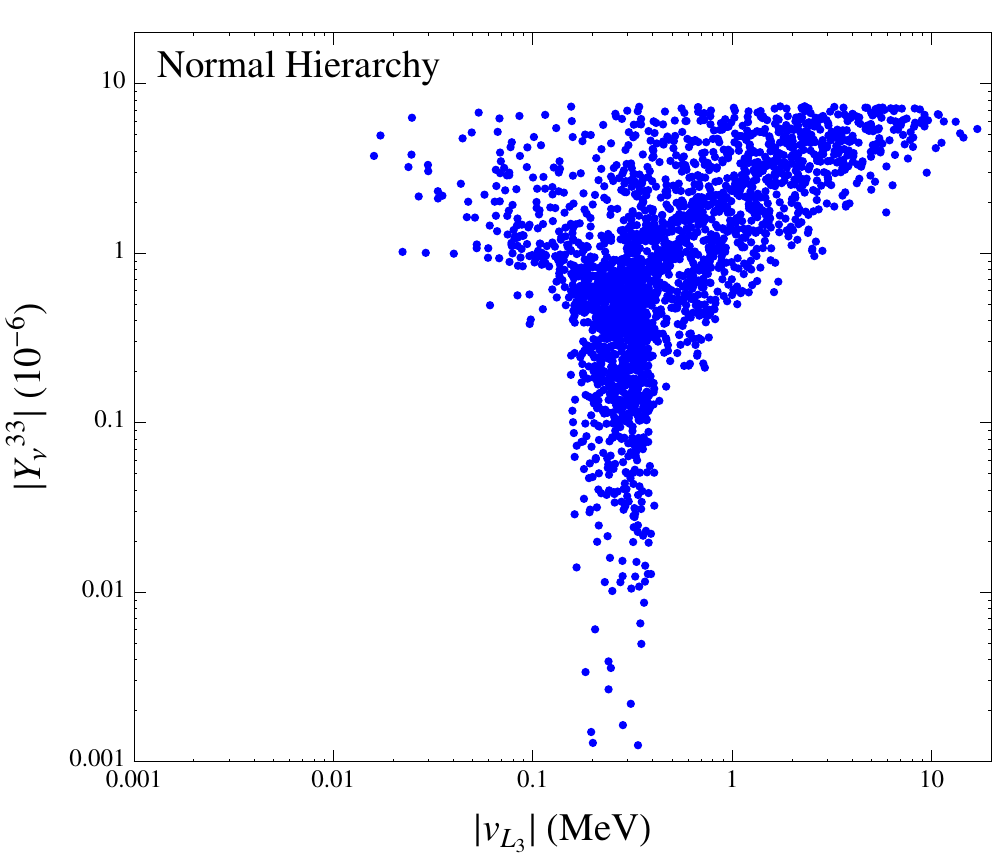}
	\put(-40,-4){(e)}
	\includegraphics[scale=0.75]{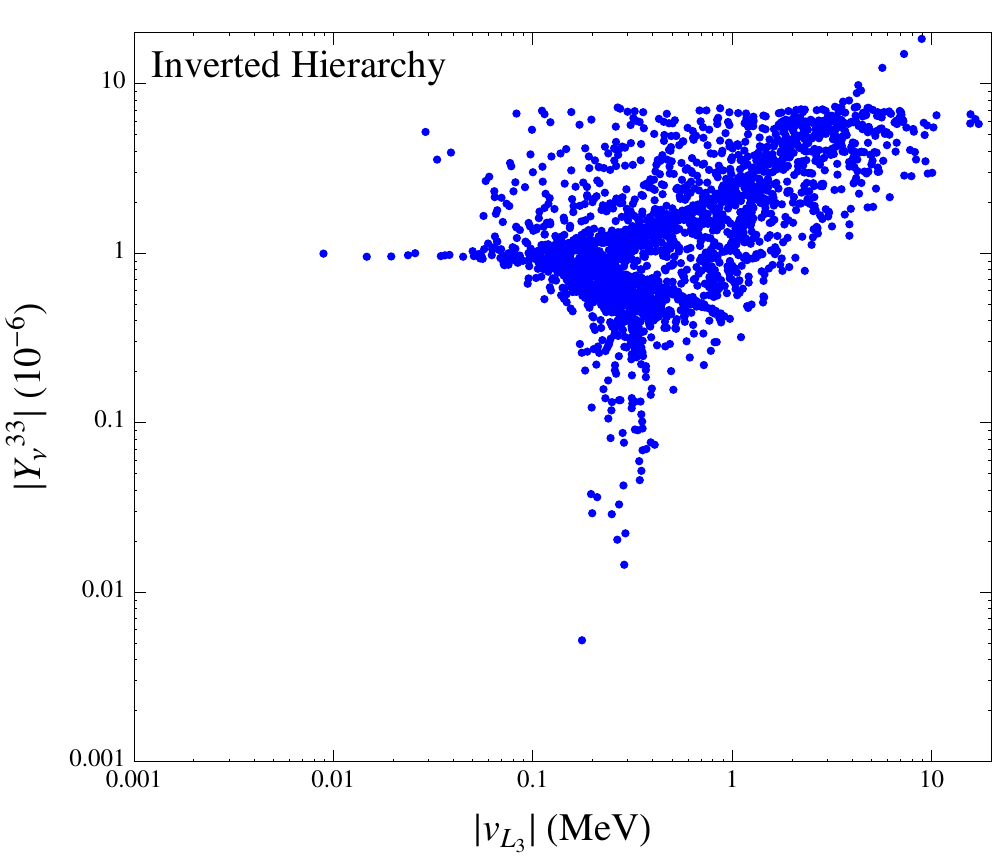}
	\put(-40,-4){(f)}
\caption
{
	Allowed values for the $v_L^i$ versus $(Y_\nu)_{i3}$ in agreement with the neutrino masses and mixings constraints in the NH (IH) in a,c and e (b,d and f) for a scan over the parameters listed in Table~\ref{tab:scan}.
}
\label{vy}
\end{figure}

In order to understand the allowed values for the vevs of the left-handed sneutrinos and the Dirac Yukawa couplings we assume for simplicity that the off-diagonal block matrices in Eq.~(\ref{Mnu}) 
are zero, hence decoupling the two light sterile neutrinos from the active ones. In this case, all neutrino masses and mixing originate from the upper-left block matrix in Eq.~(\ref{Mnu}), which we label $m_\nu$.  The flavor pattern, and hence the rank of this matrix, dictates that one neutrino will be massless.  This matrix is diagonalized by the PMNS matrix
\begin{equation}
	m_\nu = V_\text{PMNS}^T \  M_\nu \ V_\text{PMNS},
\end{equation}
where $m_\nu = \text{diag}(0, \sqrt{\Delta m_{21}^2}, \sqrt{|\Delta m_{31}^2|})$ in the Normal Hierarchy 
and $m_\nu = \text{diag}(\sqrt{|\Delta m_{31}^2|}, \sqrt{|\Delta m_{31}^2| + \Delta m_{21}^2}, 0)$ in the Inverted Hierarchy.  
This yields a system of six equations quadratic in the vevs of the right-handed sneutrinos and Yukawa couplings, although solving 
for these is not the most efficient way to proceed.  Instead, notice that the product above yields the following six terms
\begin{align}
\begin{split}
	\label{VvRelation}
	V^j & \equiv v_{L}^i \, {V_\text{PMNS}}^{ij},
	\\
	Y^j & \equiv  Y_\nu^{i3} \, V_\text{PMNS}^{ij}.
\end{split}
\end{align}
\begin{itemize}
\item{Normal Hierarchy}

In the Normal Hierarchy one obtains the following six equations
\begin{align}
	A V_1^2 + 2 B V_1 Y_1 + C Y_1^2 = & \, 0,
\\
	A V_1 V_2 + B \left(V_1 Y_2 + V_2 Y_1 \right)+ C Y_1 Y_2 = & \, 0,
\\
	A V_1 V_3 + B \left(V_1 Y_3 + V_3 Y_1 \right)+ C Y_1 Y_3 = & \, 0,
\\
	A V_2 V_3 + B \left(V_2 Y_3 + V_3 Y_2 \right) + C Y_2 Y_3 = & \, 0,
\\
	A V_2^2 + 2 B V_2 Y_2 + C Y_2^2 = &\,  m_2,
\\
	A V_3^2 + 2 B V_3 Y_3 + C Y_3^2 = & \ m_3.
\end{align}
In order for these equations to be consistent, $V_1 = Y_1 = 0$.  While this condition represents some fine-tuning between parameters, it is a result of 
the simplifying assumption that the sterile and active light states decouple.  In a more general scenario, this condition would not be necessary.

The remaining system of equations, the last three, is underdetermined with three equations and four unknowns. Solving with respect to $Y_3$ yields
\begin{align}
	Y_2 & = \epsilon_1
		\sqrt{\frac{m_2}{m_3}}
		\sqrt
		{
			\frac
			{
				- Y_3^2 R - A m_3
			}
			{
				R
			}
		},
\\
	V_3 & = \frac
		{
			-B Y_3 + \epsilon_3   \sqrt{Y_3^2 R + A m_3}
		}
		{
			A
		},
\\
V_2 & = \frac
		{
			-B Y_2 + \epsilon_2   \sqrt{Y_2^2 R + A m_2}
		}
		{
			A
		},		
\end{align}
with
\begin{equation}
	R \equiv B^2 - A C, \ \ 
	\epsilon_1  = \pm 1, \  \ 
	\epsilon_2  = \pm 1, \  \   \epsilon_3= \frac{\epsilon_1}{\epsilon_2} \frac{R Y_3}{\sqrt{R^2 Y_3^2}}.
\end{equation}
Inverting Eq.~(\ref{VvRelation}) one translates these solutions to the variables of interest.  
The result being that specifying the SUSY spectrum and B-L parameters ($M_{Z_{BL}}$ and $g_{BL}$) 
as well as $Y_3$ specifies all the values for $v_{L}^{i}$ and $Y_\nu^{i3}$.

\item Inverted Hierarchy

In the case of the inverted spectrum one can use the same procedure. However, one needs to make the following replacements:  
$m_1 \leftrightarrow m_3$, $Y_1 \leftrightarrow Y_3$, $V_1 \leftrightarrow V_3$. In this way when we solve the equations for $V_i$ and $Y_i$, one obtains 
$V_3=0$, $Y_3=0$ and the solutions above where we have made the previous substitutions.

\end{itemize}
\begin{table}[htdp]
\begin{center}
\begin{tabular}{|c|c|}
\hline
$\quad$ Parameter $\quad$ & $\quad \quad$ Range $\quad \quad$
\\
\hline
$M_1$ &	100 - 1200 GeV
\\
$M_2$ &	100 - 1200 GeV
\\
$|\mu|$ &	100 - 1200 GeV
\\
$\tan \beta$ & 3 - 50
\\
$|Y_3|$ & $10^{-7} - 10^{-5}$
\\
$M_{BL}$ & 100 - 1200 GeV
\\
$M_{Z_{BL}}$	&	1000 GeV
\\
$g_{BL}$	&	0.33
\\
\hline
\end{tabular}
\end{center}
\caption{Parameters and ranges scanned for the plots in Figs.~\ref{vy}, \ref{LSP.DL.Bino}--~\ref{LSP.BR.Higgsino}.}
\label{tab:scan}
\end{table}%
In Fig.~\ref{vy} we show the allowed values for the vevs of the left-handed sneutrinos and the Dirac Yukawa couplings for a scan over the parameters listed in Table~\ref{tab:scan}. As one can appreciate the allowed values for $v_L^i$ are in the range $(10^{-2}-10)$ MeV while the Yukawa couplings change between $10^{-7}$ and $10^{-5}$.
Now, using these results we are ready to discuss all R-parity violating decays.  

\section{Lepton Number Violation and Decays}
%
At this point, the relevant pieces of this model have be laid out and the question of interesting signals can now be tackled. Since lepton number is violated, same-sign dileptons and multijet are possible final states. Such signatures are interesting since they have no SM background. However, the final states depend critically on the nature of the LSP and since R-parity is violated the possibilities are more numerous than normal, \textit{i.e.} colored and charged fields.

\begin{figure}[h!] 
	\includegraphics[scale=1.25]{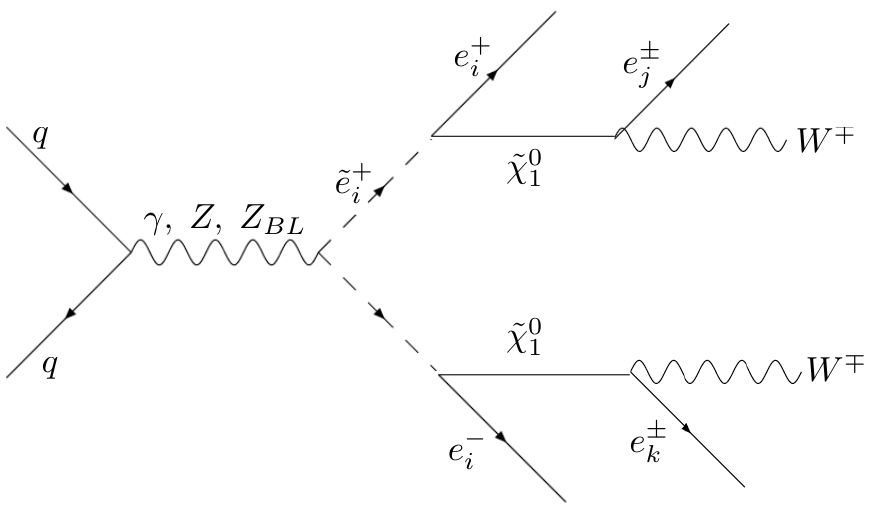}
	\put(-67,20){\Large $\tilde e^-_i$}
\caption
{
Topology of the signals with multi-leptons.
}
\label{Signal}
\end{figure}

These possibilities are briefly discuss in Appendix~\ref{LSPs}. We find that the most clear single of lepton number violation (and therefore the most interesting for us) results from the decays of a neutralino LSP through the process (see Fig~\ref{Signal})
\footnote{Throughout this paper, shorthand such as $pp \to \tilde e^* \tilde e$ represents the process $pp \to \tilde e^* \tilde e + X$, where the activity associated with $X$ has low transverse momentum and is not associated with the relevant physics of interest. Alternatively, this notation represents all possible production methods of $\tilde e^* \tilde e$ from the partons inside the proton taking their respective parton distribution functions  into account.}:
$$pp \  \to \gamma^*, Z^*, Z_{BL}^* \ \to \ \tilde{e}^*_i \tilde{e}_i \  \to  \ e^{\pm}_i \ e^{\mp}_i \ e^{\mp}_j \ e^{\mp}_k \ 4j.$$
In order to quantify this signal we will continue by investigating the decays of the $Z_{BL}$ gauge boson, the charged slepton and the neutralino and in the next section, the production mechanism for the charged sleptons at the LHC.
%
\subsection{B-L Gauge Boson Decays}
\begin{figure}[h!] 
	\includegraphics[scale=0.65]{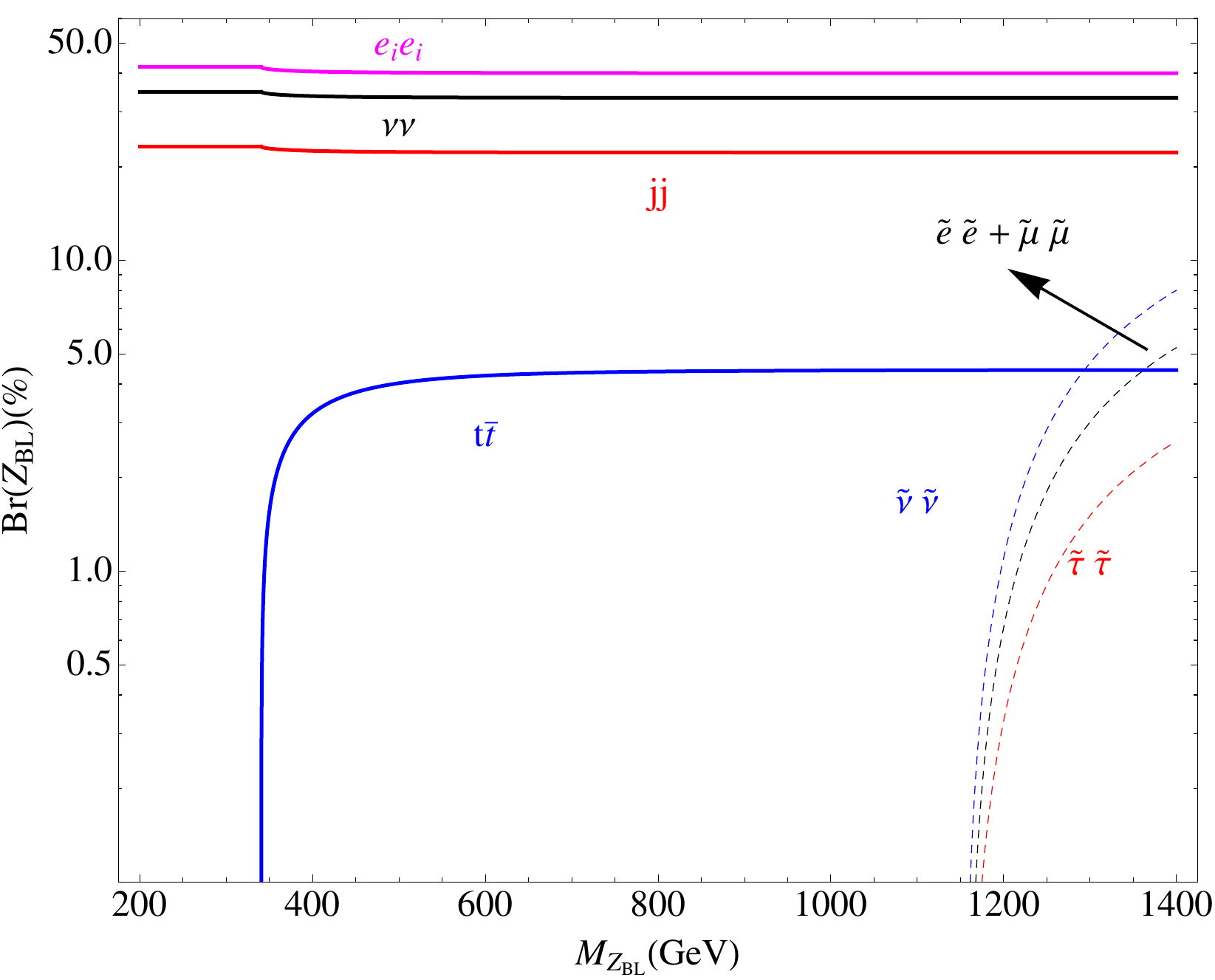}
\caption
{
$Z_{BL}$ branching ratios versus $Z_{BL}$ mass for a universal soft mass for all sfermions, $M_{SUSY}=1$ TeV.  
	Here the subscript one refers to the lightest eigenstate in each family and this case corresponds to the purely left-handed 
	slepton (zero mixing angle is assumed). Given the universal soft mass for the sfermions, only the left-handed slepton 
	channels can be open. Right-handed squark channels can open for larger values of $M_{Z_{BL}}$ but only at the 
	cost of unphysical tachyonic slepton masses.
	}
\label{ZBL.BR}
\end{figure}

%
The $Z_{BL}$ boson can decay into a pair of charged fermions, light neutrinos, and into two
sfermions. The partial widths for the decay into particles $P_1, P_2$ of masses $m_1, m_2$ are given by
\begin{equation}
	\Gamma(Z_{BL} \to P_1 P_2) = 
	\frac{1}{16 \pi M_{Z_{BL}}} \, \left| \overline{\mm} (\ZBL \to P_1 P_2) \right|^2 \, 
	\sqrt{ \left( 1 - \frac{(m_1 +m_2)^2}{M_{Z_{BL}}^2} \right) 
		 \left( 1 - \frac{(m_1 -m_2)^2}{M_{Z_{BL}^2}} \right)
	},
\end{equation}
where the squared matrix elements for specific final states are in Appendix~\ref{ZPDecay}.
The $Z_{BL}$ branching ratios are plotted in Fig.~\ref{ZBL.BR} for a fixed soft universal mass for all sfermions, $ M_{SUSY}=1$ TeV, versus the $Z_{BL}$ mass. 
Decay channels into SUSY particles only open up for a $Z_{BL}$ mass of around $1.2$ TeV and at much 1.4 TeV, the sleptons become tachyonic. 
As can be appreciated from Fig.~\ref{spectrum}, tachyonic sleptons are reached before decay channels into the squarks can open. 
The branching ratios for the sleptons are divided up into sneutrinos, smuons plus selectrons and staus in anticipation of the associated signals. 
However, each individual slepton pair has the same branching ratio, about $2.5\%$ at $m_{Z_{BL}} = 1.4$ TeV.
In Fig. 6 we show the predictions for the total decay width and the invisible decays of the $Z_{BL}$ gauge boson.
Since the new gauge boson can decay into five light neutrinos the invisible decay can be large, a few GeV when 
the mass is above 1 TeV. These properties of the $Z_{BL}$ are very important in order to discover this theory at the LHC.
In summary, one can say that the new neutral gauge boson is B-L like with branching ratios
$$ \rm{Br}( Z_{BL} \to e^+_i e^-_i) \sim 40 \%, \  \rm{Br} (Z_{BL}  \to \nu \nu) \sim 35 \% ,  \rm{Br}(Z_{BL} \to jj) \sim 20 \%, \rm{and} \ \rm{Br} (Z_{BL} \to \bar{t} t) \sim 5 \%,$$
since the branching ratios for the SUSY decays are small  and the invisible decay can be large. For example when 
$M_{Z_{BL}} = 1.2 \ \rm{TeV}$ the invisible decay width is $\Gamma_{Z_{BL}}(\rm{invisible}) \sim 3$  GeV.
\\
\begin{figure}[h!]
	\includegraphics[scale=0.7]{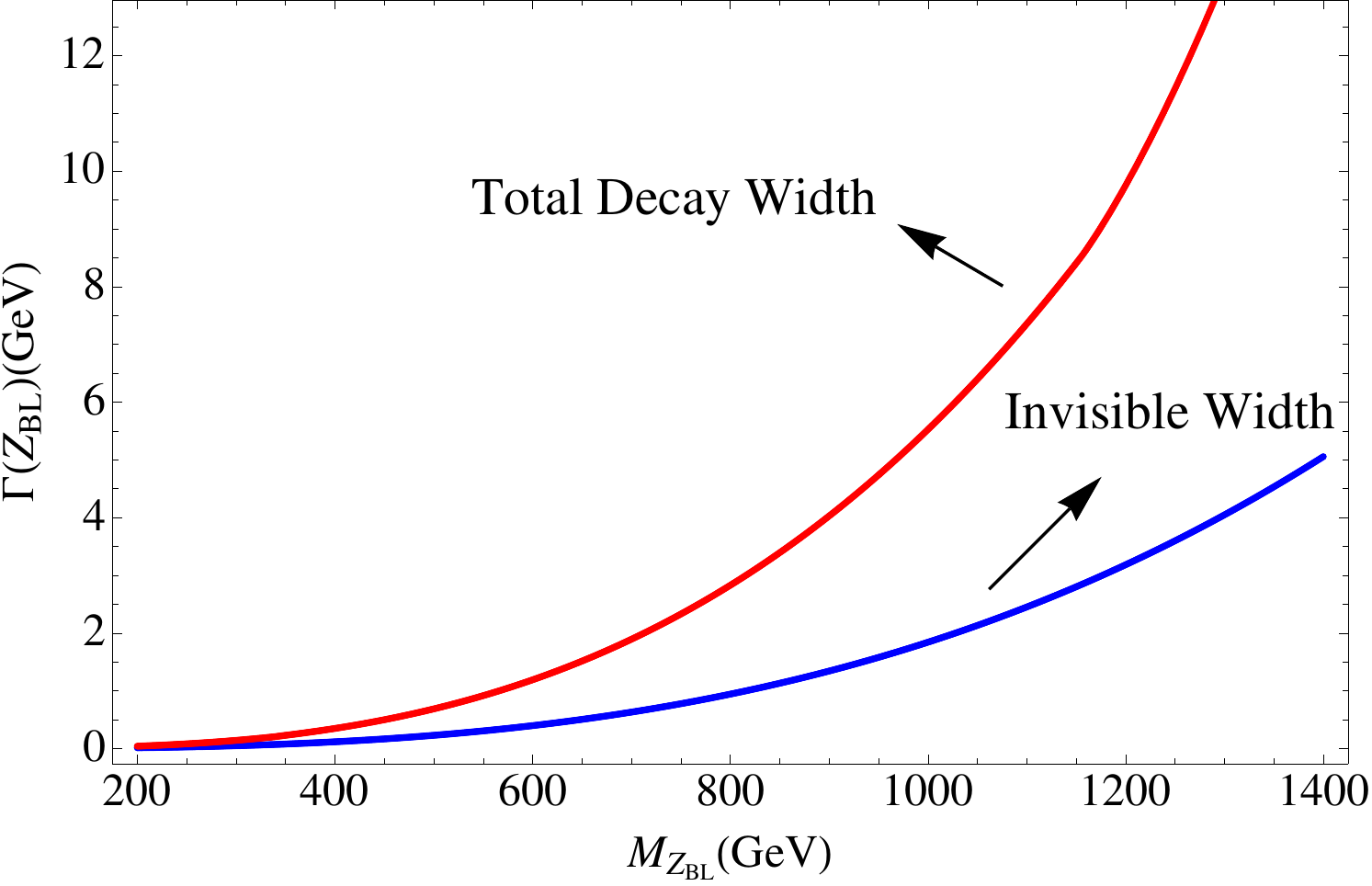}
\caption
{The total decay width, $\Gamma(Z_{BL})$, versus $Z_{BL}$ mass for a universal soft mass for all sfermions, $M_{SUSY}=1$ TeV.  
}
\label{GZBL}
\end{figure}
%
\subsection{Charged Slepton Decays}
%
\label{slep.dk}
The leading decay channels for the charged left-handed sleptons, $\tilde e_i^\pm$, are the decays into neutralinos and charginos 
\begin{equation*}
	\tilde e_i^\pm \to e_i^\pm \tilde \chi^0_a,
	\ \ \ \tilde e_i^\pm \to \nu \tilde \chi^\pm_A,
	\ \ \ \tilde e_i^\pm \to \tilde \nu_j W^\pm,
\end{equation*}
where $i,j$ are lepton generational indices, $a$ labels the neutralinos from lightest to heaviest and $A$ labels the charginos from lightest to heaviest.  In addition to these decay modes there are various $R$-parity violating decays which only dominate when the slepton is the LSP. The last channel above usually involves an off-shell product particle (a three-body decay) and is therefore suppressed.  The decay widths for the remaining two relevant channels are given by
\begin{align}
\label{slepton.neutralino}
	\Gamma(\tilde e_i^\pm \to e_i^\pm \tilde \chi^0_a) & = \frac{m_{\tilde e_i}}{ 32 \pi}
	\left(
		\left| g_{BL} N_{2a} + g_1 N_{3a} + g_2 N_{4a} \right|^2
		+ \left| \frac{2 \sqrt{2} m_{e_i}}{v_d} N_{5a}\right|^2
	\right)
		\left(1 - \frac{m_{\tilde \chi^0_a}^2} {m_{\tilde e_i}^2}\right)^2,
	\\
\label{slepton.chargino}
	\Gamma(\tilde e_i^\pm \to \nu_j \tilde \chi^\pm_A) & = \frac{m_{\tilde e_i}}{ 16 \pi}
			\left| g_2 V^-_{1A} \right|^2
		\left(1 - \frac{m_{\tilde \chi^-_A}^2} {m_{\tilde e_i}^2}\right)^2,
\end{align}
where $N_{ab}$ diagonalizes the neutralino mass matrix and in the chargino sector one has $V^- X V^+  = \text{diag} \left(m_{\tilde \chi_1}, m_{\tilde \chi_2}\right).$ Due to the presence of the right-handed neutrino and $\tilde B'$, position 2, 3, 4 and 5 in the $N$ refer to $\tilde B', \tilde B, \tilde W  \ \text{and}, \tilde{H}_d$, respectively.  While it is hard to make predictions for the branching ratios of the charged sleptons decaying into a charged lepton and LSP without knowing the details of the spectrum, we briefly outline the best case scenario. Our final results will be given both with this best case scenario in mind and with arbitrary $\text{Br}(\tilde e_i \to e_i \tilde \chi_1^0)$.

For a mostly bino LSP, it is possible that charged selectrons decay one hundred percent into the LSP since the charginos and other neutralinos could be heavier.  For a wino LSP however, the lightest chargino channel is very likely to be open.  Because of the factor of 2 difference between Eq.~(\ref{slepton.neutralino}), and Eq.(\ref{slepton.chargino}) if all other neutralinos and charginos are kinematically disallowed, one can expect
\begin{equation}
\frac{\Gamma(\tilde e_i^\pm \to e_i^\pm \tilde \chi^0_1)}{\Gamma(\tilde e_i^\pm \to \nu_j \tilde \chi^\pm_1)} \sim \frac{1}{2},
\end{equation}
meaning a $33 \%$ branching ratio for a charged slepton into a wino LSP. Meanwhile, the left-handed selectron does not couple to the charged Higgsino. 
Therefore, in the same limit where all other neutralinos and charginos are out of kinematic range, the charged sleptons decays one hundred percent into the LSP for a Higgsino LSP.

\subsection{Neutralino Decays}
%
\label{chi.decay}
The leading decay channels for the lightest neutralino, $\tilde{\chi}_1^0$, include
\begin{align}
\label{chi.decays}
	\tilde{\chi}_1^0 \to e^{\pm}_i W^{\mp},
	\  \
	\tilde{\chi}_1^0 \to \nu_i Z,
	\  \
	\tilde{\chi}_1^0  \to  \nu_i h_k,
	\ \
	\tilde{\chi}_1^0 \to e_i^\pm H^\mp.
\end{align}
The amplitude for the two first channels are proportional to the mixing between the leptons and neutralinos, while the last one is proportional to the Dirac-like Yukawa terms. While decays to all the MSSM 
Higgses are possible, typically, only the lightest MSSM Higgs, $h$ ($k = 1$), is light enough for the scenario we consider 
here and so we will only take it into account.

A naive estimation of the decay width yields
\begin{equation}
\Gamma (\tilde{\chi}_1^0) \sim \frac{g_2^2}{32 \pi} |V_{\nu \chi }|^2 M_{\chi},
\end{equation}
where $V_{\chi \nu}$ is the mixing between the neutralino and neutrino which is proportional to $\sqrt{m_\nu/M_{\chi}}$.
Assuming that $m_\nu < 0.1$ eV the decay length one finds $L(\tilde{\chi}_1^0) \gg 0.6$ mm.
Therefore, even without making a detailed analysis of the decays of the lightest neutralino one expects signals with lepton 
number violation and displaced vertices in part of the parameter space. For a recent analysis of the neutralino decays in R-parity violating models see Ref.~\cite{Porod, Bobrovskyi:2011vx}.

The specific decay width expressions are
\begin{eqnarray}
\label{lWL}
\Gamma^{e_i W_L}&\equiv&
\Gamma(\tilde{\chi}_a \to e_i^\pm W_L^\mp)= \frac{g_2^2}{64\pi
M_W^2}|V_{i a}|^2 m_{\tilde{\chi}_a}^3 \left(1-  \frac{m_W^2}{m_{\tilde{\chi}_a}^2} \right)^2,
\\
\label{lWT}
\Gamma^{e_i W_T}&\equiv&\Gamma(\tilde{\chi}_a \to e_i^\pm W_T^\mp)=
	\frac{g_2^2}{32\pi  }|V_{i a}|^2 m_{\tilde{\chi}_a}    \left(1-  \frac{m_W^2}{m_{\tilde{\chi}_a}^2} \right)^2,
\\
\label{nuZL}
\Gamma^{\nu_i Z_L}&\equiv&\Gamma(\tilde{\chi}_a\to \nu_i Z_L)=\frac{g_2^2}{
64\pi M_W^2}|V_{i a}|^2 m_{\tilde{\chi}_a}^3 \left(1-  \frac{m_Z^2}{m_{\tilde{\chi}_a}^2} \right)^2,
\\
\label{nuZT}
\Gamma^{\nu_i Z_T}&\equiv&\Gamma(\tilde{\chi}_a \to \nu_i Z_T)=\frac{g_2^2}{
32\pi c_W^2}|V_{i a}|^2 m_{\tilde{\chi}_a} \left(1-  \frac{m_Z^2}{m_{\tilde{\chi}_a}^2} \right)^2,
\\
\label{nuh}
\Gamma^{\nu_i h}&\equiv&\Gamma(\tilde{\chi}_a \to \nu_i h) = 
	\frac{g_2^2}{64\pi M_W^2}
	|V_{i a}|^2 \cos^2 \alpha \;
	m_{\tilde{\chi}_a}^3 \left(1-  \frac{m_h^2}{m_{\tilde{\chi}_a}^2} \right)^2.
\end{eqnarray}
Here $\alpha$ is the mixing angle in the Higgs sector and in the decoupling limit, $M_A^2 \gg M_Z^2$, which we assume: $\cos \alpha = \sin \beta$. The index $i$ indicates the generation of lepton and $a$ the neutralino with $a = 6$ is the heaviest and $a=1$ is the lightest.  These expressions depend on the mixing between the light neutrinos and the neutralinos, $V_{i a}$, which is derived in Appendix~\ref{Vla}.  Of course, only the decays of the LSP are relevant since the decays of the other neutralinos will be dominated by $R$-parity conserving decays, hence $a=1$ for our purposes.
\begin{table}[htdp]
\begin{center}
\begin{tabular}{|c|c|c|c|c|}
\hline
Hierarchy & LSP Personality & Bino & Wino & Higgsino
\\
\hline
&Decay Length &	1.1 mm & 0.03 mm & $1 \times 10^{-4}$ mm
\\
&2Br$(\chi_1^0 \to e^- W^+)$  & 4 $\%$	& 2 $\%$ & 13 $\%$
\\
&2Br$(\chi_1^0 \to \mu^- W^+)$  & 26 $\%$	& 12 $\%$ & 27 $\%$
\\
NH &2Br$(\chi_1^0 \to \tau^- W^+)$  & 61 $\%$	& 54 $\%$ & 30 $\%$
\\
&Br$(\chi_1^0 \to \nu Z^0)$  & 10 $\%$	& 29 $\%$ & 28 $\%$
\\
&Br$(\chi_1^0 \to \nu h)$  & 0 $\%$	& 3 $\%$ & 1 $\%$
\\
\hline
\hline
&Decay Length					& 0.6 mm	& 0.01 mm	& $1 \times 10^{-5}$ mm
\\
&2Br$(\chi_1^0 \to e^- W^+)$ 		& 17 $\%$	& 3 $\%$			& 25 $\%$
\\
&2Br$(\chi_1^0 \to \mu^- W^+)$	& 36 $\%$	& 32 $\%$		& 19 $\%$
\\
IH &2Br$(\chi_1^0 \to \tau^- W^+)$   & 38 $\%$		& 34 $\%$      & 26 $\%$
\\
&Br$(\chi_1^0 \to \nu Z^0)$		& 10 $\%$		& 29 $\%$	& 28 $\%$
\\
&Br$(\chi_1^0 \to \nu h)$			& 0 $\%$			& 3 $\%$		& 1 $\%$
\\
\hline
\end{tabular}
\end{center}
\caption
{
Values of interest for a sample point in parameter space: $\epsilon_1 = \epsilon_2 = 1, Y_3 = Y_1 = 10^{-6}, M_{Z_{BL}} = M_{\tilde{B}'} = 1 \text{ TeV}, \tan \beta =5 \text{ and } m_h = 125 \text{ GeV}$.  Here $M_1, M_2 \text{ and } \mu$ are 100 GeV, 500 GeV and 500 GeV for the bino LSP case, 500 GeV, 150 GeV and 500 GeV for the wino LSP case and 500 GeV, 500 GeV and 100 GeV for the Higgsino LSP case respectively.
}
\label{tab:test.points}
\end{table}
Table \ref{tab:scan}  displays the values of interest for a specific point in parameter space to gain an appreciation for possible values. Notice that in these scenarios the branching ratios for the channels with charged leptons can be large. 

\begin{figure}[h!] 
	\includegraphics[scale=0.85]{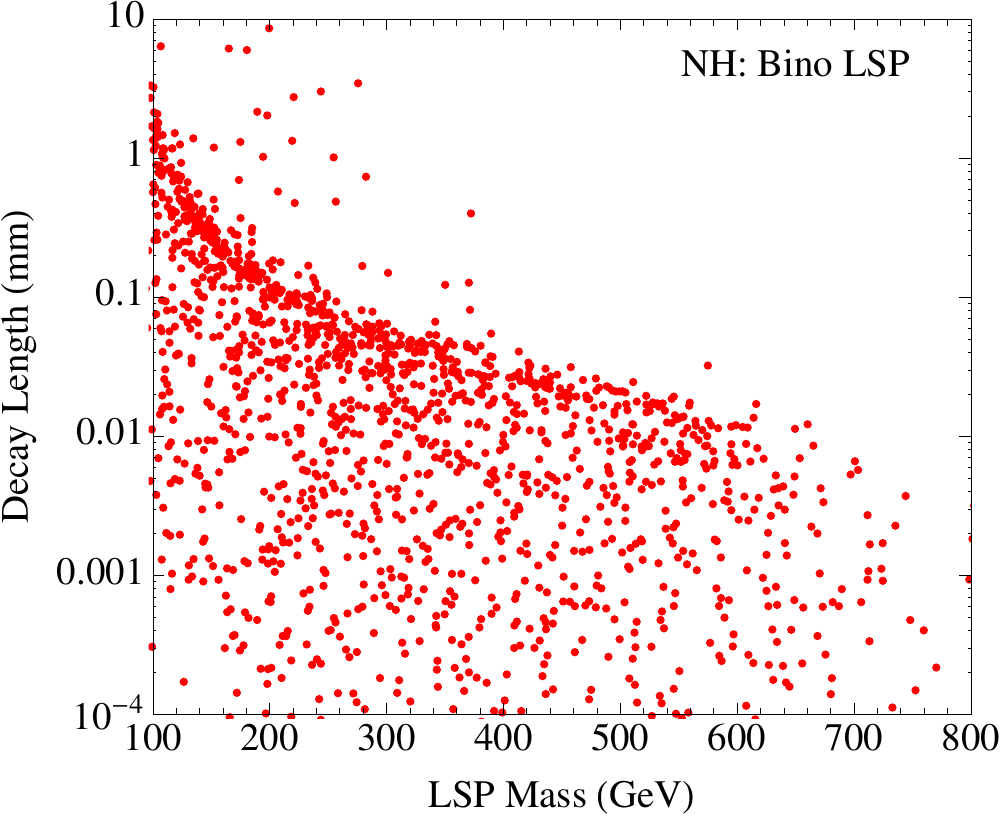}
	\put(-40,-4){(a)}
	\includegraphics[scale=0.85]{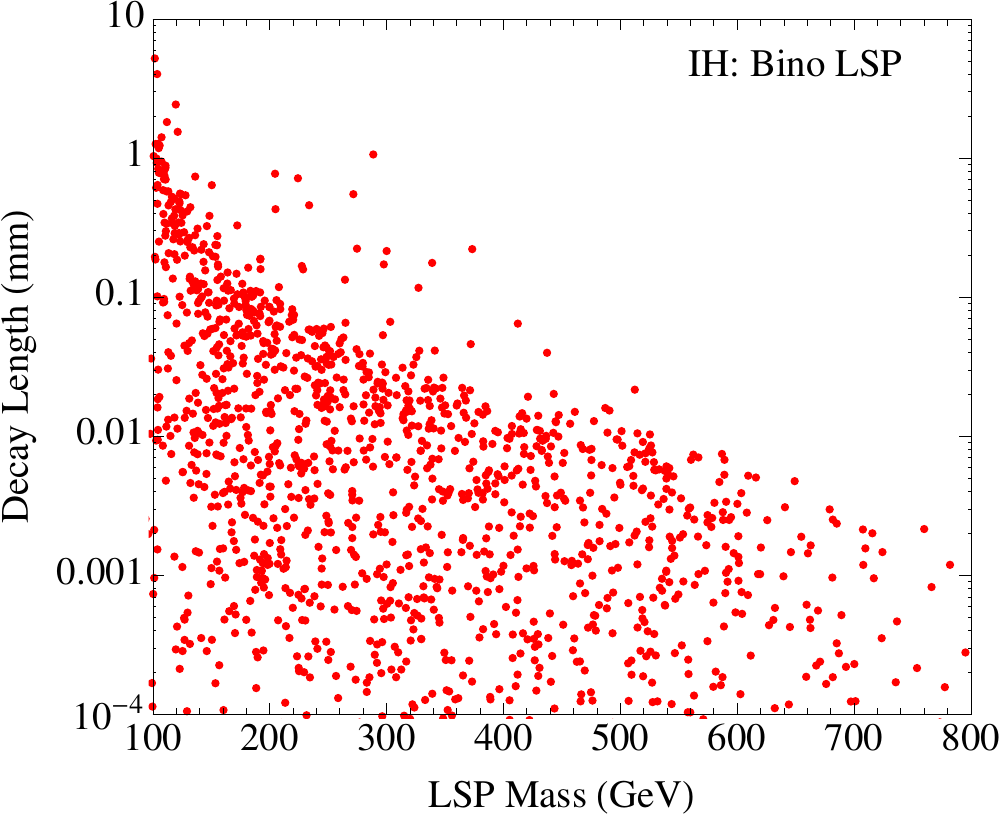}
	\put(-40,-4){(b)}
\caption{
Decay length in millimeters versus LSP mass for a dominantly bino LSP in (a) for a NH and in (b) for an IH. 
Parameters are scanned according to the ranges specified in Table~\ref{tab:scan} and over all values of $\epsilon_1$ and $\epsilon_2$.}
\label{LSP.DL.Bino}
\end{figure}

\begin{figure}
	\includegraphics[scale=0.85]{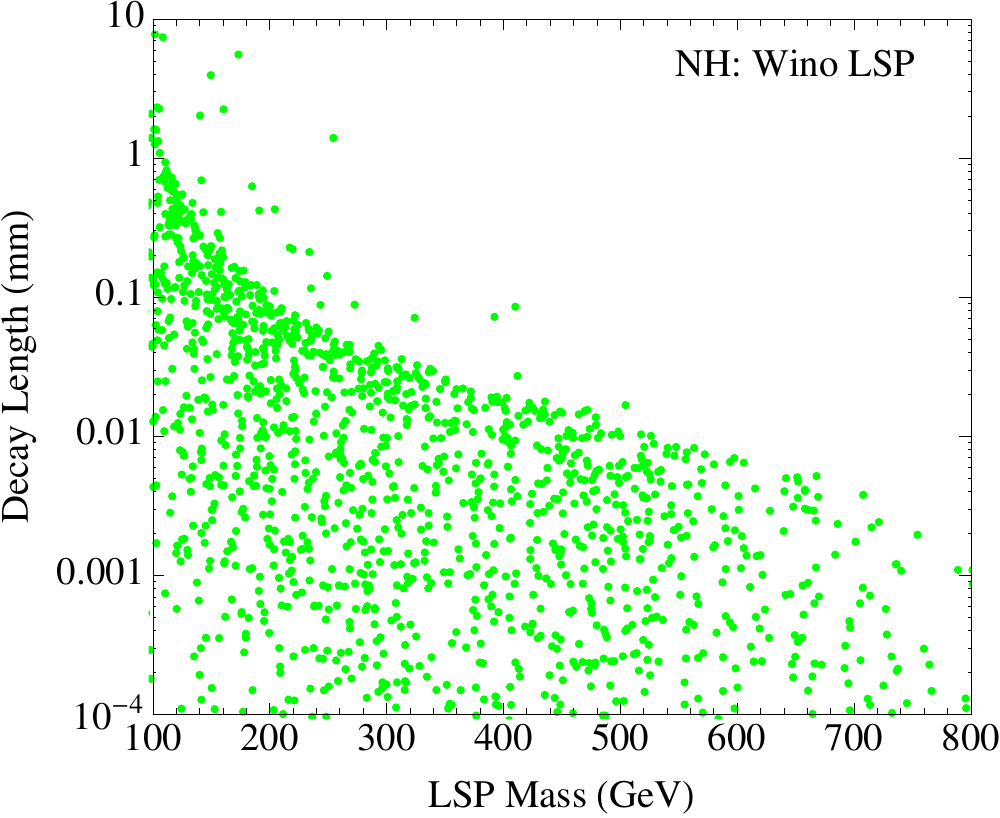}
	\put(-40,-4){(a)}
	\includegraphics[scale=0.85]{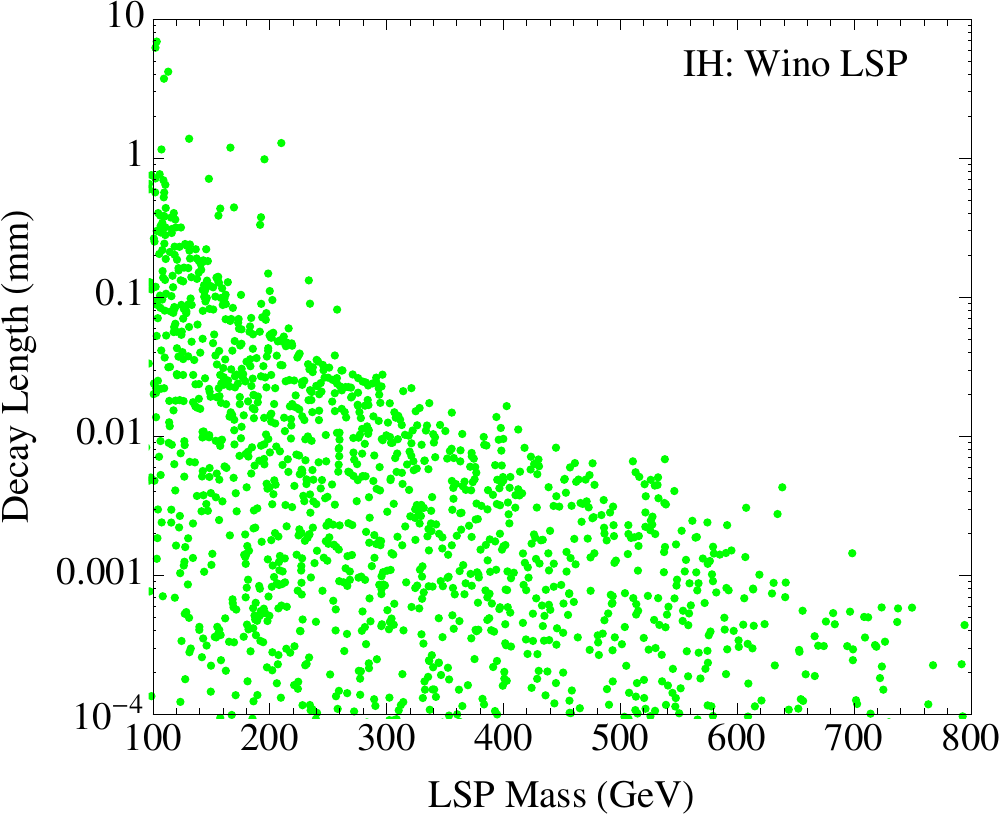}
	\put(-40,-4){(b)}
\caption{
Decay length in millimeters versus LSP mass for a dominantly wino LSP in (a) for a NH and in (b) for an IH. 
Parameters are scanned according to the ranges specified in Table~\ref{tab:scan} and over all values of $\epsilon_1$ and $\epsilon_2$.}
\label{LSP.DL.Wino}
\end{figure}

\begin{figure}
	\includegraphics[scale=0.85]{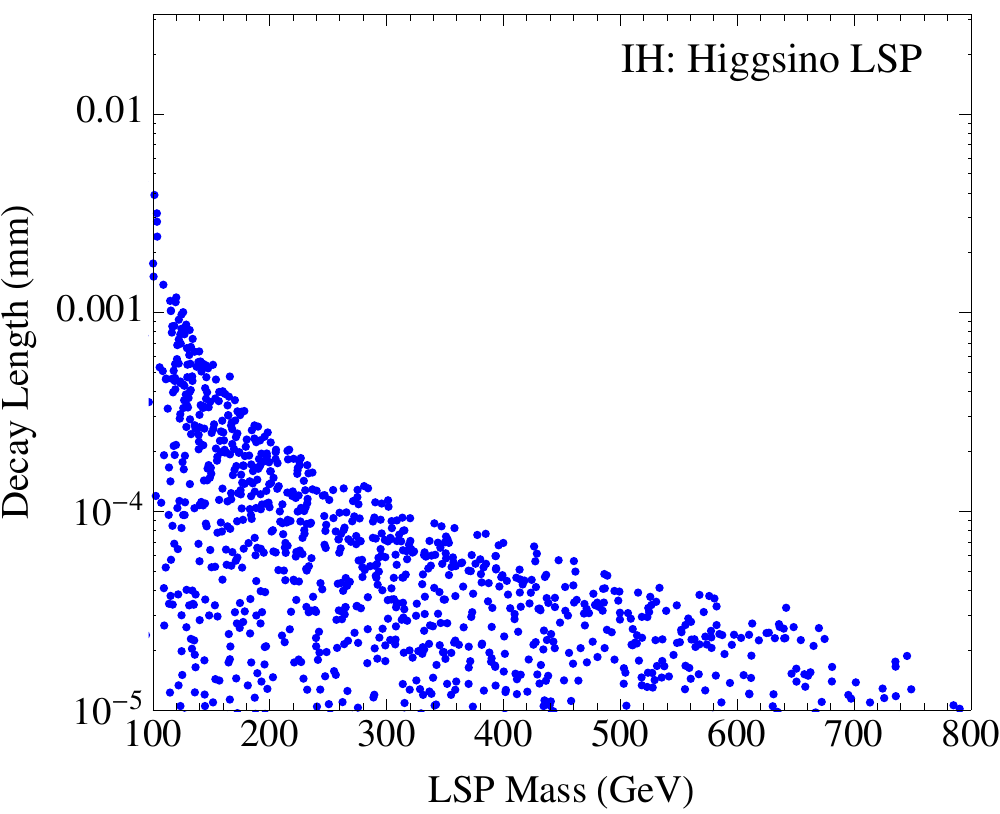}
	\put(-40,-4){(a)}
	\includegraphics[scale=0.885]{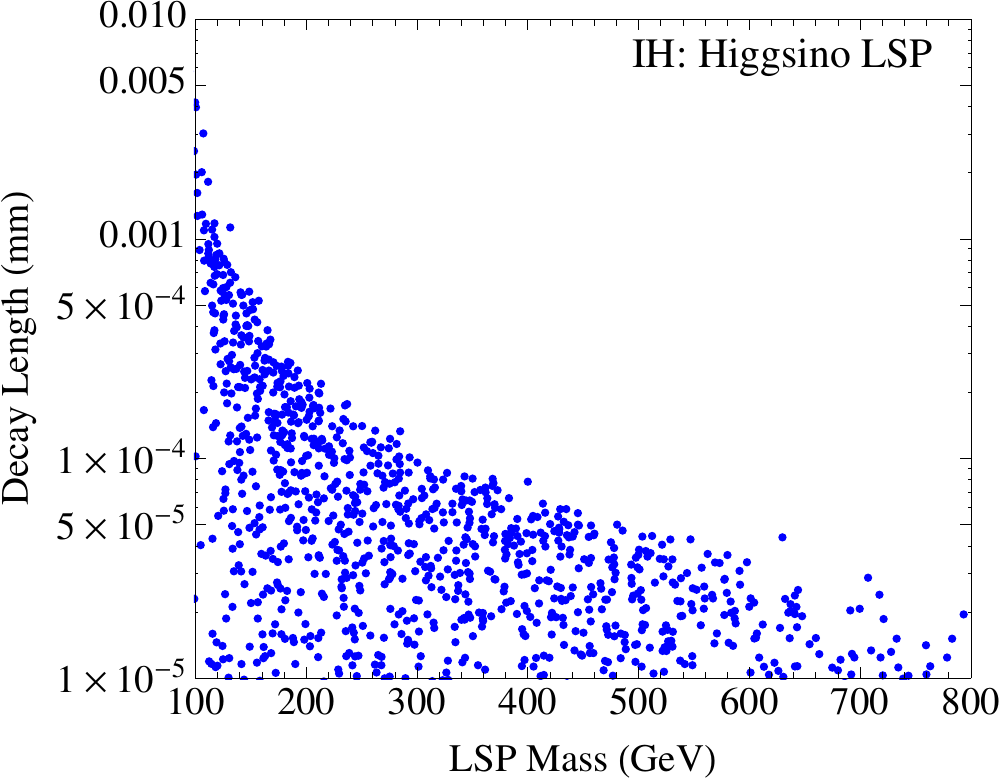}
	\put(-40,-4){(b)}
\caption
{Decay length in millimeters versus LSP mass for a dominantly Higgsino LSP in (a) for a NH and in (b) for an IH. 
Parameters are scanned according to the ranges specified in Table~\ref{tab:scan} and over all values of $\epsilon_1$ and $\epsilon_2$.}
\label{LSP.DL.Higgsino}
\end{figure}

In Figs.~\ref{LSP.DL.Bino}-\ref{LSP.DL.Higgsino} are the decay lengths versus LSP mass resulting from a scan over all the possible values of $\epsilon_1$ and $\epsilon_2$ and over the parameters and ranges specified in Table~\ref{tab:scan}.  The points are divided according to the largest component of the LSP and the neutrino hierarchy with a dominantly bino, wino and Higgsino LSP in the NH shown in (a) and for an IH in (b), respectively. The relevant decay lengths can be understood by studying the mixings in Eq.~(\ref{neutralino}).  Since the higgsino-neutrino decay strength is the largest, $\sim Y_\nu v_R$, the Higgsino LSP has the shortest decay length.  It is followed by the wino LSP with mixing $\sim g_2 v_L$ and finally the bino with coupling $\sim g_1 v_L$ and therefore the largest possible decay lengths.  Displaced vertices associated with the lifetime of the LSP will only be discernible in a very limited part of the parameter space.


\begin{figure}[t!] 
	\includegraphics[scale=0.85]{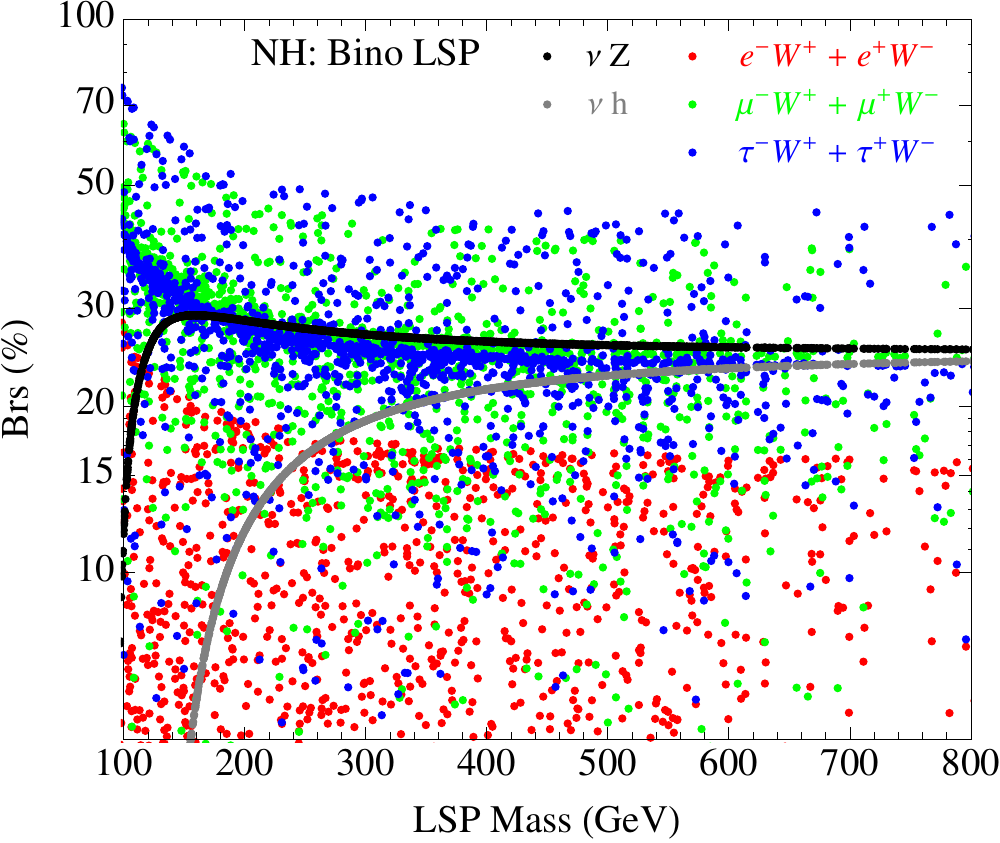}
	\put(-40,-4){(a)}
	\includegraphics[scale=0.85]{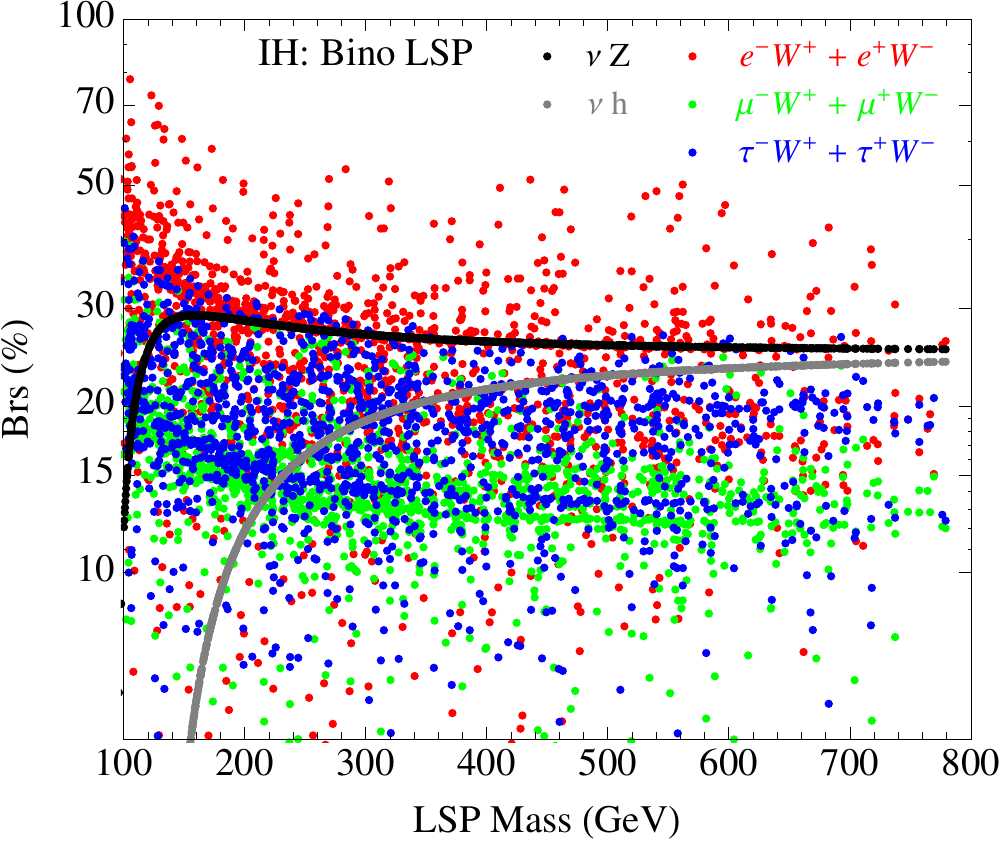}
	\put(-40,-4){(b)}
\caption{
LSP branching ratios versus LSP mass for a dominantly bino LSP in (a) for a NH and in (b) for an IH. 
	Parameters are scanned according to the ranges specified in Table~\ref{tab:scan} and over all values of $\epsilon_1$ and $\epsilon_2$.
	}
\label{LSP.BR.Bino}
\end{figure}

\begin{figure}
	\includegraphics[scale=0.85]{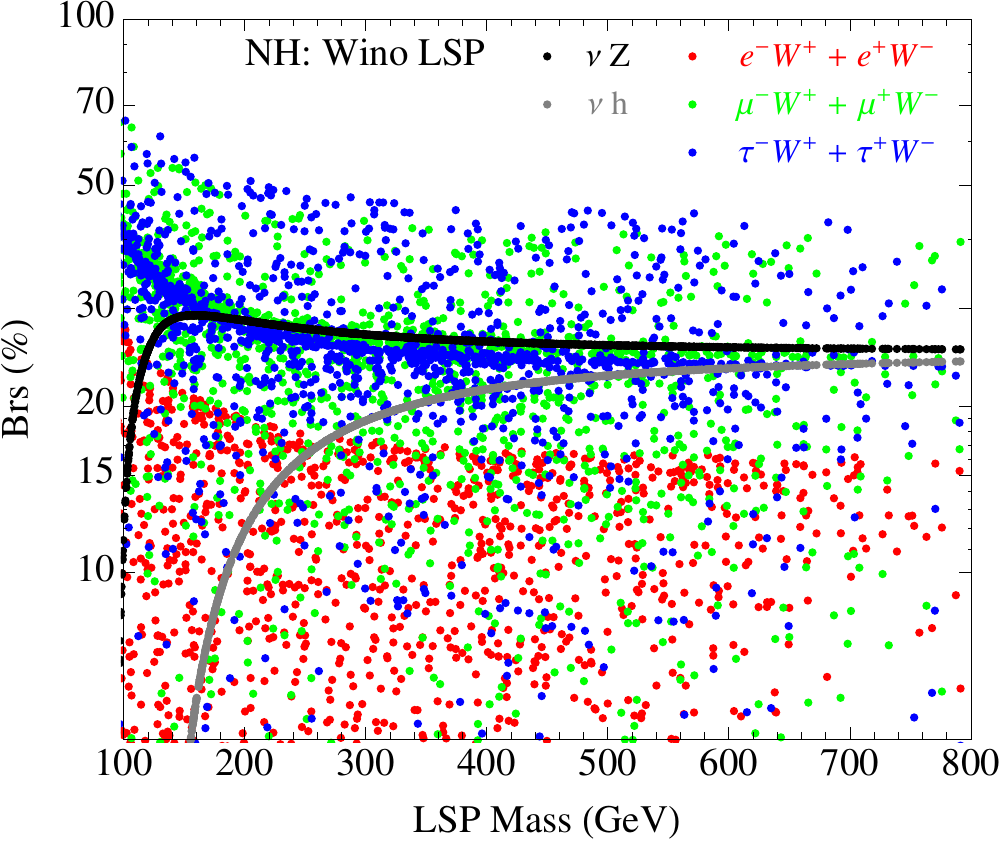}
	\put(-40,-4){(a)}
	\includegraphics[scale=0.85]{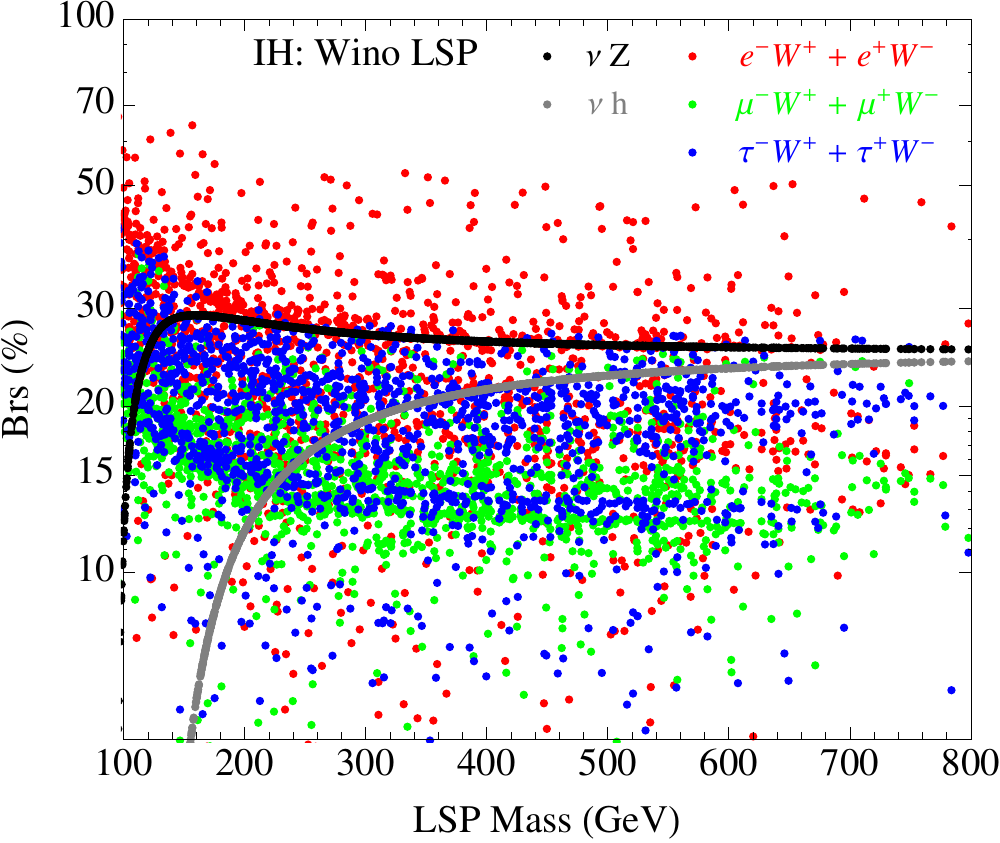}
	\put(-40,-4){(b)}
\caption{LSP branching ratios versus LSP mass for a dominantly wino  LSP in (a) for a NH and in (b) for an IH. 
Parameters are scanned according to the ranges specified in Table~\ref{tab:scan} and over all values of $\epsilon_1$ and $\epsilon_2$.}
\label{LSP.BR.Wino}
\end{figure}

\begin{figure}
	\includegraphics[scale=0.85]{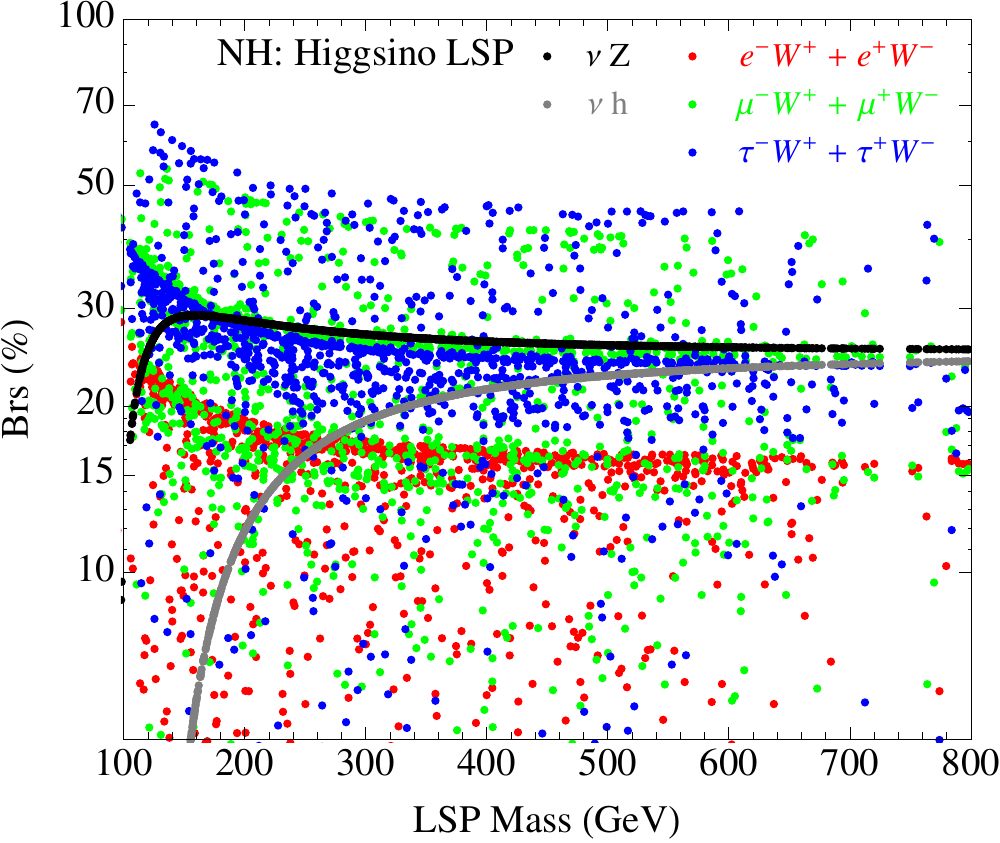}
	\put(-40,-4){(a)}
	\includegraphics[scale=0.85]{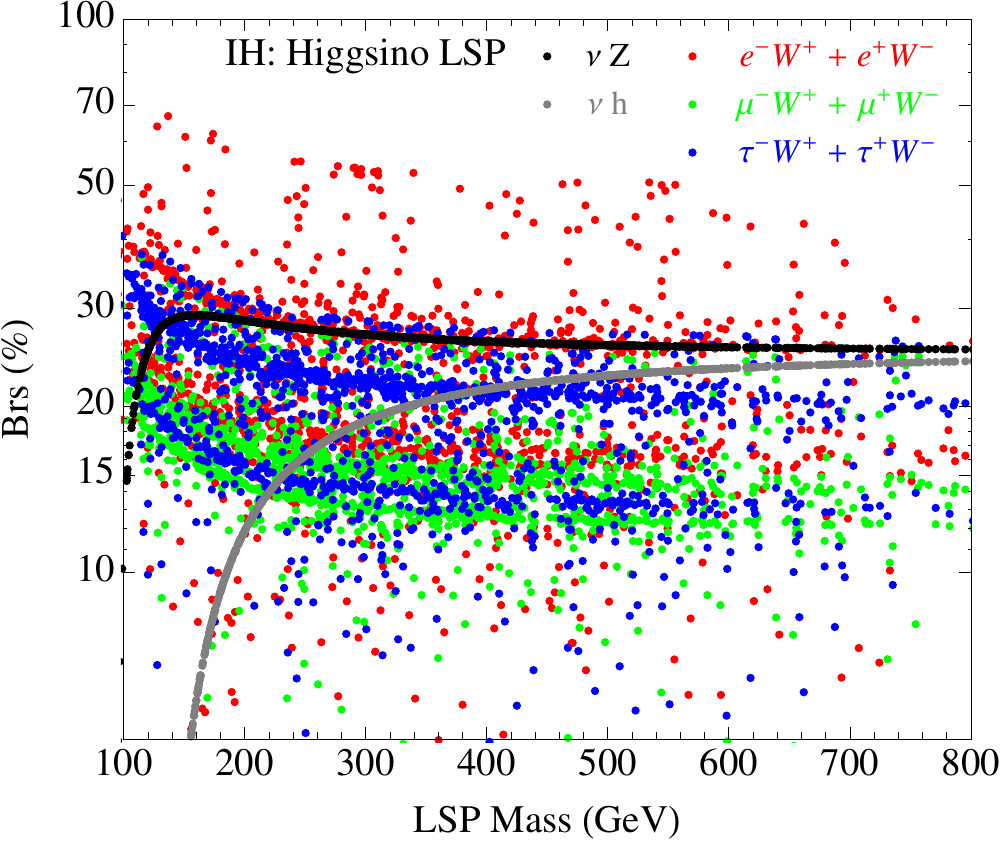}
	\put(-40,-4){(b)}
\caption
{
	LSP branching ratios versus LSP mass for a dominantly Higgsino LSP in (a) for a NH and in (b) for an IH. 
	Parameters are scanned according to the ranges specified in Table~\ref{tab:scan} and over all values of $\epsilon_1$ and $\epsilon_2$.
}
\label{LSP.BR.Higgsino}
\end{figure}

The LSP branching ratios into the various possible channels versus the LSP mass are displayed in Figs.~\ref{LSP.BR.Bino}-\ref{LSP.BR.Higgsino} scanning over the parameters in Table~\ref{tab:scan} and plotting the dominantly bino, wino and Higgsino LSP in (a) for a NH and in (b) for an IH.  The lack of variance with scanned parameters displayed in the $\nu \; Z$ and $\nu \; h$ channels are due to the sum over all three flavors of neutrinos and  also exists for the sum over the three charged lepton plus $W^\pm$ channels which total about $50 \%$ (or more if the other channels hadn't fully turned on yet).  Although it is not obvious from the Figs.~\ref{LSP.BR.Bino}-\ref{LSP.BR.Higgsino}, the branching ratio to the electron $W^\pm$ channel is always smaller then either the $\mu^\mp W^\pm$ or the $\tau^\mp W^\pm$ in the NH. Since we now know the properties of the neutralinos and selectrons decays we are ready to study the production channels. 


\section{Production Mechanisms and Signals}
The lepton number violating signal discussed in the beginning of the previous section proceeds from the pair production of charged sleptons. While the MSSM contributions to this production drop rapidly with charged slepton mass (since that mass must be above the $Z$ threshold), in this model further contributions due to the $Z'$ resonance can significantly increase the cross section. These enhancement are discussed first followed by a study of the expected number of events at the LHC.

\subsection{Sleptons Production Mechanisms}
%
The main production channel for the charged sleptons is through the photon, the Z gauge boson and the $Z_{BL}$ boson
\begin{displaymath}
q(p_1) \bar{q}(p_2) \  \to \ \gamma, Z^*, Z^*_{BL} \  \to \  \tilde{e}^* (p_3) \tilde{e}  (p_4).
\end{displaymath}
The hadronic cross section is given by 
\begin{equation}
d \sigma_{pp \to \tilde{e}^* \tilde{e}} (s) = \sum_{q=u,d,c,s}  \int_{\tau_0}^{1} d \tau \ \frac{d {\cal{L}}^{pp}_{q \bar{q}}}{d \tau}  d \hat{\sigma}_{q \bar{q} \to \tilde{e}^* \tilde{e}} (\hat{s}),
\end{equation}
where $\tau_0=4 M_{\tilde{e}}^2 / s$ and the differential partonic cross section is
\begin{equation}
d \hat{\sigma}_{q \bar{q} \to \tilde{e}^* \tilde{e}} (\hat{s}) = | \overline{{\cal M}}_{q \bar{q} \to \tilde{e}^* \tilde{e}} (\hat{s})|^2 \frac{\rm{d PS}^{(2)}}{2 \hat{s}}.
\end{equation}
Here $\rm{d PS}^{(2)}=d \hat{t}/ 8 \pi \hat{s}$ is the two particle phase-space element and $\hat{s}= \tau s$, where $s$ is the hadronic center-of-mass energy squared.
As it is well-known the parton luminosities are given by
\begin{equation}
\frac{\rm{d} {\cal L}_{ab}^{AB} }{\rm{d} \tau}= \frac{1}{1 + \delta_{ab}} \int_{\tau}^1 \left(   f_{a/A} (x,\mu)  f_{b/B} (\frac{\tau}{x}, \mu) +  f_{a/B} (\frac{\tau}{x}, \mu) f_{b/A} (x, \mu) \right),
\end{equation}
where the functions $f_{a/A} (x,\mu)$ are the particle distribution functions (PDFs).
The amplitude squared for these processes can be written as
\begin{equation}
 | \overline{{\cal M}}_{q \bar{q} \to \tilde{e}^* \tilde{e}} (\hat{s})|^2 = \frac{2}{3} \left( \hat{u} \hat{t} - M_{\tilde{e}}^4 \right)   | {{\cal A}}_{q \bar{q} \to \tilde{e}^* \tilde{e}} (\hat{s})|^2,
\end{equation}
with $\hat{s}=(p_1 + p_2)^2$, $\hat{t}=(p_1-p_3)^2$, $\hat{u}=(p_1-p_4)^2$ and
\begin{equation}
{{\cal A}}_{q \bar{q} \to \tilde{e}^* \tilde{e}} (\hat{s})=\frac{C_{q\bar{q}\gamma} C_{\gamma \tilde{e}^* \tilde{e} }}{\hat{s}} \ + \ \frac{2C_{q\bar{q} Z} C_{Z \tilde{e}^* \tilde{e} }}{\hat{s}-M_Z^2 + i M_Z \Gamma_Z}
\ + \ \frac{C_{q\bar{q} Z_{BL}} C_{Z_{BL} \tilde{e}^* \tilde{e} }}{\hat{s}-M_{Z_{BL}}^2 + i M_{Z_{BL}} \Gamma_{Z_{BL}}},
\end{equation}
where
\begin{eqnarray}
C_{\bar{q} q \gamma}&=&e_q \ e, \  \  C_{\gamma \tilde{e}^*_L \tilde{e}_L}=e_l \ e, \ \ C_{Z \tilde{e}^*_L \tilde{e}_L} = \frac{e}{\sin 2 \theta_W} L_e, \\
C_{\bar{q}_L q_L Z} &=& \frac{e L_q}{\sin 2 \theta_W}, \  \   C_{\bar{q}_R q_R Z} = \frac{e R_q}{\sin 2 \theta_W}, \ \ 
C_{\bar{f} f Z_{BL}}= g_{BL} \frac{n_{BL}^f}{2}, \ \ C_{\tilde{f}^* \tilde{f} Z_{BL}}= g_{BL} \frac{n_{BL}^f}{2}.
\end{eqnarray}
Here $L_f=I_f^3 -  e_f \sin^2 \theta_W$ and $R_f=-e_f \sin^2 \theta_W$, where $I_f^3$ is the isospin of the fermion f.
Now, using the equations
\begin{eqnarray}
\hat{u}&=& 2 M_{\tilde{e}}^2 - \hat{t}-\hat{s}, \\
\hat{t} &=& M_{\tilde{e}}^2 - \frac{\hat{s}}{2} + y \sqrt{\hat{s} \left( \frac{\hat{s}}{4} - M_{\tilde{e}}^2 \right)} ,
\end{eqnarray}
we can compute the cross section
\begin{equation}
\sigma_{pp \to \tilde{e}^*_L \tilde{e}_L} (s)= \sum_{q=u,d,s} \ \int_{-1}^1 dy  \int_{\tau_0}^1 d \tau \frac{d {\cal L}^{pp}_{q \bar{q}}}{d \tau} \ \sigma (M_{\tilde{e}}, y, \tau, s). 
\end{equation}

\begin{figure}[h!] 
	\includegraphics[scale=0.8]{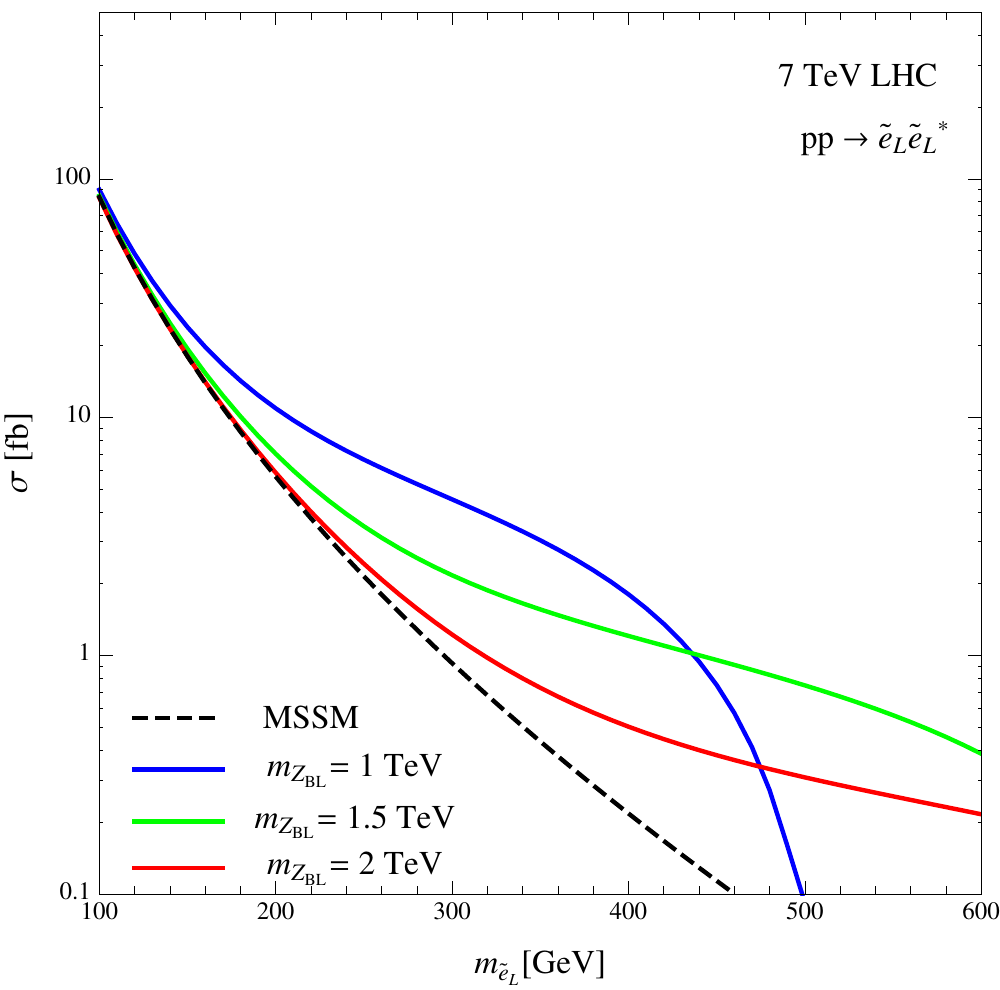}
	\put(-40,-4){(a)}
	\includegraphics[scale=0.8]{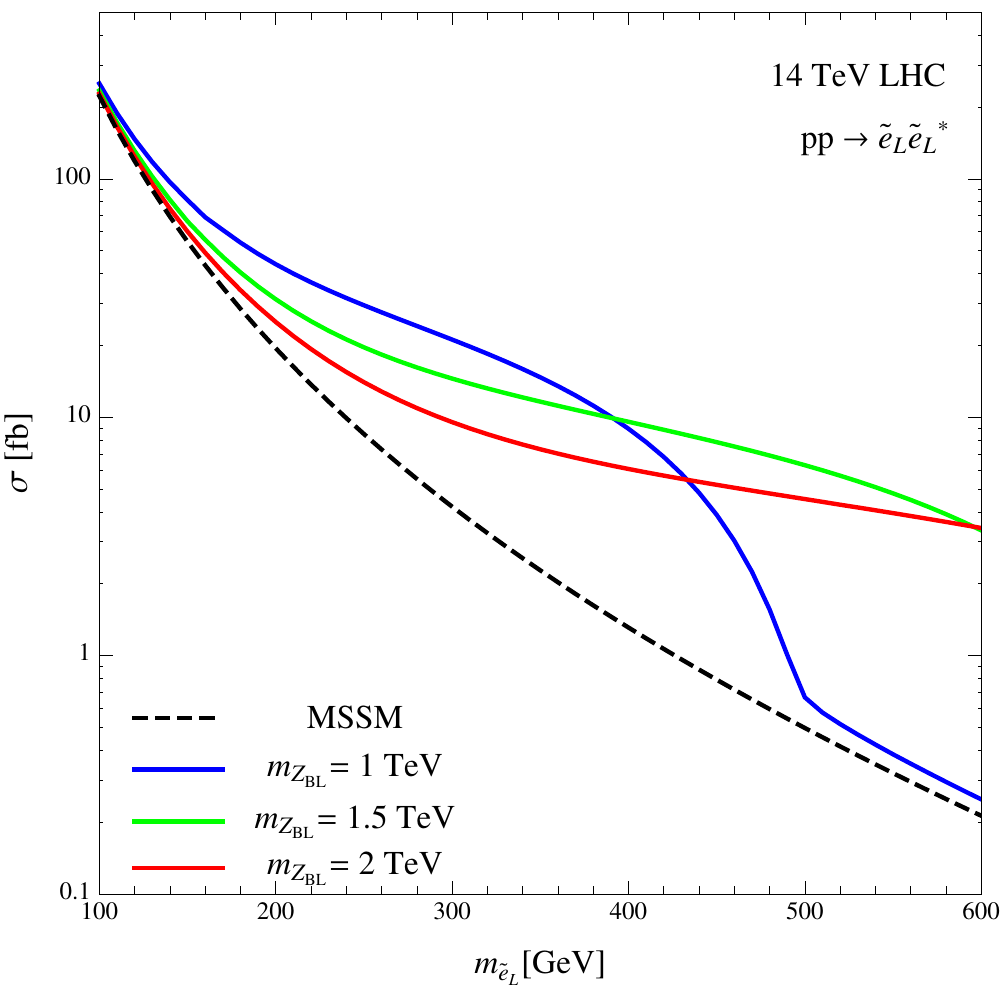}
	\put(-40,-4){(b)}
\caption
{Drell-Yan production cross sections for the charged sleptons in our model. The dashed line corresponds to the prediction in the MSSM and the solid lines show the results in our model for different values of the $Z_{BL}$ mass. The gauge coupling $g_{BL}$ is assumed to be at the maximum value allowed by the experimental constraints: Eq~(\ref{ZBL.const}).  
}
\label{Slepton.sigma}
\end{figure}

The numerical results for the selectron production cross sections are shown in Fig.~\ref{Slepton.sigma} for different scenarios, with $g_{BL}$ assumed to be the maximum value allowed by the experimental constraints: Eq~(\ref{ZBL.const}). 
We have compared our analytical results for the cross section with the results in Ref.~\cite{Dawson} and 
found the same result in the case of the MSSM. In Fig.~\ref{Slepton.sigma} we can see that even in the MSSM the cross section can be large and when the $Z_{BL}$ is included 
the cross section can be even larger due to the resonance enhancement. For example when the $M_{Z_{BL}}=1$ TeV one can have a cross section above 
1 fb where the selectron mass is below 450 GeV and $\sqrt{s}=7$ TeV. 
%
\subsection{Signals with Multi-Leptons}
%
In this paper we wish to investigate the most promising signals associated with lepton number violation, through the process
\begin{equation}
	q \bar{q} \  \to
	\ \gamma, Z^*, Z^*_{BL} \ \to
	\  \tilde{e}_i^* \tilde{e}_i \to e_i^+ e_i^- \tilde{\chi}^0_1 \tilde{\chi}^0_1 \  \to
	\  e_i^+ e_i^- e^{\pm}_j e^{\pm}_k \ 4j,
\end{equation}
%
%
where $i,j,k=1..3$ are generational indices. See Fig.~\ref{Signal} for the illustration of these signals. However, we will focus on the electron and muon channels since hadronic activity associated with the tau would blemish the signal of lepton number violation, namely three same-sign leptons, 
one lepton with opposite sign, four jets and no missing energy. Taking a cue from Eq.~(\ref{Selectron.Mass}), which shows that the left-handed (right-handed) sleptons receive a negative (positive) contribution to their mass from the $B-L$ $D$-term, we assume that only the left-handed sleptons are producible through this process.

We begin by giving an estimate for the number of events in the limit of mostly bino, wino and Higgsino LSP.  
We present results for a 7 TeV LHC with $10 \text{ fb}^{-1}$ of data. The combinatorics factor for the channels of interest are given by
\begin{align}
\begin{split}
	\mathcal F_{jk}  = &
	2 \left(2 - \delta_{jk} \right) \, \text{Br}\left(\tilde e^\pm_i \to e^\pm_i \tilde \chi_1^0\right)^2 \times
	\text{Br}\left(\tilde \chi_1^0 \to e^\pm_j W^\mp \right)
\\
	& \times \text{Br}\left(\tilde \chi_1^0 \to e^\pm_k W^\mp \right) 
	\times \text{Br}\left(W^\pm \to jj\right)^2,
\end{split}
\end{align}
so that the final states are
$$
e^\pm e^\mp e^\mp e^\mp 4j, \quad e^\pm e^\mp e^\mp \mu^\mp 4j, \quad e^\pm e^\mp \mu^\mp \mu^\mp 4j,
$$
$$
\mu^\pm \mu^\mp e^\mp e^\mp 4j, \quad \mu^\pm \mu^\mp e^\mp \mu^\mp 4j, \quad \mu^\pm \mu^\mp \mu^\mp \mu^\mp 4j.
$$
\begin{table}[htdp]
\begin{center}
\begin{tabular}{|c|c|c|c|c|c|c|}
\hline
Hierarchy& LSP & $\text{Br}(\tilde \chi_1^0 \to e^\pm W^\mp)$	& $\text{Br}(\tilde \chi_1^0 \to \mu^\pm W^\mp)$ &
$\quad \quad \ \mathcal{F}_{ee} \quad \quad \ $ & $\ \quad \quad \mathcal{F}_{e \mu} \quad \quad \ $ & $\quad \quad \ \mathcal{F}_{\mu \mu} \quad \quad \ $
\\
\hline
NH 	& Bino & 1-20\%	& 10-50\% & 0.0001-0.036 & 0.002-0.18 & 0.01-0.22
\\
NH	& Wino&	1-20 \%	& 10-50\% & 0.00001-0.004 & 0.0002-0.02 & 0.001-0.024
\\
NH	& $\quad$ Higgsino $\quad$&	2-25 \%	& 10-50\% & 0.0004-0.056 & 0.004-0.22 & 0.008-0.24
\\
\hline
\hline
IH 	& Bino & 10-60\%	& 10-30 \% & 0.009-0.32 & 0.018-0.32 & 0.009-0.08
\\
IH 	& Wino & 10-60\%	& 13-30\% & 0.001-0.035 & 0.003-0.034 & 0.002-0.009
\\
IH 	& Higgsino & 10-60\%	& 13-35\% & 0.008-0.32 & 0.024-0.37 & 0.016-0.11
\\
\hline
\end{tabular}
\end{center}
\caption{Ranges for the branching ratios of the LSP to charge lepton and $W$ boson taken from the corresponding dense regions in Figs.~\ref{LSP.BR.Bino}~-~\ref{LSP.BR.Higgsino}. These are used to calculate the overall combinatorics factor, $\mathcal{F}_{jk}$ for the final state $e_i^\pm e_i^\mp e^\pm_j e^\pm_k 4j$.  Values are separated by the composition of the LSP: mostly bino, wino and Higgsino and for both the normal and inverted hierarchies.}
\label{Comb}
\end{table}
We assume that $\text{Br}\left(\tilde{e}^\pm_i \to e^\pm_i \tilde{\chi}^0_1 \right) \sim 100 \%, 33 \%, 100 \%$ for a bino, wino and Higgsino LSP respectively, following the discussion in Section~\ref{slep.dk}.  The branching ratio of the $W$ boson into jets is about 67\%. For the RPV neutralino decays we pick the most prominent regions from Figs.~\ref{LSP.BR.Bino}~-~\ref{LSP.BR.Higgsino} and display these values along with $\mathcal{F}_{jk}$ in Table~\ref{Comb}.  Values are shown for both the normal and inverted hierarchies. 

\begin{figure}[h!] 
	\includegraphics[scale=0.85]{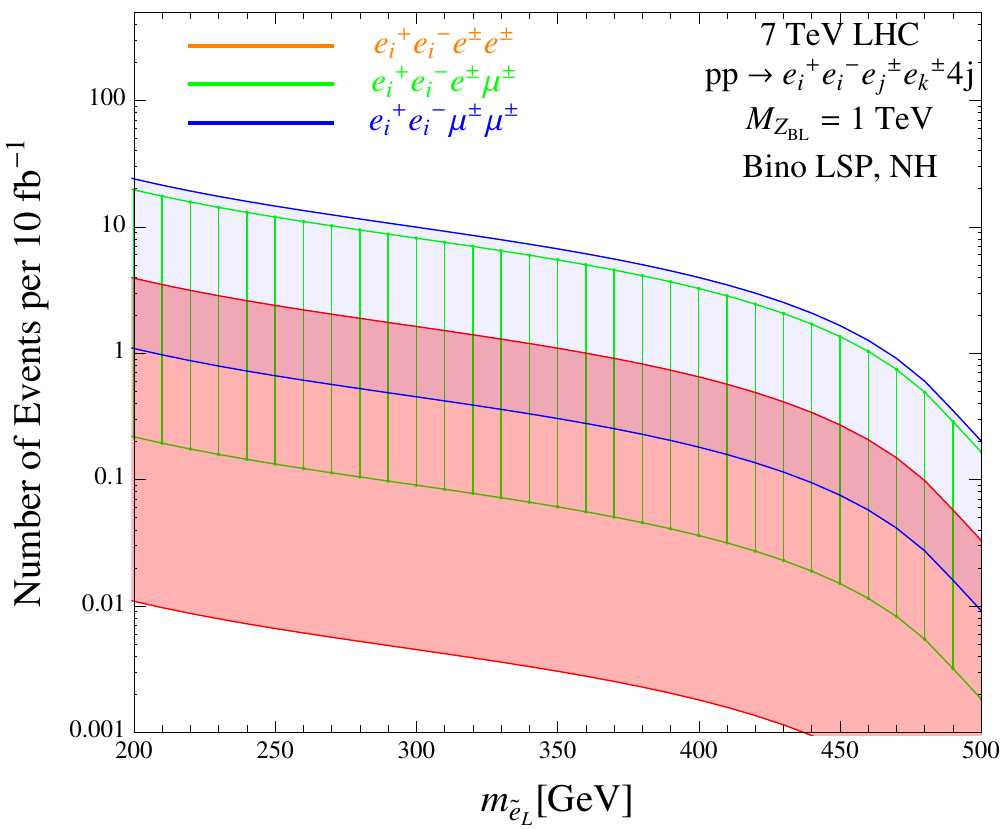}
	\put(-40,-4){(a)}
	\includegraphics[scale=0.85]{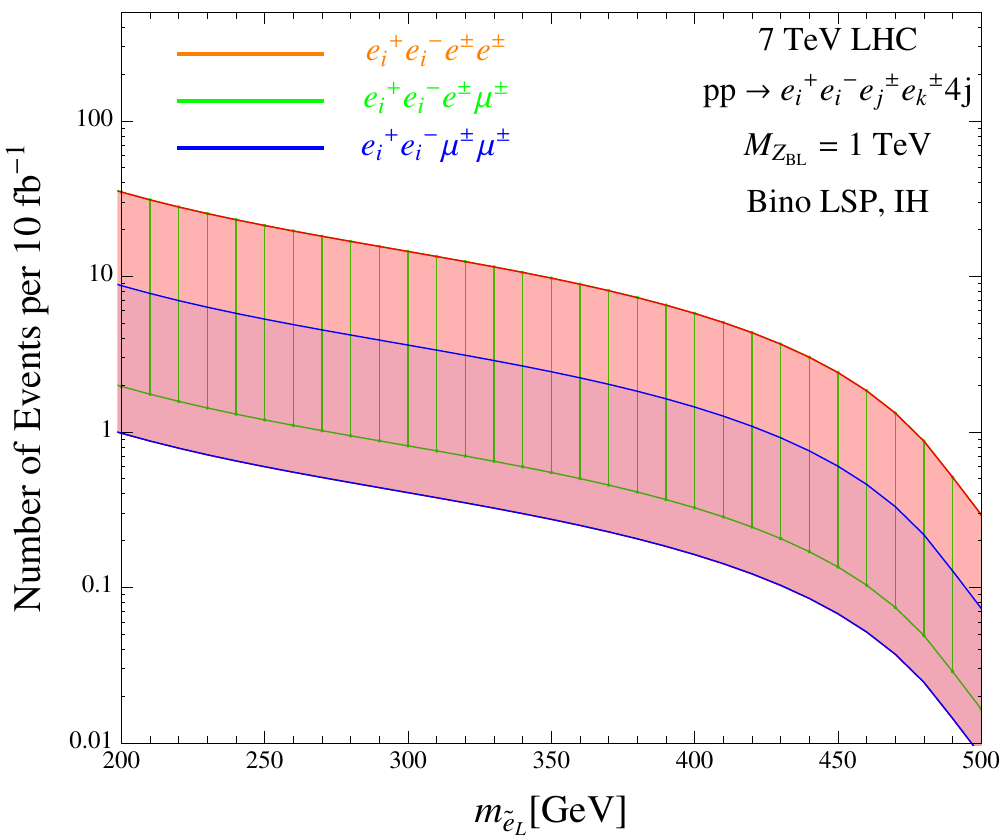}
	\put(-40,-4){(b)}
\caption{
Number of $e_i^\pm e_i^\mp e_j^\pm e_k^\pm 4j$ at a 7 TeV LHC for 10 $\text{fb}^{-1}$ of data for a bino LSP. Branching ratio values are shown in Table~\ref{Comb}, while cross section values are taken from Fig~\ref{Slepton.sigma}. 
Data is divided into (a) for the NH, and (b) for the IN.}
\label{Num.Events.Bino}
\end{figure}

\begin{figure}
	\includegraphics[scale=0.85]{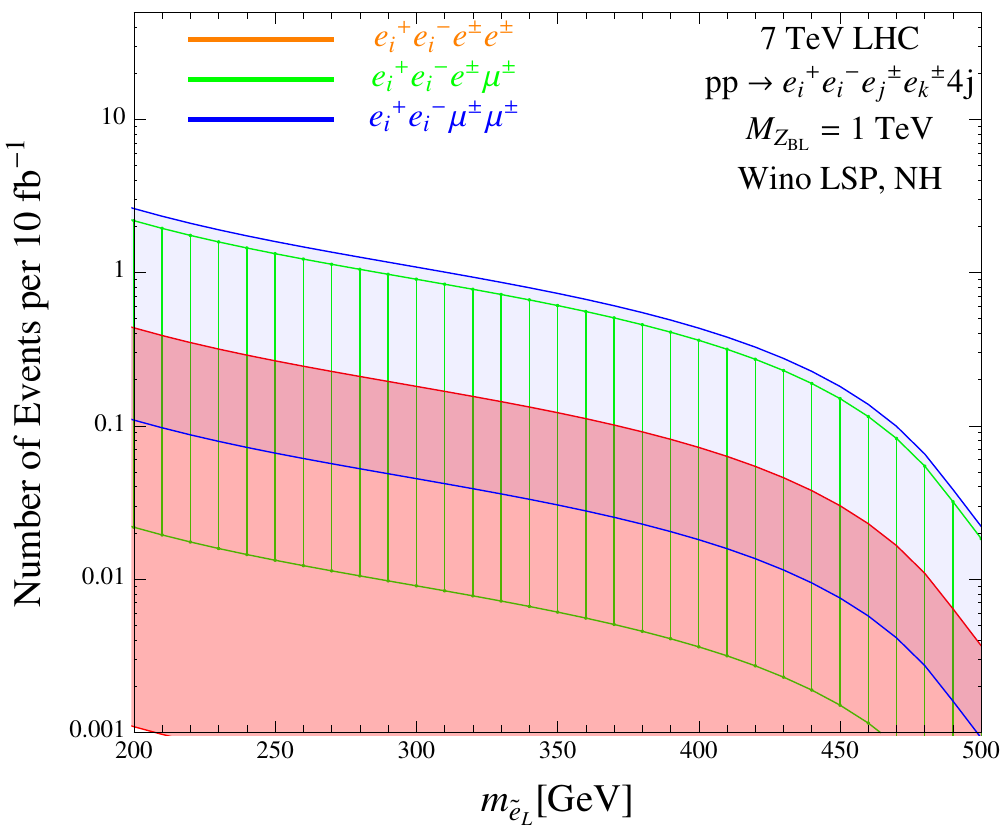}
	\put(-40,-4){(a)}
	\includegraphics[scale=0.85]{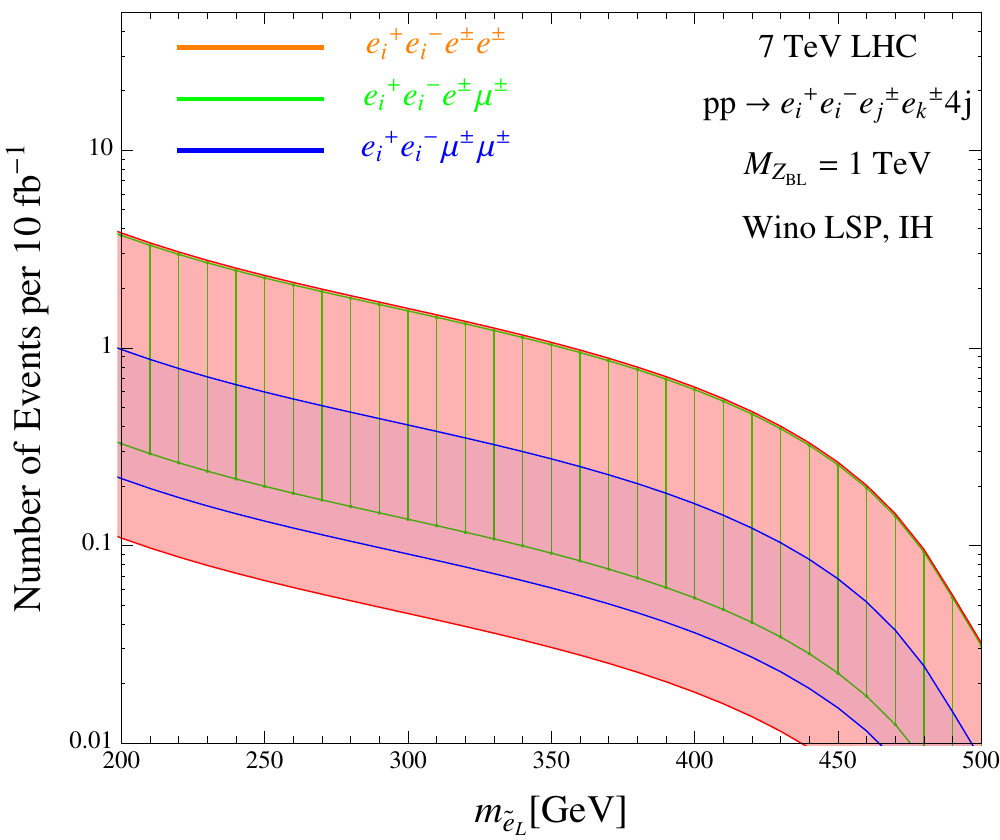}
	\put(-40,-4){(b)}
\caption{
Number of $e_i^\pm e_i^\mp e_j^\pm e_k^\pm 4j$ at a 7 TeV LHC for 10 $\text{fb}^{-1}$ of data for a wino LSP. Branching ratio values are shown in Table~\ref{Comb}, while cross section values are taken from Fig~\ref{Slepton.sigma}. 
Data is divided into (a) for the NH, and (b) for the IN.
}
\label{Num.Events.Wino}
\end{figure}

\begin{figure}
	\includegraphics[scale=0.85]{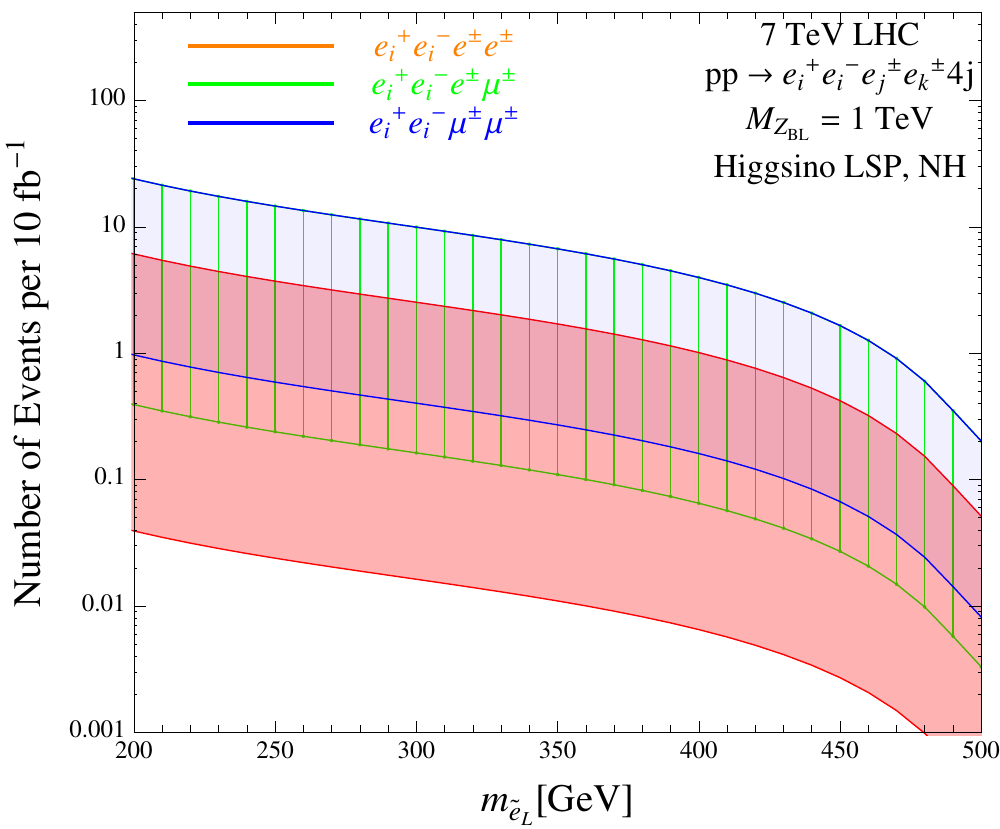}
	\put(-40,-4){(a)}
	\includegraphics[scale=0.85]{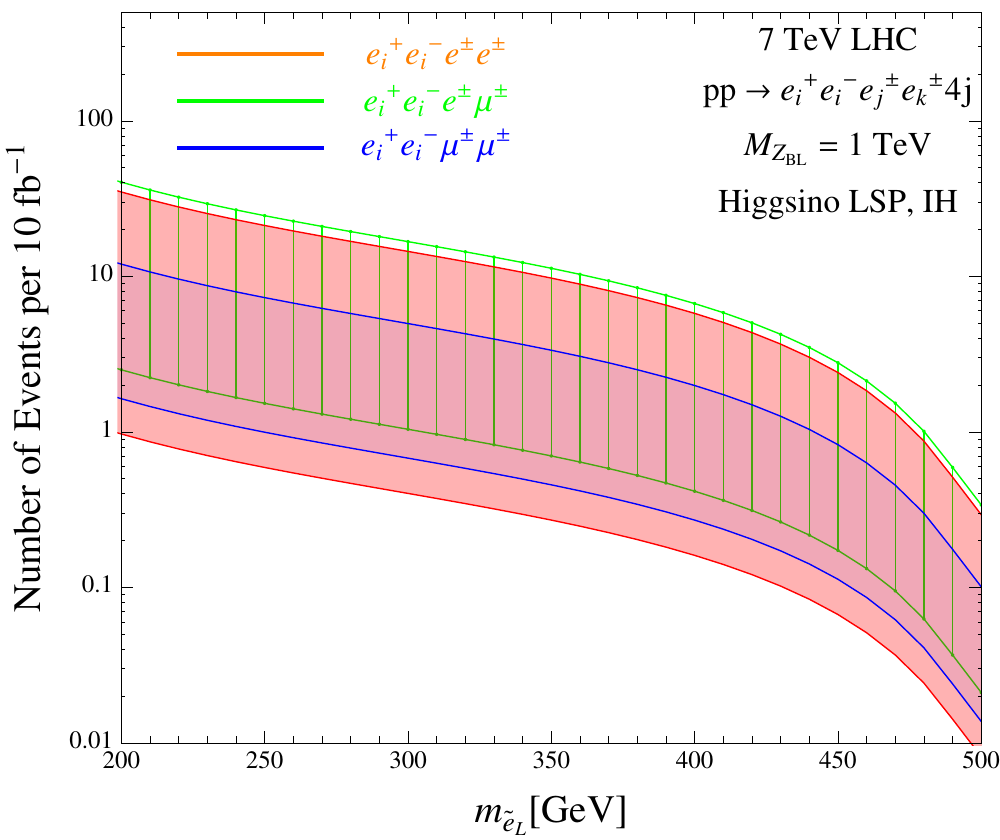}
	\put(-40,-4){(b)}
\caption
{Number of $e_i^\pm e_i^\mp e_j^\pm e_k^\pm 4j$ at a 7 TeV LHC for 10 $\text{fb}^{-1}$ of data for a Higgsino LSP. Branching ratio values are shown in Table~\ref{Comb}, while cross section values are taken from Fig~\ref{Slepton.sigma}. 
Data is divided into (a) for the NH, and (b) for the IN.
	}
\label{Num.Events.Higgsino}
\end{figure}

To calculate the number of events expected after 10 $\text{fb}^{-1}$ of data, we convolute the combinatorics in Table~\ref{Comb} with the cross sections for a 1 TeV $Z_{BL}$ shown in Fig.~\ref{Slepton.sigma} and multiply by ten; the results are displayed in Figs.~\ref{Num.Events.Bino}~-~\ref{Num.Events.Higgsino}. Notice that in most of the scenarios one can have several events which are basically background free. 
The main backgrounds coming from $t\bar{t} W Z$ and $jjjj W^{\pm} W^{\pm} Z$ are very suppressed.

\begin{figure}[h!] 
	\includegraphics[scale=0.9]{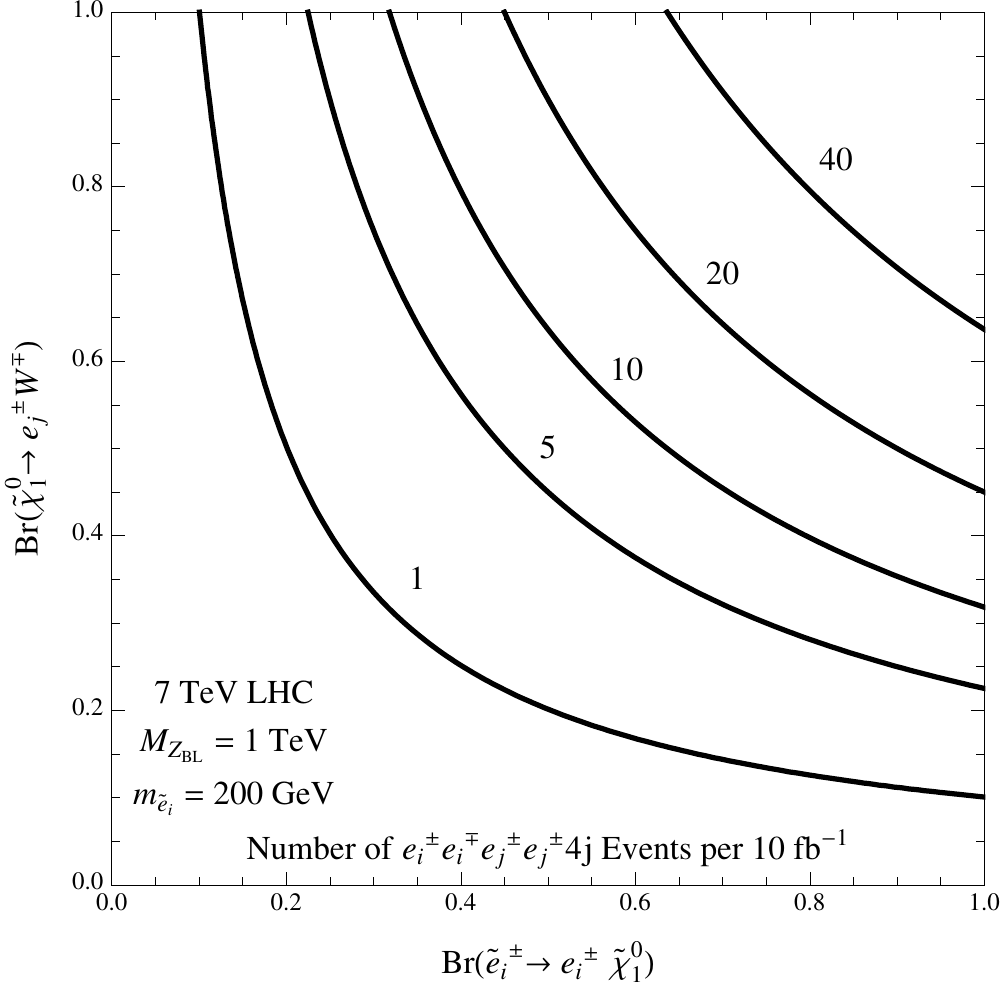}
\caption
{
	Curves of constant number of events for the final state $e_i^\pm e_i^\mp e_j^\pm e_j^\pm 4j$ in the  $\text{Br}\left(\tilde \chi^0_1 \to e_j^\pm W^\mp \right)-\text{Br}\left(\tilde e_i^\pm \to e_i^\pm \tilde \chi^0_1 \right)$ plane.  Values are shown for a seven TeV LHC, $M_{Z_{BL}}=1$ TeV and $m_{\tilde e_i}=200$ GeV.
}
\label{Const.Num.Events}
\end{figure}

In order to understand the testability of the model at the LHC we show curves of constant number of $e_i^\pm e_j^\mp e_j^\pm e_j^\pm 4j$ events per 10 $\text{fb}^{-1}$ of data in Fig.~\ref{Const.Num.Events} 
in the $\text{Br}(\tilde \chi^0_1 \to e_j^\pm W^\mp)-\text{Br}(\tilde e_i^\pm \to e_i^\pm \tilde \chi^0_1)$ plane. Values are shown for a seven TeV LHC, with a 1 TeV $Z_{BL}$ 
and $m_{\tilde e_i}=200$ GeV. In the case of the observation of such events a the LHC, the $Z_{BL}$ mass can be reconstructed from electron-electron and muon-muon events and the selectron mass may be reconstructible from its decay into two leptons and two jets. Therefore allowing the calculation of the cross section for charged slepton pair-production. A plot such as Fig.\ref{Const.Num.Events} can be used to get a better handle on the two unknown branching ratios and shed further light on the model.

\begin{figure}[h!] 
	\includegraphics[scale=0.9]{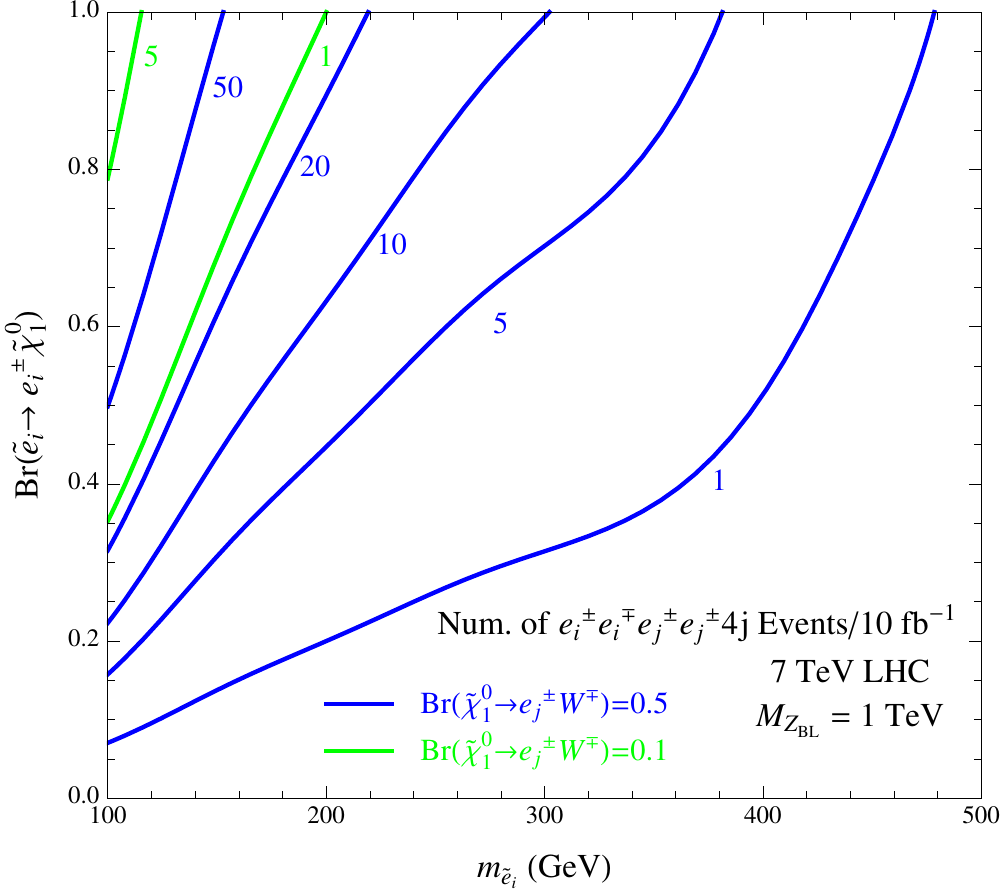}
\caption
{
	Curves of constant number of events for the final state $e_i^\pm e_i^\mp e_j^\pm e_j^\pm 4j$ in the  $\text{Br}\left(\tilde e_i^\pm \to e_i^\pm \tilde \chi^0_1 \right)-m_{\tilde e_i}$ plane. 
	Values are shown for a seven TeV LHC, $M_{Z_{BL}}=1$ TeV and for two different values of $\text{Br}\left(\tilde \chi^0_1 \to e_j^\pm W^\mp \right)$, representing the upper (lower) part of that range in blue (green).
}
\label{Const.Num.Events2}
\end{figure}

In order to estimate the reach of the LHC we also present curves of constant number of $e_i^\pm e_i^\mp e_j^\pm e_j^\pm 4j$ events per 10 $\text{fb}^{-1}$ of data in the $\text{Br}\left(\tilde e_i^\pm \to e_i^\pm \tilde \chi^0_1 \right)-m_{\tilde e_i}$ plane in Fig.~\ref{Const.Num.Events2}. This is again for a seven TeV LHC, with a 1 TeV $Z_{BL}$ and we show two possible values for $\text{Br}\left(\tilde \chi^0_1 \to e_j^\pm W^\mp \right)$, representing the upper (lower) part of that range in blue (green). One can see that even if the slepton mass is around 450 GeV one could observe a few events with multileptons and four jets. It is important to mention we satisfy the recent bounds coming from ATLAS~\cite{Atlas1, Atlas2}.

%

\section{Summary and Outlook}
%
In this article we have studied in detail the theory proposed in Ref.~\cite{Rp2} which we consider the simplest gauge theory for R-parity violation.
This theory makes a prediction for the LHC since in order to break the B-L gauge symmetry the right-handed sneutrino must get a vacuum 
expectation value and one should observe lepton number violation at colliders. We have found the following results:

\begin{itemize}

\item In Fig. 2 we have illustrated in a simple way that one can have a realistic scenario for all sfermion masses even 
if some of the sfermion masses have a negative and large contribution from the B-L D-term in the theory. We have shown 
that in order to avoid tachyonic masses one should satisfy the condition $M_{Z_{BL}} < \sqrt{2} m_{\tilde{L}}$. This is a 
simple result which helps us to understand the constraints on the spectrum.

\item The full spectrum of the theory and the constraints coming from neutrino masses were analyzed in detail. 
The spectrum for neutrinos is interesting since it contains five light neutrinos: three active neutrinos and two sterile neutrinos.
Using the experimental constraints on the masses and mixing for the active neutrinos we show in Fig. 4 the allowed 
values for the vacuum expectations of the left-handed neutrinos and the Yukawa couplings. As we have discussed in the text, 
these results are crucial to understand the decays of the lightest supersymmetric particle in the theory.

\item In Figs. 5 and 6 we have shown the properties of the new neutral gauge boson in the theory, the B-L gauge boson. 
Since one has two extra light neutrinos in the theory the invisible decay width is larger in this case. 
The contributions of the supersymmetric particle to the decay width are small and so the $Z_{BL}$ 
is like the B-L gauge boson in the non-SUSY scenarios. 

\item We have investigated the neutralinos decays in great detail. In Figs. 7-9 we have shown the results for the decay 
length in the different cases. As one can appreciate in Figs. 7-9 there are some scenarios in the Bino limit where one could expects 
displaced vertices. The branching ratios have been investigated in Figs.10-12 and we can summarize the results in the following way
$$\rm{Br}( \tilde{\chi}^0_1 \to \tau W), \rm{Br}(\tilde{\chi}^0_1 \to \mu W)> \rm{Br}(\tilde{\chi}^0_1 \to e W),$$  in the Normal Hierarchy
and
$$\rm{Br}( \tilde{\chi}^0_1 \to e W), \rm{Br}(\tilde{\chi}^0_1 \to \mu W)> \rm{Br}(\tilde{\chi}^0_1 \to \tau W),$$ in the Inverted Hierarchy,
in majority of the parameter space.

\item We have studied the main production channels for the charged sleptons at the LHC. In this case one can produce the charged sleptons 
through the photon and the Z as in the MSSM, and through the new neutral gauge boson, $Z_{BL}$, in our model. 
As we have shown in Fig. 13 the production cross section can be large and thanks to the presence of the $Z_{BL}$ one 
can have even larger values for the cross section due to the resonance enhancement. We should point out that this production channel (throught the photon and $Z$)
is very important to understand the signals in any model for R-parity violation.

\item The most striking signals for lepton number violation in this context  are the channels with three leptons with the same electric 
charge and four jets. In Fig.~\ref{Num.Events.Bino}-\ref{Const.Num.Events2} we have shown that one can have a large number of events at the LHC with only 10 $\rm{fb}^{-1}$. 
The background for these channels is suppressed, therefore there is a hope to test or rule out this theory in the near future. 

\end{itemize}

{\textit{Acknowledgments}}:
The work of P.F.P. is supported by the James Arthur Fellowship, CCPP-New York University. P. F. P. thanks A. Haas for a discussion about 
the searches for multi-leptons at the LHC. 

\appendix

\section{Mass Matrices}
In the case of the CP-odd neutral scalars, in the basis $(A_L, A_R, A_d, A_u)$, 
one finds that the mass matrix reads as
\begin{equation}
\label{CP-odd}
{\cal M}_{odd}^2
	=
	\begin{pmatrix}
	  	\frac{v_R}{v_L} B_\nu
	&
		B_\nu
	&
		-\frac{1}{\sqrt{2}} Y_\nu \mu v_R
	&
		-\frac{1}{\sqrt{2}} a_\nu v_R
\\
		B_\nu
	&
		\frac{v_L}{v_R} B_\nu
	&
		-\frac{1}{\sqrt{2}} Y_\nu \mu v_L
	&
		-\frac{1}{\sqrt{2}} a_\nu v_L
\\
		-\frac{1}{\sqrt{2}} Y_\nu \mu v_R
	&
		-\frac{1}{\sqrt{2}} Y_\nu \mu v_L
	&
		\frac{v_u}{v_d} B\mu \ + \frac{Y_\nu \mu v_L v_R}{\sqrt{2} v_d}
	&
		B\mu
\\
		-\frac{1}{\sqrt{2}} a_\nu v_R
	&
		-\frac{1}{\sqrt{2}} a_\nu v_L
	&
		B\mu
	&
		\frac{v_d}{v_u} B\mu \ - \frac{a_\nu \mu v_L v_R}{\sqrt{2} v_u}
	  \end{pmatrix},
\end{equation}
while for the CP-even scalars, in the basis $(h_L, h_R, h_d, h_u)$, one finds:
\begin{equation}
\label{CP-even}
	{\cal M}_S^2	=
	\begin{pmatrix}
		S_{\nu}^2
		&
		S_{\nu H}^2
	\\
		\left(S_{\nu H}^{2}\right)^T
		&
		S_{H}^2
	\end{pmatrix},
\end{equation}
where
\begin{align}
	S_\nu^2	\equiv	&
	\begin{pmatrix}
		\frac{1}{4} \left(g_1^2 + g_2^2 + g_{BL}^2 \right) v_L^2 + \frac{v_R}{v_L} B_\nu
		&
		-\frac{1}{4} \left(g_{BL}^2 - 2 Y_\nu^{2}\right) v_L v_R - B_\nu
	\\
		-\frac{1}{4} \left(g_{BL}^2 - 2 Y_\nu^{2}\right) v_L v_R - B_\nu
		&
		\frac{1}{4} g_{BL}^2 v_R^2 + \frac{v_L}{v_R} B_\nu
	\end{pmatrix},
\end{align}
\begin{align}
	S_{\nu H}^2 \equiv	&
	\begin{pmatrix}
		\frac{1}{4} \left(g_1^2 + g_2^2 \right) v_L v_d - \frac{1}{\sqrt{2}} Y_\nu \mu v_R
		&
		-\frac{1}{4} \left(g_1^2 + g_2^2 - 4 Y_\nu^{2} \right) v_L v_u + \frac{1}{\sqrt{2}} a_\nu v_R
	\\
		-\frac{1}{\sqrt{2}} Y_\nu \mu v_L
		&
		Y_\nu^{2} v_R v_u + \frac{1}{\sqrt{2}} a_e v_L
	\end{pmatrix},
\\
\nonumber
\\
	S_{H}^2	\equiv &
	\begin{pmatrix}
		\frac{1}{4} \left( g_1^2 + g_2^2 \right)v_d^2 + \frac{v_u}{v_d} B\mu  + \frac{Y_\nu \mu v_R v_L}{\sqrt{2} v_d}
		&
		 - \frac{1}{4} \left(g_1^2 + g_2^2 \right) v_u v_d - B\mu
	\\
		- \frac{1}{4} \left(g_1^2 + g_2^2 \right) v_u v_d - B\mu
		&
		\frac{1}{4} \left(g_1^2 + g_2^2 \right) v_u^2 + \frac{v_d}{v_u} B\mu - \frac{a_\nu v_L v_R}{\sqrt{2} v_u}
	\end{pmatrix}.	
\end{align}
In the case of the charged scalars, in the basis $(\tilde{e}, (\tilde{e}^c)^*, H_u^-, H_d^-)$ the mass matrix reads as:
\begin{equation}
\label{Charged}
	M_C^2 =
	\begin{pmatrix}
		C_e^2
		&
		C_{e C}^2
	\\
		\left(C_{e H}^2 \right)^T
		&
		C_{H}^2
	\end{pmatrix},
\end{equation}
with
\begin{flushleft}
\begin{align}
	C_e^2 \equiv &
	\begin{pmatrix}
		C_{11}^2
		&
		B_e
	\\
		B_e
		&
		C_{22}^2
	\end{pmatrix},
\end{align}
\end{flushleft}
\begin{flushleft}
\begin{align}
	C_{e H}^2 \equiv
	&
	\begin{pmatrix}
		\frac{1}{4} g_2^2 v_L v_d - \frac{1}{2} Y_e^2 v_L v_d - \frac{1}{\sqrt{2}} Y_\nu \mu v_R
		&
		\frac{1}{4} g_2^2 v_L v_u - \frac{1}{2} Y_\nu^2 v_L v_u - \frac{1}{\sqrt{2}} a_\nu v_R
	\\
		\frac{1}{2} Y_e Y_\nu v_R v_u + \frac{1}{\sqrt{2}} a_e v_L
		&
		\frac{1}{2} Y_e Y_\nu v_R v_d + \frac{1}{\sqrt{2}} Y_e \mu v_L
	\end{pmatrix},
\\
\nonumber
\\
	C_{H}^2 \equiv
	&
	\begin{pmatrix}
		\frac{1}{4} g_2^2\left(v_u^2-v_L^2 \right) + B\mu \frac{v_u}{v_d} + \frac{1}{2} Y_e^2 v_L^2 + \frac{Y_\nu \mu v_R v_L}{\sqrt{2} v_d}
		&
		B\mu + \frac{1}{4} g_2^2 v_u v_d
	\\
		B\mu + \frac{1}{4} g_2^2 v_u v_d
		&
		\frac{1}{4} g_2^2 \left(v_d^2 + v_L^2 \right) + \frac{v_d}{v_u}B\mu - \frac{1}{2} Y_\nu^{2} v_L^2 - \frac{a_\nu v_L v_R}{\sqrt{2} v_u}
	\end{pmatrix}.
\end{align}
\end{flushleft}
In the above equations $C_{11}^2$ and $C_{22}^2$ are given by
\begin{eqnarray}
C_{11}^2 & = & \frac{1}{4} g_2^2 \left(v_u^2 - v_d^2 \right) + \frac{1}{2}Y_e^2 v_d^2 - \frac{1}{2} Y_\nu^{2} v_u^2 + \frac{v_R}{v_L} B_e,
\end{eqnarray}
and
\begin{eqnarray}
C_{22}^2 &=& M_{\tilde e^c}^2 + \frac{1}{4} g_1^2 \left(v_u^2 - v_d^2 -v_L^2 \right) + \frac{1}{8} g_{BL}^2 \left(v_R^2 - v_L^2 \right) + \frac{1}{2} Y_e^2 \left( v_d^2 + v_L^2 \right).
\end{eqnarray}
We also define for convenience: $B_\nu = \frac{1}{\sqrt{2}} \left(Y_\nu \mu v_d - a_\nu v_u \right)$ and
$B_e = \frac{1}{\sqrt{2}} \left(Y_e \mu v_u - a_e v_d \right)$. 

\section{Decay Amplitude}
\label{ZPDecay}

The amplitude for $Z_{BL}$ are
\begin{eqnarray}
\label{Zff}
 \left| \overline{\mm} (Z_{BL} \to f_i \bar{f_i}) \right|^2 &=&~
	\frac{4}{3} c_{f} \left( \frac{g_{BL} }{2}n_{BL}^{f}\right)^2 \mzbl^2 
		\left(1+\frac{2 m_{f_i}^2}{\mzbl^2} \right),
	 \  \  \ f_i = u,d,c,s,b, t, e, \mu, \tau;	
\\
 \left| \overline{\mm} (Z_{BL} \to \nu_i \bar{\nu_i}) \right|^2 &=&~
	 \frac{2}{3} \left( \frac{g_{BL} }{2}n_{BL}^{\nu}\right)^2 \mzbl^2,
\\
 \left| \overline{\mm} (Z_{BL} \to  \bar{N} N) \right|^2 &=&~
	 \frac{2}{3} \left( \frac{g_{BL} }{2}n_{BL}^{\nu_R}\right)^2
	\, \mzbl^2 \,  
	\left(1 - 4\frac{m_{N}^2}{\mzbl^2}\right),
\\
 \left| \overline{\mm} (Z_{BL} \to \tilde{f}_{\alpha} \tilde{f}^\ast_{\beta}) \right|^2  &=&~
	\frac{1}{3} c_{\tilde{f}}
	\left( \frac{g_{BL} }{2}n_{BL}^{\tilde{f}}\right)^2 \mzbl^2
	\left( 1 - \frac{2 m_{\tilde{f}_{\alpha}}^2 + 2 m_{\tilde{f}_{\beta}}^2}{\mzbl^2}
		+   \frac{\left(m_{\tilde{f}_{\alpha}}^2 - m_{\tilde{f}_{\beta}}^2\right)^2}{\mzbl^4}
	\right)
\\&&
	\times~
	\left( U_{\alpha 1}^{\tilde{f}} U_{\beta 1}^{\tilde{f}} + 
		U_{\alpha 2}^{\tilde{f}} U_{\beta 2}^{\tilde{f}} \right)^2,
	\qquad\quad \tilde{f}_{\alpha} \tilde{f}_{\beta}^* = 
		\tilde{q}_{i\alpha} \tilde{q}_{i\beta}^{*},
		\tilde{l}_{i\alpha} \tilde{l}_{i\beta}^{*},
		\tilde{\nu}_i \tilde{\nu}_{i}^{*},
		\tilde{\nu}_{Ri} \tilde{\nu}_{Ri}^{*}.
\end{eqnarray}
Here, $i$ is a generation index, $c_f$ are color factors ($c_{q_i} = 3$, $c_{l_i}=1$) and $U^{\tilde{f}}$ 
are the unitary sfermion mixing matrices.
%
\section{Neutrino-Neutralino Mixing Matrix}
\label{Vla}
In the basis $\psi^T = \left(\nu, \chi\right)$, the mass matrix has the general form
\begin{equation}
	\mathcal{M} =
	\begin{pmatrix}
		0_{3\times3}
		&
		m_D
		\\
		m_D^T
		&
		M_\chi
	\end{pmatrix},
\end{equation}
where $\mathcal{M}$ is diagonalized by $\mathcal{N}$
\begin{equation}
\label{diagonalM}
	\mathcal N^\dagger \mathcal{M} \mathcal{N}^* = 
	\begin{pmatrix}
		m_\nu^D
		&
		0
	\\
		0
	&
		M_\chi^D
	\end{pmatrix},
\end{equation}
$m_\nu^D$ is the diagonal mass matrix for the light neutrinos, $M_\chi^D$ is the diagonal mass matrix for the neutralinos and
\begin{equation}
	\mathcal{N} = 
	\begin{pmatrix}
		U
		&
		V
	\\
		V_c
		&
		U_c
	\end{pmatrix}.
\end{equation}
Eq.~(\ref{diagonalM}) yields
\begin{align}
\label{nu.mass}
	m_\nu^D & = U^\dagger m_D V_c^* + V_c^\dagger m_D^T U^* + V_c^\dagger M_\chi V_c,
\\
\label{chi.mass}
	M_\chi^D & = V^\dagger m_D^T V_c^* + U_c^\dagger M_\chi U_c^* + U_c^\dagger m_D^T V^*,
\\
\label{zero.mass}
	0 & = U^\dagger m_D U_c^* + V_c^\dagger M_\chi U_c^* + V_c^\dagger m_D^T V^*,
\end{align}
and
\begin{equation}
	U \sim U_c \sim \mathcal{O}(1); \quad \quad \quad V \sim V_c \sim \mathcal{O}(\frac{m_\nu^D}{m_D}).
\end{equation}
The unitarity condition yields the following expressions
\begin{align}
\begin{split}
	& U U^\dagger + V V^\dagger = U^\dagger U + V_c^\dagger V_c = V_c V_c^\dagger + U_c U_c^\dagger =
		V^\dagger V + U_c^\dagger U_c = I,
\\
	& U V_c^\dagger + V U_c^\dagger = U^\dagger V + V_c^\dagger U^c = 0.
\end{split}
\end{align}
In Eq.~\ref{zero.mass}, the $V_c^\dagger m_D^T V^*$ is negligible whereas in Eq.~\ref{chi.mass} the $M_\chi$ term dominates.  Therefore
\begin{align}
\label{um}
	U^\dagger m_D + V_c^\dagger M_\chi & = 0,
\\
\label{chi.mass.diag}
	U_c^\dagger M_\chi U_c & = M_\chi^D,
\\
	m_D U_c^* & = V M,
\end{align}
where the last expression is a result of inverting Eq.~\ref{chi.mass.diag},  substituting it into Eq.~\ref{um} and making use of the unitarity conditions.  Substituted Eq.~\ref{um} into Eq.~\ref{nu.mass} yields
\begin{equation}
\label{mnuU}
	V_c^\dagger m_D^T = m_\nu^D U^T.
\end{equation}
These results can be use to manipulate the seesaw relation:
\begin{equation}
	m_\nu = U m_\nu^D U^T = m_D M_\chi^{-1} m_D^T,
\end{equation}
where $m_\nu$ is the nondiagonal light neutrino mass matrix diagonalized by $U$.  Substituting Eq.~(\ref{mnuU}) for $m_\nu^D U^T$, rearranging using the unitarity condition yields and solving for V yields
\begin{equation}
	V  = m_D M_\chi^{-1} U_c
\end{equation}
This can be rewritten by inverting Eq.~(\ref{chi.mass}):
\begin{equation}
	V  = m_D U_c^* \left(M_\chi^D\right)^{-1}.
\end{equation}
Where $V$ can be identified with $V_{i a}$, the matrix that describes the mixing between the neutrinos and the neutralinos and is necessary for computing the neutralino decay properties.  This result agrees with the naive expectation from the mass insertion approximation.  While in the decay widths, Eq.~(\ref{lWL}-\ref{nuh}), factors of $U$ and $E$ (the matrix that diagonalizes the charged lepton mass matrix) appear, they do so as sums of unitarity quantities and therefore are either zeroes or ones.
%
\section{LSP Candidates and Their Final States}
%
\label{LSPs}
The violation of R-parity increases the space of possible LSPs, which is now no longer restricted to chargeless fields. We therefore take the time here to make a quick survey of the possible final states in this model. For each possible LSP, we consider its production, if lepton number violation is observable in principle and if there are any obstruction to this observation. Finally, we judge which LSP leads to the most interesting signals. For us, these are the signals where lepton number violation is explicit: same-sign leptons with no missing energy (which might be due to neutrinos thereby confounding the counting of lepton number). Of course this can only arise from neutral LSPs.

\begin{itemize}

\item Gluino $(\tilde{g})$ LSP:

Gluino pairs are produced through strong cross sections at the LHC. Their possible decays are
$$pp \  \to \  \tilde{g} \tilde{g}  \  \to \  t t \ \bar{b} \bar{b}  \ e^{-}_i e^{-}_j, \ t \bar t \, t \bar t \, \nu \nu,$$
$$pp \  \to  \tilde{g} \tilde{g} \  \to  4j  \ e^{\pm}_i \ e^{\pm}_j,\ 4j \, \nu \nu,$$ 
where the former is favored if the third generation squarks are lighter than the first two. The gluino decay width can be estimated as
$$\Gamma (\tilde{g} \to f^{'} \bar{f} e^{\pm}_i) \sim \alpha_s \frac{M_{\tilde{g}}^5 (v_L^i)^2}{ M_{\tilde{q}}^4 M_{\tilde{\chi}^+}^2 64 \pi^2}.$$ 
For $M_{\tilde{g}} =100$ GeV, $M_{\tilde{q}}=500$ GeV, $M_{\tilde{\chi}^+}=500$ GeV and $v_L^i=10$ MeV, one finds that the decay width is smaller than $10^{-13}$ GeV: a long enough lifetime for the gluino to form bound states but short enough so that it decays within the detector.

In principle, these channels can yield spectacular signals at the LHC. However a recent inclusive analysis in the search for isolated same-sign muons published by the ATLAS collaboration has placed a model independent upper bound on the gluino pair production cross section of 58 fb~\cite{ATLAS1}. Imposing this bound translate into a lower bound on the gluino mass of around 1 TeV indicating a heavy SUSY spectrum. Since we are interested in the scenarios with low energy supersymmetry we do not pursue this scenario further.

\item Squark $(\tilde{q})$ LSP:

A stop LSP allows for final states with two third generation quarks of the same type and two leptons:
$$pp \  \to \  \tilde{t}^* \tilde{t} \  \to \ \bar{b} b \ e^{\pm}_i \ e^{\mp}_j, \  \rm{or} \ \bar t t \, \nu \nu,$$
while a first or second generation squark LSP has channels with two jets and two leptons:
$$pp \  \to \  \tilde q^* \tilde q \ \to \ 2j \ e^{\pm}_i \ e^{\mp}_j, \ 2j \, \nu \nu.$$
These channels have strong cross sections but do not provide information on the violation of the total lepton number.
Since squarks in this case act as leptoquarks (each decaying into a quark and lepton) bounds on this scenario can be derived from leptoquark searches.

\item Charged slepton $(\tilde{e}_i)$ LSP:

Charged sleptons can be pair produced through the $Z$ and $Z_{BL}$, with signals
$$pp \  \to \  \tilde{e}^*_i  \tilde{e}_i \  \to \  \bar{t} t \ b \bar{b}, \ e_i^+ e_i^- \, \nu \nu,$$
where once more, lepton number violation is not discernible. The $t \bar b$ finals state is due to the mixing of the charged sleptons with the charged Higgs boson which typically decays in this way. The leptonic channel arises due to the R-parity violating mixing between the charged leptons (neutrinos) and the charginos (neutralinos).

\item Sneutrino $(\tilde{\nu}_i)$ LSP:

Sneutrino pair production also proceeds through the $Z$ and $Z_{BL}$ with the following possible final states:
$$pp \  \to \  \tilde{\nu}^* \tilde{\nu} \ \to \ b \bar{b} \ b \bar{b}, \ \nu \nu \nu \nu, \ e_i^+ e_j^- e_i^+ e_k^-$$
The first final state results from the R-parity violating mixing of the sneutrino with the Higgs boson, while the latter two states are due to the R-parity violating mixing between the charged leptons (neutrinos) and the charginos (neutralinos).

\item Chargino $(\tilde{\chi}^{\pm})$ LSP:

Charginos pair production is possible through the $Z$ and leads to channels with two charged leptons due to the R-parity violating mixing between the charged leptons (neutrinos) and the charginos (neutralinos).
$$pp \  \to \  \tilde{\chi}^+ \tilde{\chi}^- \  \to \  e_i^+ e_j^- Z Z, \ \nu \nu \, W^+ W^-.$$
While in this case lepton flavor violation is observable, total lepton number cannot be probed.

\item Neutralino $(\tilde{\chi}_1^0)$ LSP:

This scenario allows for several interesting channels with lepton number violation. If the neutralino is Higgsino-like the pair production (See Section~\ref{chi.decay} for information on neutralino decays):
$$pp \  \to  \  \tilde{\chi}_1^0 \tilde{\chi}^0_1 \  \to \ 4j \ e^{\pm}_i \ e^{\pm}_j,$$
is possible through the $Z$, as well as associated production which gives rise to channels with three charged leptons:
$$pp \  \to \  \tilde{\chi}_1^0 \tilde{\chi}^{\pm}_1 \  \to \ 4j \  \nu \ e^{\pm}_i \ e^{\pm}_j \ e^{\pm}_k.$$
Unfortunately, these channels are interesting only in the Higgsino-like neutralino scenario and in general the cross sections can be small. However, striking channels with three same-sign charged leptons, multijets and no missing energy through the pair production of selectrons are generally present:
$$pp \  \to \ \tilde{e}^*_i \tilde{e}_i \  \to  \ e^{\pm}_i \ e^{\mp}_i \ e^{\mp}_j \ e^{\mp}_k \ 4j.$$
Such striking signals maybe the signatures that help test this model at the LHC. As we show above, the production cross section can be large and there are no relevant backgrounds.

\end{itemize}



\begin{thebibliography}{000}

\bibitem{MSSM1}
  P.~Fayet,
  ``Supersymmetry and Weak, Electromagnetic and Strong Interactions,''
  Phys.\ Lett.\ B {\bf 64} (1976) 159.
  
\bibitem{MSSM2}
  P.~Fayet,
  ``Spontaneously Broken Supersymmetric Theories of Weak, Electromagnetic and Strong Interactions,''
  Phys.\ Lett.\ B {\bf 69} (1977) 489.

\bibitem{MSSM3}
  S.~Dimopoulos and H.~Georgi,
  ``Softly Broken Supersymmetry and SU(5),''
  Nucl.\ Phys.\ B {\bf 193} (1981) 150.
  
\bibitem{R1}
  A.~Salam and J.~A.~Strathdee,
  ``Supersymmetry and Fermion Number Conservation,''
  Nucl.\ Phys.\ B {\bf 87} (1975) 85.
  
\bibitem{R2}
  P.~Fayet,
  ``Supergauge Invariant Extension of the Higgs Mechanism and a Model for the electron and Its Neutrino,''
  Nucl.\ Phys.\ B {\bf 90} (1975) 104.

\bibitem{Hayashi}
  M.~J.~Hayashi and A.~Murayama,
  ``Radiative Breaking of SU(2)-R x U(1)-(B-L) gauge symmetry induced by broken N=1 supersymmetry in a left-right symmetry model,''
  Phys.\ Lett.\ B {\bf 153} (1985) 251.


\bibitem{Mohapatra}
  R.~N.~Mohapatra,
  ``Mechanism For Understanding Small Neutrino Mass In Superstring Theories,''
  Phys.\ Rev.\ Lett.\  {\bf 56} (1986) 561.
  
\bibitem{Martin}
  S.~P.~Martin,
  ``Some simple criteria for gauged R-parity,''
  Phys.\ Rev.\ D {\bf 46} (1992) 2769
  [hep-ph/9207218].
  
  
\bibitem{AM}
  C.~S.~Aulakh and R.~N.~Mohapatra,
  ``Neutrino as the Supersymmetric Partner of the Majoron,''
  Phys.\ Lett.\ B {\bf 119} (1982) 136.
  
\bibitem{Martin2}
  S.~P.~Martin,
  ``Implications of supersymmetric models with natural R-parity conservation,''
  Phys.\ Rev.\ D {\bf 54} (1996) 2340
  [hep-ph/9602349].
  
\bibitem{Masiero}
  A.~Masiero and J.~W.~F.~Valle,
  ``A Model For Spontaneous R Parity Breaking,''
  Phys.\ Lett.\ B {\bf 251} (1990) 273.
  
\bibitem{Goran1}
  C.~S.~Aulakh, A.~Melfo, A.~Rasin and G.~Senjanovic,
  ``Seesaw and supersymmetry or exact R-parity,''
  Phys.\ Lett.\ B {\bf 459} (1999) 557
  [hep-ph/9902409].
 
\bibitem{Goran2}
  C.~S.~Aulakh, B.~Bajc, A.~Melfo, A.~Rasin and G.~Senjanovic,
  ``SO(10) theory of R-parity and neutrino mass,''
  Nucl.\ Phys.\ B {\bf 597} (2001) 89
  [hep-ph/0004031].
  
\bibitem{Rp1}
  P.~Fileviez Perez and S.~Spinner,
  ``Spontaneous R-Parity Breaking and Left-Right Symmetry,''
  Phys.\ Lett.\ B {\bf 673} (2009) 251
  [arXiv:0811.3424 [hep-ph]].
  
\bibitem{Rp2}
  V.~Barger, P.~Fileviez Perez and S.~Spinner,
  ``Minimal gauged U(1)(B-L) model with spontaneous R-parity violation,''
  Phys.\ Rev.\ Lett.\  {\bf 102} (2009) 181802
  [arXiv:0812.3661 [hep-ph]].
  

\bibitem{Rp5}
  P.~Fileviez Perez, M.~Gonzalez-Alonso and S.~Spinner,
  ``Gauge Origin of M-Parity and the mu-Term in Supersymmetry,''
  Phys.\ Rev.\ D {\bf 84} (2011) 095014
  [arXiv:1109.1823 [hep-ph]].
  
\bibitem{Rp6}
  D.~Feldman, P.~Fileviez Perez and P.~Nath,
  ``R-parity Conservation via the Stueckelberg Mechanism: LHC and Dark Matter Signals,''
  arXiv:1109.2901 [hep-ph].

\bibitem{Ibanez:1982fr}
L.~Alvarez-Gaume, J.~Polchinski and M.~B.~Wise,
  ``Minimal Low-Energy Supergravity,''
  Nucl.\ Phys.\  B {\bf 221}, 495 (1983);
  L.~E.~Ibanez and G.~G.~Ross,
  ``SU(2)-L X U(1) Symmetry Breaking As A Radiative Effect Of Supersymmetry
  Breaking In Guts,''
  Phys.\ Lett.\  B {\bf 110}, 215 (1982).
  
\bibitem{Ambroso:2009jd}
  M.~Ambroso and B.~Ovrut,
  ``The B-L/Electroweak Hierarchy in Heterotic String and M-Theory,''
  JHEP {\bf 0910} (2009) 011
  [arXiv:0904.4509 [hep-th]].
  
\bibitem{Ambroso:2009sc}
  M.~Ambroso and B.~A.~Ovrut,
  ``The B-L/Electroweak Hierarchy in Smooth Heterotic Compactifications,''
  Int.\ J.\ Mod.\ Phys.\ A {\bf 25} (2010) 2631
  [arXiv:0910.1129 [hep-th]].
 
\bibitem{Ambroso:2010pe}
  M.~Ambroso and B.~A.~Ovrut,
  ``The Mass Spectra, Hierarchy and Cosmology of B-L MSSM Heterotic Compactifications,''
  arXiv:1005.5392 [hep-th].
 
\bibitem{Fate}
  P.~Fileviez Perez and S.~Spinner,
  ``The Fate of R-Parity,''
  Phys.\ Rev.\ D {\bf 83} (2011) 035004
  [arXiv:1005.4930 [hep-ph]].
  
\bibitem{Carena:2004xs}
  M.~S.~Carena, A.~Daleo, B.~A.~Dobrescu and T.~M.~P.~Tait,
  ``Z-prime gauge bosons at the Tevatron,''
  Phys.\ Rev.\  D {\bf 70} (2004) 093009
  [arXiv:hep-ph/0408098].
  

\bibitem{Barger:2010iv}
  V.~Barger, P.~Fileviez Perez, S.~Spinner,
  ``Three Layers of Neutrinos,''
  Phys.\ Lett.\  {\bf B696 } (2011)  509-512.
  [arXiv:1010.4023 [hep-ph]].

\bibitem{Ghosh:2010hy}
  D.~K.~Ghosh, G.~Senjanovic and Y.~Zhang,
  ``Naturally Light Sterile Neutrinos from Theory of R-parity,''
  Phys.\ Lett.\ B {\bf 698} (2011) 420
  [arXiv:1010.3968 [hep-ph]].

\bibitem{Hamann:2010bk}
  J.~Hamann, S.~Hannestad, G.~G.~Raffelt, I.~Tamborra and Y.~Y.~Y.~Wong,
  ``Cosmology seeking friendship with sterile neutrinos,''
  Phys.\ Rev.\ Lett.\  {\bf 105} (2010) 181301
  [arXiv:1006.5276 [hep-ph]].
  
\bibitem{proton}
  P.~Nath and P.~Fileviez Perez,
  ``Proton stability in grand unified theories, in strings and in branes,''
  Phys.\ Rept.\  {\bf 441} (2007) 191
  [hep-ph/0601023].
  
\bibitem{Yuval}
  C.~Csaki, Y.~Grossman and B.~Heidenreich,
  ``MFV SUSY: A Natural Theory for R-Parity Violation,''
  arXiv:1111.1239 [hep-ph].

\bibitem{Takayama:2000uz} 
S.~Borgani, A.~Masiero and M.~Yamaguchi,
  ``Light gravitinos as mixed dark matter,''
  Phys.\ Lett.\ B {\bf 386} (1996) 189
  [hep-ph/9605222];
  F.~Takayama and M.~Yamaguchi,
  ``Gravitino dark matter without R-parity,''
  Phys.\ Lett.\ B {\bf 485}, 388 (2000)
  [hep-ph/0005214].

\bibitem{Buchmuller:2007ui} 
  W.~Buchmuller, L.~Covi, K.~Hamaguchi, A.~Ibarra and T.~Yanagida,
  ``Gravitino Dark Matter in R-Parity Breaking Vacua,''
  JHEP {\bf 0703}, 037 (2007)
  [hep-ph/0702184 [HEP-PH]];
  W.~Buchmuller,
  ``Gravitino Dark Matter,''
  AIP Conf.\ Proc.\  {\bf 1200} (2010) 155
  [arXiv:0910.1870 [hep-ph]].
  
  
  
\bibitem{arXiv:1103.0734}
  T.~Schwetz, M.~Tortola and J.~W.~F.~Valle,
  ``Global neutrino data and recent reactor fluxes: status of three-flavour oscillation parameters,''
  New J.\ Phys.\ \ {\bf 13} (2011) 063004
  [arXiv:1103.0734 [hep-ph]].

\bibitem{Porod}
  F.~Thomas and W.~Porod,
  ``Determining R-parity violating parameters from neutrino and LHC data,''
  JHEP {\bf 1110} (2011) 089
  [arXiv:1106.4658 [hep-ph]].
  
\bibitem{Bobrovskyi:2011vx}
  S.~Bobrovskyi, W.~Buchmuller, J.~Hajer and J.~Schmidt,
  ``Quasi-stable neutralinos at the LHC,''
  JHEP {\bf 1109} (2011) 119
  [arXiv:1107.0926 [hep-ph]].

\bibitem{ATLAS1}
  G.~Aad {\it et al.}  [ATLAS Collaboration],
  ``Search for anomalous production of prompt like-sign muon pairs and constraints on physics beyond the Standard Model with the ATLAS detector,''
  arXiv:1201.1091 [hep-ex].
  
\bibitem{Dawson}
  S.~Dawson, E.~Eichten and C.~Quigg,
  ``Search for Supersymmetric Particles in Hadron - Hadron Collisions,''
  Phys.\ Rev.\ D {\bf 31} (1985) 1581.
  
\bibitem{Atlas1}
	The ATLAS Collaboration
	"Search for New Phenomena in Events with Four Charged Leptons,"
	http://cdsweb.cern.ch/record/1388601/files/ATLAS-CONF-2011-144.pdf

\bibitem{Atlas2}
	The ATLAS Collaboration
	"Search for New Phenomena in Events with Three or more Charged Leptons,"
	http://cdsweb.cern.ch/record/1399618/files/ATLAS-CONF-2011-158.pdf
\end{thebibliography}
\end{document}